\documentclass{aa}  

\usepackage{graphicx}
\usepackage{tabularx}
\usepackage[dvipsnames]{xcolor}
\usepackage{tikz}
\usetikzlibrary{fit} % fitting shapes to coordinates
\usetikzlibrary{backgrounds} % drawing the background after the foreground
\usepackage{subcaption}
\usepackage{txfonts}
\usepackage{libertine}
\usepackage{gensymb}
\usepackage{xfrac}
\begin{document}

\title{Filamentary Baryons and Where to Find Them}
   %\title{Deep physical constraints on the synchrotron cosmic web - I}
  
  \subtitle{A forecast of synchrotron radiation from merger and accretion shocks in the local Cosmic Web}
  %\subtitle{A probabilistic forecast of synchrotron radiation from accretion shocks in the nearby Cosmic Web}
  %Probability distributions on the sky
   %\subtitle{The best sky regions to look for synchrotron radiation from accretion shocks in the nearby Cosmic Web}

   %\author{M.S.S.L. Oei\inst{1}\thanks{\email{oei@strw.leidenuniv.nl}}
   \author{Martijn S.S.L. Oei\inst{1}\thanks{E-mail: \textcolor{blue}{oei@strw.leidenuniv.nl}. \emph{In dearest memory of Grandma Maria, who brightened my universe with so much love and laughter.}}
   %affection --> love, loving --> dearest
   %She brought light to my universe, and will remain a beacon to guide me through today and tomorrow.}}
   %Her light shone bright in my universe - a light that will guide me through today and tomorrow.}}%\emph{In loving memory of Grandma Maria, a bright light in my universe.}} Her light shone bright in my universe; her presence is still felt today.
          \and
          Reinout J. van Weeren\inst{1}%\thanks{\email{rvweeren@strw.leidenuniv.nl}}
          \and
          Franco Vazza\inst{2,3,4}
          \and
          Florent Leclercq\inst{5}
          \and
          Akshatha Gopinath\inst{6}
          \and
          Huub J.A. R\"ottgering\inst{1}%\thanks{\email{rottgering@strw.leidenuniv.nl}}
          }

   \institute{Leiden Observatory, Leiden University, Niels Bohrweg 2, NL-2300 RA Leiden, the Netherlands
        \and
            Department of Physics and Astronomy, University of Bologna, Via Gobetti 93/2, I-40129 Bologna, Italy
        \and
            INAF, Istituto di Radioastronomia, Via Gobetti 101, I-40129 Bologna, Italy
        \and
            Hamburg Observatory, Hamburg University, Gojensbergweg 112, 21029 Hamburg, Germany
        \and
            Imperial Centre for Inference and Cosmology, Imperial College London, 1012 Blackett Laboratory, Prince Consort Road, London SW7 2AZ, United Kingdom
        \and
            Anton Pannekoek Institute for Astronomy, University of Amsterdam, Science Park 904, NL-1098 XH Amsterdam, the Netherlands}

   \date{Received February 1, 12021 H.E.; accepted March 7, 12022 H.E.}

% \abstract{}{}{}{}{} 
% 5 {} token are mandatory

%The detection of synchrotron radiation from accretion shocks onto filaments of the cosmic web constitutes a new frontier to attack the missing baryon problem, constrain the origin and evolution of extragalactic magnetic fields, and test physical models of shocks and their radiation mechanisms.
   %Pending the arrival of the Square-Kilometre Array (SKA), the Low-Frequency Array (LOFAR) may well be the telescope best-positioned to open the radio window into the magnetised cosmic web. Direct or indirect detection of synchrotron emission from accretion shocks in the warm-hot intergalactic medium (WHIM) would not only provide a new method to address the missing baryon problem, but would also allow observational research into the structure and origin of the magnetic fields that permeate filaments - the largest thought to exist in the Universe.}%: those that permeate the filaments of the cosmic web.}
   
   %The era of direct detection of radiation from the filaments of the cosmic web is imminent.
  \abstract
  {The detection of synchrotron radiation from the intergalactic medium (IGM) that pervades the filaments of the Cosmic Web, constitutes an upcoming frontier to test physical models of astrophysical shocks and their radiation mechanisms, trace the missing baryons, and constrain magnetogenesis: the origin and evolution of extragalactic magnetic fields.}
  %{The detection of synchrotron radiation from accretion shocks onto filaments of the Cosmic Web constitutes a futuristic frontier to attack the missing baryon problem, test physical models of astrophysical shocks and their radiation mechanisms, and constrain the origin and evolution of extragalactic magnetic fields.} %futuristic/new/impending/imminent?
   %{The advent of low-frequency radio observations of the magnetised cosmic is imminent.}
  %{Hitherto, no detection of synchrotron radiation has been made that can unambiguously }
  {At the moment of writing, the first synchrotron detections of the IGM within filaments have been claimed. Now is the time to develop a rigorous statistical framework to predict sky regions with the strongest signal, and to move from mere detection to inference: the identification of the most plausible physical models and parameter values from the observations.}
  %rigorously infer the most probable physical models and their parameter values from the observations.}
  % in a principled manner
  %\textcolor{red}{Hitherto, no unambiguous detection of the IGM of typical members of the filament population has been made in synchrotron light, and it is expected that sensitive observations with the world's premier low-frequency radio telescopes are necessary to obtain such signals. To aid this search, this article attempts to answer the question: where on the sky should we point our instruments to maximise the chances of discovery?}}
  %{Hitherto, there have not been any detections of representative members of the filament population in synchrotron light, and it is expected that sensitive observations with the world's premier low-frequency radio telescopes are necessary to establish unambiguous signals. Here we answer the question: where on the sky should we point our instruments to maximise the chances of discovery?}
  %, in a fully probabilistic manner, a probability distribution for the synchrotron radiation from accretion shocks} We describe a novel approach to generating
  %The current leading theory posits that the filament IGM lights up predominantly through shocks that originate from large-scale structure formation.
  {Current theory posits that the filament IGM lights up through shocks that originate from large-scale structure formation.
  With Bayesian inference, we generate a probability distribution on the set of specific intensity functions that represent our view of the merger- and accretion-shocked synchrotron Cosmic Web (MASSCW).
  We combine the BORG SDSS total matter density posterior, which is based on spectroscopic observations of galaxies within SDSS DR7, snapshots of Enzo MHD cosmological simulations, Gaussian random fields (GRFs) and a ray tracing approach to arrive at the result.}
  %the specific intensity nearby Cosmic Web's accretion shock synchrotron }
   %{We search for a signal from the magnetised cosmic web via synchrotron emission in deep low-frequency interferometric radio observations, and present limits on the gas density and magnetic field strength of the filaments of the cosmic web.}
   %{From 100 hours of \textit{LOFAR Deep Fields: Lockman Hole} observations, we construct the most sensitive residual map yet of the 150 MHz sky at 20 arcsecond resolution, revealing unidentified diffuse emission. Secondly, using galactic redshifts from the Sloan Digital Sky Survey (SDSS) and state-of-the-art Enzo magnetohydrodynamics (MHD) simulations, we construct a physically-informed prediction of the synchrotron cosmic web for the Northern Galactic Cap (NGC). Next, we inject LOFAR observational signatures into the Lockman Hole prediction, and correlate the resulting predictive map with the residual map. From a statistical analysis, we constrain the properties of the filaments of the cosmic web.}
   %The prior thus obtained provides a physics-based prediction of the MASSCW signal
   %It can be used to identify the most promising targets in the Local Universe
   %that is informative over; total sky
   {We present a physics-based prediction of the MASSCW signal, including principled uncertainty quantification, for a quarter of the sky and up to cosmological redshift $z_\mathrm{max} = 0.2$.
   The super-Mpc 3D resolution of the current implementation limits the resolution of the predicted 2D imagery, so that individual merger and accretion shocks are \emph{not} resolved.
   The MASSCW prior can be used to identify the most promising fields to target with low-frequency radio telescopes, and to conduct actual detection experiments.
   We furthermore calculate a probability distribution for the flux-density-weighted mean (i.e. sky-averaged) redshift $\bar{\bar{z}}$ of the MASSCW signal up to $z_\mathrm{max}$, and find a median of $\bar{\bar{z}} = 0.077$.
   We construct a low-parametric analytic model that produces a similar distribution for $\bar{\bar{z}}$, with a median of $\bar{\bar{z}} = 0.072$.
   Extrapolating the model, we can calculate $\bar{\bar{z}}$ for \emph{all} large-scale structure in the Universe (including what lies beyond $z_\mathrm{max}$) and show that, if one only considers filaments, $\bar{\bar{z}}$ depends on virtually one parameter.
   %Extrapolating the model, we can calculate $\bar{\bar{z}}$ for \emph{all} large-scale structure in the Universe (including what lies beyond $z_\mathrm{max}$) and show that, in case only filaments are considered, $\bar{\bar{z}}$ depends on virtually one parameter.
   %\textcolor{red}{We furthermore calculate the redshift properties of the predicted signal, and derive an insightful parametric model that reproduces these findings.
   %We show that the filament-only $\bar{\bar{z}}$ for future reconstructions that incorporate \emph{all} large-scale structure (including what lies beyond $z > 0.2$), or equivalently, for the \emph{actual} Universe, is a function of just one influential parameter.}
   %Its value is hitherto uncertain but determines the 
   %By extrapolation, this model can predict $\bar{\bar{z}}$ for the actual Universe too, 
   %We furthermore calculate the redshift distribution of the predicted signal, whose mean is $\bar{\bar{z}} \sim 0.07$, and derive a three-parameter geometric model that accurately reproduces this finding.
   %By extrapolation, this model predicts $\bar{\bar{z}} \sim 0.13$ for the actual Universe, or equivalently, for future reconstructions that incorporate \emph{all} large-scale structure (including what lies beyond $z > 0.2$).
   %This model predicts that future reconstructions that incorporate \emph{all} large-scale structure (including what lies beyond $z > 0.2$) have $\bar{z} \sim 0.13$.
   %analytical or geometric? Analytical maybe more pretentious, geometric more precise/descriptive
   As case studies, we finally explore the predictions of our MASSCW specific intensity function prior in the vicinity of three galaxy clusters: the Hercules Cluster, the Coma Cluster, and Abell 2199; and in three deep LOFAR HBA fields: the Lockman Hole, Abell 2255, and the Ursa Major Supercluster.} %We furthermore provide a lower limit for the sky-filling fraction of the ASS CW specific intensity caused by shocks at a given cosmological redshift and above.
   %{We put upper limits unto the mean filamentary magnetic field strength of ... .}
   %{We provide the deepest constrains yet on the properties of the magnetised cosmic web.}
   %framework for
   {We describe and implement a novel, flexible and principled framework for predicting the low-frequency, low-resolution specific intensity function of the Cosmic Web due to merger and accretion shocks that arise during large-scale structure formation.
   The predictions guide Local Universe searches for filamentary baryons through half of the Northern Sky.
   %We foresee framework extensions that predict 
   Once cosmological simulations of alternative emission mechanisms have matured, our approach can be extended to predict additional physical pathways that contribute to the elusive synchrotron Cosmic Web signal.}
   %Framework extensions that allow for formal model selection are foreseen once cosmological simulations of alternative emission mechanisms have matured.}

   %\keywords{large-scale structure -- Cosmic Web -- clusters -- filaments -- intergalactic medium -- intra-cluster medium -- warm-hot intergalactic medium -- extragalactic magnetic fields -- merger shocks -- accretion shocks -- diffusive shock acceleration -- synchrotron radiation -- low-frequency radio astronomy -- observational cosmology -- LOFAR -- BORG SDSS -- Enzo MHD cosmological simulations}

   \maketitle

\section{Introduction}
\subsection{The Cosmic Web}
Just after inflation, the Universe’s dark and baryonic matter density functions resembled realisations of nearly constant, isotropic and stationary Gaussian random fields \textcolor{blue}{\citep{Linde12008}}.
By the influence of gravity alone, these fields evolved into the highly inhomogeneous, network-like large-scale structure (LSS) present today \textcolor{blue}{\citep{Springel12005}}.
%Gravitational collapse of dark and baryonic matter transformed the nearly constant Gaussian random density fields as emerged from inflation into the starkly inhomogeneous, intricate morphology present today.
The late-time Universe consists of two components: voids, and the \emph{Cosmic Web}, which can be further partitioned into sheets, filaments and (galaxy) clusters.\footnote{Some authors use `Cosmic Web' to refer exclusively to filaments, but in this work, the term is used to refer to \emph{all} of the late-time Universe excluding the voids.
We furthermore differentiate between `cosmic web' (for the concept in an arbitrary universe), and `Cosmic Web' (for the concept in ours).}
%(Because the Cosmic Web is a subset of spacetime and therefore a topographic entity, we choose to capitalise the term, breaking with common practice.)}
The initial density conditions, through the morphology of the Cosmic Web, determine the spatial distribution of galaxies and some of their internal properties, such as magnetic field and spin.
%The initial conditions determine the morphology of the Cosmic Web, which in turn sets the spatial distribution of galaxies and some internal properties such as magnetic field and spin.
%The spatial distribution of galaxies, and some internal properties such as magnetic field and spin, are determined by the initial conditions
%This \emph{Cosmic Web}, which is usually partitioned into (galaxy) clusters, filaments, sheets and voids, determines the spatial distribution of galaxies, and even some of their internal properties such as magnetic field and spin.
%\footnote{Some authors use `Cosmic Web' to refer exclusively to filaments, but in this work (following \textcolor{blue}{\citet{Hahn12007}}), the term is used as a synonym for \emph{all} of the late-time Universe. (Because the Cosmic Web is a subset of spacetime and therefore a topographic entity, we choose to capitalise the term, breaking with common practice.)}
%This \emph{Cosmic Web}, which is usually classified - and thus partitioned - into (galaxy) clusters, filaments, sheets and voids, dictates the spatial distribution of galaxies, and even some of their internal properties such as magnetic field and spin.\footnote{Some authors use `Cosmic Web' to refer exclusively to filaments, but in this work (following \textcolor{blue}{\citet{Hahn12007}}), the term is used as a synonym for \textit{late-time Universe}. As this (strict) subset of all spacetime clearly forms a topographic entity, we choose to capitalise the term, breaking with common practice.}
\textcolor{blue}{\citet{Hahn12007}}, inspired by the seminal work of \textcolor{blue}{\citet{Zeldovich11970}}, have provided a\footnote{Many other cosmic web classifiers exist, such as DIVA \textcolor{blue}{\citep{Lavaux12010}}, V-web \textcolor{blue}{\citep{Hoffman12012}}, and LICH \textcolor{blue}{\citep{Leclercq12017}}.} rigorous definition (known as the `T-web') of the four canonical structure types (voids, sheets, filaments and clusters) based on the number of positive eigenvalues (0, 1, 2 or 3, respectively) of the tidal field tensor.
%Either "tidal field tensor", or "gravitational potential's Hessian".
%Furthermore, they show via $N$-body simulations that the prevalence of the structure types (quantified by e.g. their volume filling factors) has changed continuously over time, with clusters and voids growing in prominence at the expense of sheets and filaments at the present day.\\
Their $N$-body simulations reveal that the prevalence of the structure types (quantified by e.g. volume-filling factors) is evolving, with filaments and sheets disappearing in favour of clusters and voids at the present day.\footnote{\textcolor{blue}{\citet{Forero-Romero12009}} subsequently refined Hahn's definition, by counting eigenvalues above a tuneable threshold (rather than zero) related to the gravitational collapse timescale. The partitioning of the cosmic web into the four structure types depends sensitively on the choice of this threshold, and it should therefore always be mentioned when quantitative structure-type properties are stated, such as volume- and mass-filling fractions.}
%Furthermore, they show via $N$-body simulations that the prevalence of the structure types (quantified by e.g. their volume-filling factors) is evolving, with filaments and sheets disappearing in favour of clusters and voids at the present day.\footnote{\textcolor{blue}{\citet{Forero-Romero12009}} subsequently refined Hahn's definition, by counting eigenvalues above a tuneable threshold (rather than zero) related to the gravitational collapse timescale. The partitioning of the Cosmic Web into the four structure types depends sensitively on the choice of this threshold, and it should therefore always be mentioned when quantitative structure-type properties are stated, such as volume- and mass-filling fractions.}
%The prevalence of each structure type (quantified by e.g. its volume filling factor) has changed continuously over time, with sheets and filaments disappearing in favour of clusters and voids at the present day \textcolor{blue}{\citep{Hahn12007}}.\\
%has been evolving continuously over the Universe's lifetime, and will continue to do so in the future as galaxy clusters assemble further.\\
\subsection{Baryons in the ICM and WHIM}
Of the four structure types, galaxy clusters are most easily studied: in the X-ray, optical and radio bands, e.g. through thermal Bremsstrahlung, gravitational lensing and synchrotron radiation.
As the most massive gravitationally bound structures in the Universe thus far, they weigh up to ${\sim}10^{15}\ M_\odot$, contain up to ${\sim}10^3$ galaxies, and are pervaded by a dilute (${\sim}10^3\ \mathrm{m}^{-3}$), hot (${\sim}10^8\ \mathrm{K}$) and magnetised (${\sim}1\ \mathrm{\mu G}$) hydrogen- and helium-dominated plasma: the intra-cluster medium (ICM) \textcolor{blue}{\citep{Cavaliere12011}}.\\
%The ICM can been studied in the X-ray (via thermal Bremsstrahlung), optical (via gravitational lensing) and radio (via synchrotron radiation) bands.\\
%The ICM can be studied in the X-ray, optical and radio bands, via thermal Bremsstrahlung, gravitational lensing and synchrotron radiation, respectively.\\
At the outskirts of clusters, the ICM transitions into the warm--hot intergalactic medium (WHIM) --- the dominant baryonic constituent of filaments.
The WHIM is a plasma of nearly primordial chemical composition, but is less dense (${\sim}1 - 10\ \mathrm{m}^{-3}$), cooler (${\sim}10^5-10^7\ \mathrm{K}$) and less magnetised (${\sim}10^{-3}-10^{-1}\ \mathrm{\mu G}$) than the ICM.
Compared to clusters, filaments are therefore harder to detect in all three wavelength bands.
Despite this, filaments are cosmologically relevant, as simulations predict that the WHIM harbours up to $\sim 90\%$ of the Universe's baryons \textcolor{blue}{\citep{Cen11999, Eckert12015, deGraaff12019, Tanimura12019}}.%(e.g.  \textcolor{blue}{\citet{Cen11999}}, \text and \textcolor{blue}{\citet{deGraaff12019}}).
%Whereas galaxy clusters are detected routinely by telescopes across the electromagnetic spectrum, the filaments connecting them have remained mostly elusive.
%Filaments are pervaded by the warm-hot intergalactic medium (WHIM), a dilute ($1$ - $10\ \mathrm{m}^{-3}$) magnetised ($10^{-11}$ - $10^{-10}\ \mathrm{T}$) plasma ($10^5$ - $10^7\ \mathrm{K}$) of nearly primordial chemical composition (hydrogen and helium).
%Cosmological simulations predict that the WHIM harbours half of the Universe's baryons.\\
\subsection{Synchrotron radiation from merger and accretion shocks}
The formation of filaments (and, indirectly, the galaxy clusters they fuel) has occurred primarily through the influx of dark matter (DM) and cold gas from sheets and voids.
% baryonic formation?
Once these free-falling pockets of gas reach a filament's surface, they generate supersonic accretion shocks, with upstream Mach numbers $\mathcal{M}_\mathrm{u} \sim 10^0 - 10^2$ \textcolor{blue}{\citep{Ryu12003}}.
%The model proposed in \textcolor{blue}{\citet{Hoeft12007}} to explain merger shocks in clusters, 
Following \textcolor{blue}{\citet{Ensslin11998, Miniati12001}}, \textcolor{blue}{\citet{Hoeft12007}} have proposed that \emph{merger} shocks in \emph{clusters} boost electrons in the high-energy tail of the ICM's Maxwell--Boltzmann distribution to ultrarelativistic velocities via diffusive shock acceleration (DSA).
By extension, the \textcolor{blue}{\citet{Hoeft12007}} model could also describe (again via DSA) how \emph{accretion} shocks in \emph{filaments} boost electrons in the high-energy tail of the WHIM's Maxwell--Boltzmann distribution to ultrarelativistic velocities.
%By analogy, the \textcolor{blue}{\citet{Hoeft12007}} model is also often taken to explain how \emph{accretion} shocks in \emph{filaments} boost electrons in the high-energy tail of the WHIM's Maxwell--Boltzmann distribution to ultrarelativistic velocities via DSA.
%, and by extension accretion shocks in filaments, then boost the electrons in the high-energy tail of the WHIM's Maxwell--Boltzmann distribution to ultrarelativistic velocities via diffusive shock acceleration (DSA).
%\textcolor{blue}{\citet{Hoeft12007}} have proposed that these shocks then boost the electrons in the high-energy tail of the WHIM's Maxwell--Boltzmann distribution to ultrarelativistic velocities via diffusive shock acceleration (DSA).
The DSA mechanism details how charges diffuse back and forth across the shock front, trapped in a magnetic mirror, and gain speed accordingly \textcolor{blue}{\citep{Krymskii11977, Axford11977, Bell11978_1, Bell11978_2, Blandford11978, Drury11983, Blandford11987, Jones11991, Baring11997, Malkov12001}}.
%The theory of diffusive shock acceleration (DSA), which has been largely successful in explaining shock emission in galaxy clusters \textcolor{blue}{\citep{vanWeeren12019}}, posits that the shock surface elevates the pre-existent population of suprathermal electrons to ultrarelativistic velocities by trapping them in a magnetic mirror \textcolor{blue}{\citep{Hoeft12007}}.
Once released, these high-energy electrons subsequently spiral along the magnetic field lines of the intergalactic medium (IGM), glowing in synchrotron light.
It is thought that this accretion-shock-based radiation mechanism provides the dominant contribution to the filaments' synchrotron Cosmic Web (SCW) signal.\\
Although the Mach numbers of accretion shocks in filaments are higher than those of merger shocks in clusters (where they are $\mathcal{M}_\mathrm{u} \sim 1 - 5$) \textcolor{blue}{\citep{Ryu12003}}, shocks in filaments remain fainter due to the aforementioned adverse density, temperature and magnetic field strength conditions.\footnote{Precisely how much fainter shocks in filaments are compared to those in clusters, is unknown, because the typical filament IGM magnetic field strength $B_\mathrm{IGM}$ and the electron acceleration efficiency $\xi_e$ of shocks --- and especially weak ones --- is highly uncertain.}
This is why observing filaments through synchrotron radiation constitutes an ambitious, futuristic frontier.%In observational studies of large-scale structure, then, filaments and sheets are the next frontier. is challenging, and why - if radio astronomy wants to advance cosmology through observations of large-scale structure -
%However, studying the filaments of the cosmic web, has proven to be more challenging.
%This illustrates why observing the WHIM through the radio window is challenging, and why - if we want to advance cosmology through observations of large-scale structure - filaments and sheets constitute the next frontier.\\
\subsection{Scientific prospects of detecting synchrotron radiation from the filament IGM}
\subsubsection{Physics of astrophysical shocks}
\begin{figure}
    \centering
    \includegraphics[width=\columnwidth]{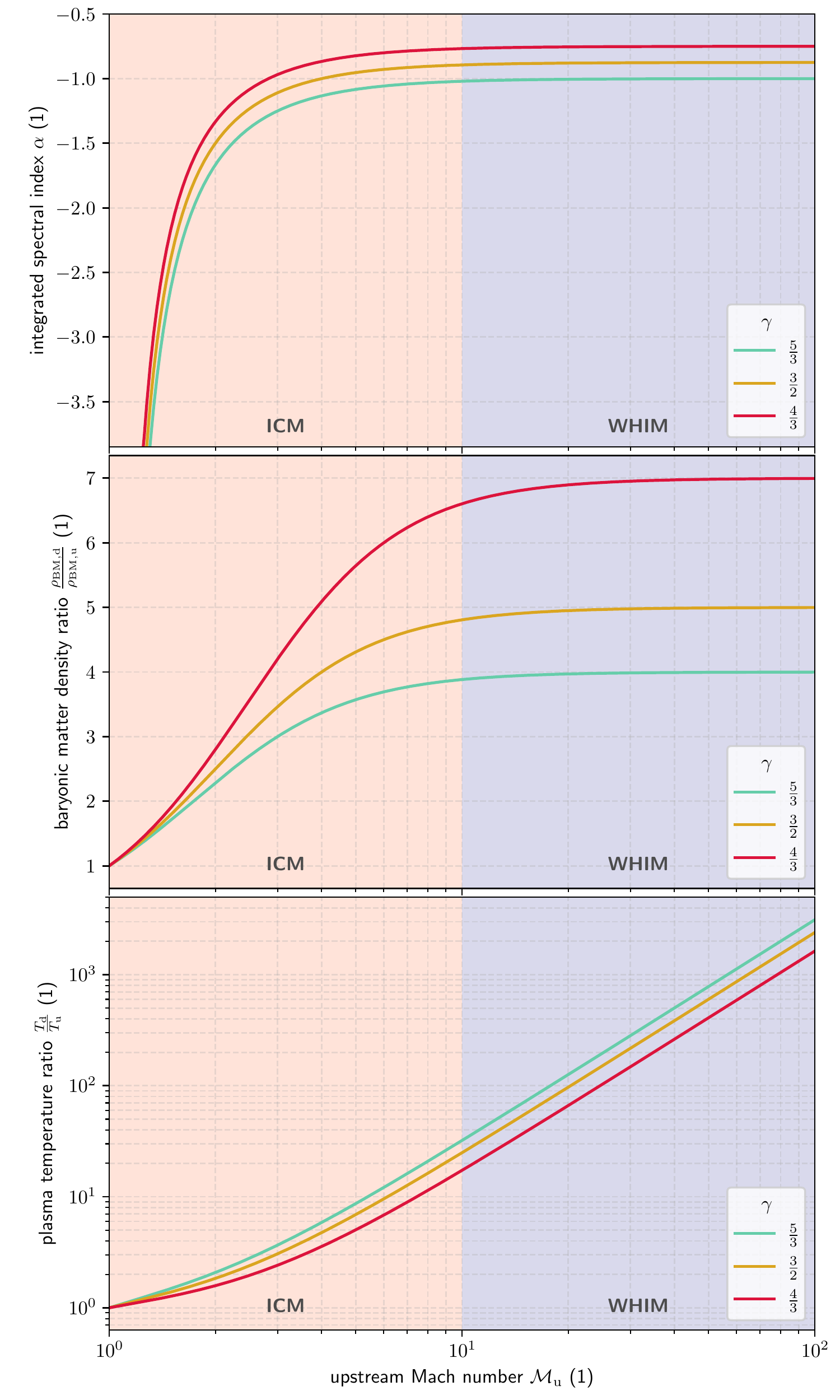}
    \vspace{-20pt}
    \caption{The dependence of three central (merger or accretion) shock quantities on upstream Mach number $\mathcal{M}_\mathrm{u}$ and adiabatic index $\gamma$, as derived from the Rankine--Hugoniot jump conditions in ideal gases.
    Radiation from $\mathcal{M}_\mathrm{u} < 10$ shocks is predominantly generated by electrons that stem from the ICM, whilst radiation from $\mathcal{M}_\mathrm{u} > 10$ shocks is predominantly generated by electrons that stem from the WHIM.
    %We differentiate between broad ICM and WHIM Mach number regimes.
    \textbf{Top:} the spectral index $\alpha$ of angularly unresolved shocks, assuming standard DSA.
    \textbf{Middle:} the downstream-over-upstream plasma density ratio.
    \textbf{Bottom:} the downstream-over-upstream plasma temperature ratio.}
    \label{fig:spectralIndex}
\end{figure}
Modern radio telescopes, such as the upgraded Giant Metrewave Radio Telescope (uGMRT), the Expanded Very Large Array (EVLA) and the Low-frequency Array High-band Antennae (LOFAR HBA), have enabled detailed studies of particle acceleration in the cluster IGM (e.g. \textcolor{blue}{\citet{diGennaro12018, Kale12020, Locatelli12020, Mandal12020}}) by the detection of radio halos, phoenices and relics.
%\textcolor{blue}{\citet{Kale12020}}, \textcolor{blue}{\citet{Locatelli12020}} and \textcolor{blue}{\citet{Mandal12020}}
In contrast, no single shock in filaments has hitherto been observed.
Doing so would open up density, temperature, Mach number and magnetic field strength regimes different by orders of magnitude via which astrophysical shock models could be held to the test.
For example, the top panel of \textbf{Figure}~\ref{fig:spectralIndex} shows that DSA predicts an (angularly unresolved) synchrotron spectral index $\alpha = -1$ for virtually all filament shocks (with a slight dependency on the adiabatic index $\gamma$).
Significant deviations from $\alpha = -1$ would falsify standard DSA.\footnote{\textcolor{blue}{\citet{Caprioli12019}} describe recent advances in DSA theory, which suggest steeper electron energy and synchrotron spectra for the filaments' strong shocks (i.e. $\alpha < -1$ for $\gamma = \sfrac{5}{3}$ and $\mathcal{M}_\mathrm{u} > 10$).}
A better understanding of astrophysical shocks has ramifications beyond the study of large-scale structure, as possibly similar shocks are found in, amongst other places, accreting X-ray binaries, stellar and pulsar winds, and supernova remnants.
\subsubsection{Missing baryons}
Routine detection of the filament IGM in synchrotron would provide a novel way to address the missing baryon problem --- the possible discrepancy between today's mean baryon density as inferred from galaxy surveys versus that predicted by CMB-constrained $\Lambda$CDM models.\\
Direct imaging of the filament IGM in the low-frequency radio window adds spectral diversity to a growing list of methodologies that trace the WHIM, complementing X-ray observations of ionised oxygen (O VII) absorption along the line of sight to quasars \textcolor{blue}{\citep{Nicastro12018}}, microwave measurements of the thermal Sunyaev--Zel'dovich effect due to hot gas between adjacent galaxies \textcolor{blue}{\citep{deGraaff12019}}, millimetre searches for the hyperfine spin-flip transition of single-electron nitrogen ions (N VII) \textcolor{blue}{\citep{Bregman12007}} and dispersion measurements of localised fast radio bursts (FRBs) \textcolor{blue}{\citep{Macquart12020}}.
Just as direct imaging of the WHIM in the X-ray band \textcolor{blue}{\citep{Eckert12015}}, detecting baryons through synchrotron emission does not necessitate a special (line-of-sight) geometry.
%, nor a (local) abundance of pairs of adjacent galaxies.
%This method would add the low-frequency radio window to the missing baryon problem toolkit 
%changed change geometry to special geometry
Telescopes such as the LOFAR could thus corroborate the current baryon census, in which still ${\sim}34\%$ is not identified conclusively \textcolor{blue}{\citep{deGraaff12019}}.
\begin{figure}[t!]
    \centering
    \includegraphics[width=\columnwidth]{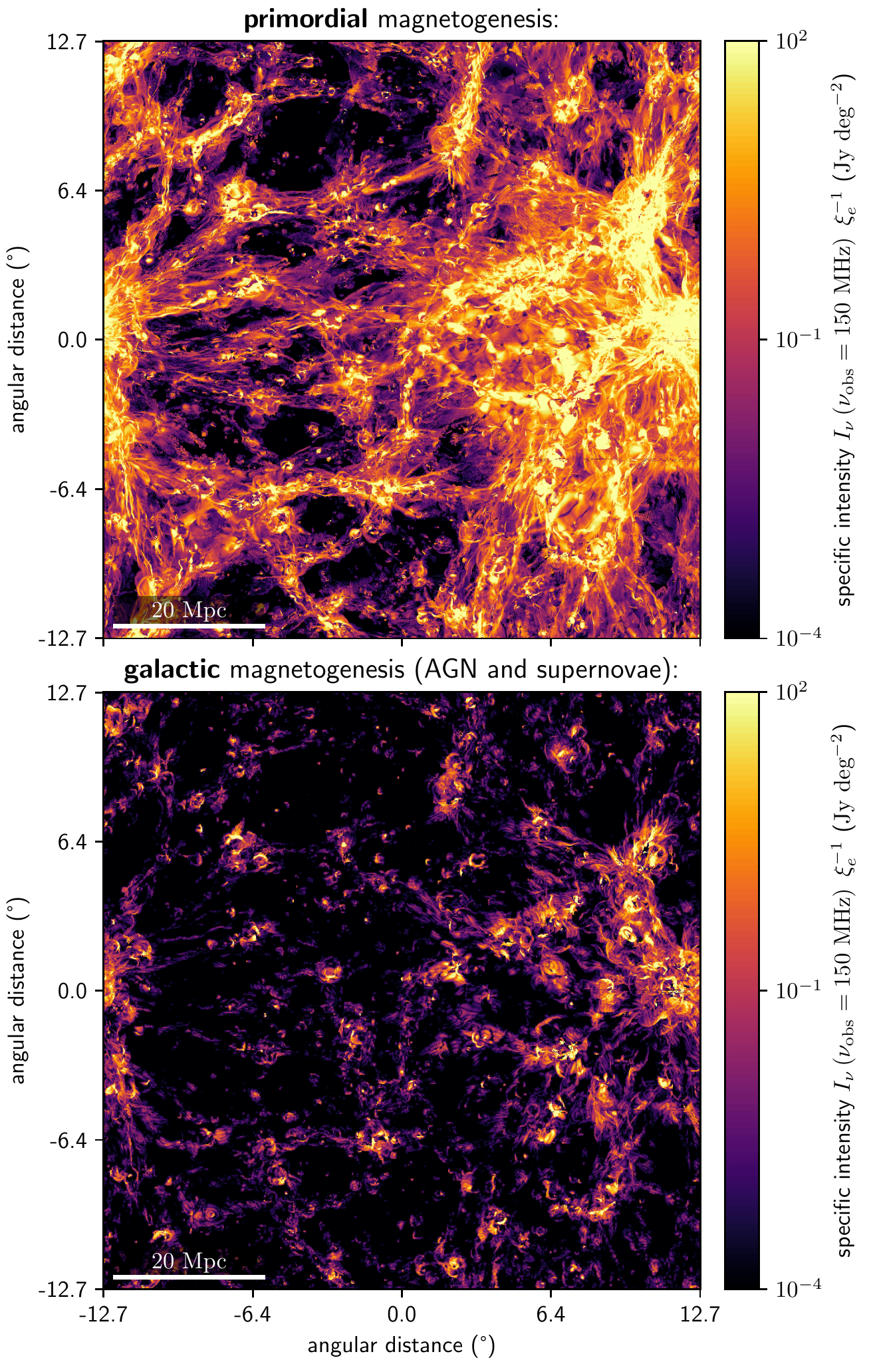}
    \vspace{-10pt}
    \caption{Two simulated specific intensity functions at $\nu_\mathrm{obs} = 150\ \mathrm{MHz}$, assuming synchrotron radiation from merger and accretion shocks (as in \textcolor{blue}{\citet{Hoeft12007}}) in LSS at $z = 0.045$. \textbf{Top:} primordial scenario for magnetogenesis, starting from $B_\mathrm{IGM} = 1\ \mathrm{nG}$ in the Early Universe. (In contrast, the simulation underlying this article's predictions starts at $B_\mathrm{IGM} = 0.1\ \mathrm{nG}$.) \textbf{Bottom:} galactic scenario for magnetogenesis, in which seeding occurs through AGN outflows and supernova winds. See \textcolor{blue}{\citet{Gheller12020}} for details on these scenarios.}
    %\caption{Two predictions for radio emission from filaments at low redshifts, under different scenarios of magnetogenesis, from \textcolor{blue}{\citet{Vazza12015b}}. In the primordial model, magnetic fields smoothly trace the cosmic web, whilst the astrophysical model predicts clumpy peaks around galaxies.}
    \label{fig:magneticFields}
\end{figure}
\subsubsection{Magnetogenesis}
\label{sec:magnetogenesis}
%Moreover
Thirdly, the low galaxy number density in filaments means that its IGM, if largely untouched by galactic feedback, can retain information on its initial conditions for billions of years.
In particular, MHD simulations \textcolor{blue}{\citep{Vazza12015a, Vazza12017}} demonstrate that different assumptions for the dominant physical process that drove the growth of magnetism in filaments, lead to different strengths and morphologies of the IGM's magnetic fields today.
% \begin{figure*}[t!]
%     \centering
%     \includegraphics[width=\textwidth]{BORGSDSSRayTracingOverview43Power23:6.png}
%     \vspace{-10pt}
%     \caption{Prediction for the low-frequency specific intensity function for filamentary accretion shocks, derived by ray tracing through the BORG SDSS posterior mean. Three massive nearby galaxy clusters are shown, alongside the 3C196 EoR field, the Lockman Hole, Bo\"otes and ELAIS-N1, of which the Lockman Hole is the brightest. We demonstrate that the HETDEX field is expected to contain the brightest nearby cosmic web of the SDSS DR7 sky.}
%     \label{fig:BORGSDSSFlat}
% \end{figure*}
The authors evaluate three different scenarios for cosmic magnetogenesis: primordial, dynamo, and galactic models, and calculate their evolution from a cosmological redshift of 38 to the present day.
The models are calibrated by magnetic field strength measurements of the ICM.
In primordial models, seed field fluctuations grow as LSS formation compresses and rarefies the magnetic field lines.
In dynamo models, the seed field is much weaker ($10^{-22}\ \mathrm{T}$ vs. $10^{-13}\ \mathrm{T}$), but grows in strength as energy in solenoidal, turbulent gas motion is converted into magnetic energy through small-scale dynamos \textcolor{blue}{\citep{Ryu12008}}.
%kinetic energy in solenoidal
%within the emerging Cosmic Web
In galactic (or `astrophysical') models, finally, no seed field is assumed, with star formation and outflows from the jets of supermassive black holes (SMBHs) at galactic centres being the dominant contributors to the emergence of the magnetised cosmic web.
%The authors survey three different scenarios for cosmic magnetogenesis: primordial, dynamo, and astrophysical models, and calculated the evolution from a redshift of 38 to the present day.
%In the primordial (family of) models, the magnetic field is initialised as a $10^{-13}\ \mathrm{T}$ seed field, and changes as large-scale structure (LSS) formation compresses and rarefies the magnetic field.
%In the dynamo (family of) models, the seed field is much weaker ($10^{-22}\ \mathrm{T}$), but grows in strength as kinetic energy in solenoidal, turbulent gas motion within the emerging Cosmic Web is converted into magnetic energy.
%In the astrophysical (family of) models, finally, no seed field is assumed, with star formation and outflows from the jets of supermassive black holes (SMBHs) at the centres of galaxies being the dominant contributors to the emergence of the magnetised Cosmic Web.
\textbf{Figure}~\ref{fig:magneticFields} illustrates that different magnetisation scenarios give rise to morphologically different low-frequency specific intensity functions.
%different WHIM magnetic field morphologies, and therefore
Thus, SCW observations could rule out at least some models of magnetogenesis --- a puzzle widely considered to be amongst the most significant open problems in cosmology.\\
Hitherto, several groups have constrained the magnetic field strength of the filament IGM by means of Stokes I low-frequency radio observations.
Using Murchison Widefield Array (MWA) Epoch of Reionisation Field 0 data and the equipartition energy condition, \textcolor{blue}{\citet{Vernstrom12017}} derived a parameter-dependent upper limit of $B_\mathrm{IGM} < 0.03 - 1.98\ \mathrm{\mu G}$, while \textcolor{blue}{\citet{Brown12017}} used Parkes 64m Telescope S-PASS data to find a density-weighted upper limit of $B_\mathrm{IGM} < 0.13\ \mathrm{\mu G}$ at the present epoch.
Finally, using MWA GLEAM and ROSAT RASS data, \textcolor{blue}{\citet{Vernstrom12021}} suggest an average magnetic field strength of $30\ \mathrm{nG} < B_\mathrm{IGM} < 60\ \mathrm{nG}$, using both equipartition and inverse Compton arguments.
%Hitherto, a handful of searches for the SCW have been performed.
%The first two, by \textcolor{blue}{\citet{Vernstrom12017}} and \textcolor{blue}{\citet{Brown12017}}, used MWA data.
%The first \emph{LOFAR} search of the SCW was initiated by R.J. van Weeren and H.T. Jense in 12018 H.E., following the strategy of \textcolor{blue}{\citet{Vernstrom12017}}.
In an upcoming publication \textcolor{blue}{(Oei et al., in prep.)}, we describe and present a LOFAR search, that includes new methodology.\\
Alternatively, \textcolor{blue}{\citet{O'Sullivan12019, O'Sullivan12020, Stuardi12020}} constrain the properties of the IGM's magnetic fields via rotation measure (RM) synthesis applied to LOFAR Two-metre Sky Survey (LoTSS) observations of the lobes of (giant) radio galaxies, finding $B_\mathrm{IGM} < 4\ \mathrm{nG}$ in filaments.
\textcolor{blue}{\citet{Vernstrom12019}} have performed a similar analysis with NVSS data, finding $B_\mathrm{IGM} < 40\ \mathrm{nG}$ in filaments.
\subsection{Article structure}
The goal of this article is, first and foremost, to explain a new method for SCW prediction, that yields a probability distribution over the set $\mathrm{Map}\left(\mathbb{S}^2, \mathbb{R}_{\geq 0}\right)$ of specific intensity functions on the 2-sphere $\mathbb{S}^2$.
The secondary goal is to demonstrate the method's potential, by showing results for the modern, nearby Universe ($z < 0.2$) over half of the Northern Hemisphere --- $25\%$ of the full sky.\\
In \textbf{Section}~\ref{sec:methods}, we first provide a general outline of our SCW signal prediction method.
We then detail the methodology step-by-step, providing a background to the input data as we proceed.
In \textbf{Section}~\ref{sec:model}, we develop a simple geometric model of the cosmic web that yields SCW redshift predictions.
In \textbf{Section}~\ref{sec:results}, we analyse predictions of the main method, and compare these with predictions of the geometric model.
Finally, in \textbf{Section}~\ref{sec:discussion}, we discuss caveats of the current work, and give recommendations for future extensions, before we present conclusions in \textbf{Section}~\ref{sec:conclusions}.\\
%give recommendations for future observational campaigns,
We adopt a\footnote{Rather fortunately, the BORG SDSS and Enzo data products combined in this work have been made assuming almost identical cosmological parameters; the set reported is of the BORG SDSS. Authors of future work who strive to achieve \emph{full} self-consistency should use the same cosmological model for the LSS reconstructions as for the MHD simulations.} concordance inflationary $\Lambda$CDM cosmological model $\mathfrak{M} = (\Omega_{\Lambda,0} = 0.728, \Omega_{\mathrm{DM},0} = 0.227,$ $ \Omega_{\mathrm{BM},0} = 0.045, h = 0.702, \sigma_8 = 0.807, n_s = 0.961)$, so that $\Omega_{\mathrm{M},0} \coloneqq \Omega_{\mathrm{DM},0} + \Omega_{\mathrm{BM},0} = 0.272$ and the Hubble constant $H_0 \coloneqq h \cdot 100\ \mathrm{km\ s^{-1}\ Mpc^{-1}}$ \textcolor{blue}{\citep{Jasche12015}}.

\section{Methods}
\label{sec:methods}
\subsection{Overview}
For structural clarity, we begin with an overview of our synchrotron Cosmic Web prediction approach.
\begin{enumerate}
\item Our starting point is a probability distribution over the total (i.e. baryonic and dark) matter density fields of the modern, nearby Universe.
\item Then, using a cosmological MHD simulation, we establish the connection between the total matter density $\rho$ and the merger and accretion shock synchrotron monochromatic emission coefficient (MEC) $j_\nu$, in the form of the conditional probability distribution $P\left(j_\nu\left(\mathbf{r}\right)\ \vert\ \rho\left(\mathbf{r}\right)\right)$.
\item Next, to create a MEC field, one could independently realise the random variables (RVs) $j_\nu\left(\mathbf{r}\right)\ \vert\ \rho\left(\mathbf{r}\right)$ for all $\mathbf{r}$ in the volume of the inferred density fields.
However, such an approach would disregard the fact that in reality, adjacent locations have similar physical conditions and thus similar MECs (i.e. the MEC field exhibits spatial correlation).
To account for spatial correlation, we generate realisations of an isotropic, stationary, three-dimensional Gaussian random field (GRF) with zero mean and unit variance.
By applying the cumulative density function (CDF) of the standard normal RV in point-wise fashion, the GRF transforms into a `percentile random field' --- with values strictly between 0 and 1.
Note that this new field inherits the spatial correlation present in the GRF.
We then let the percentile random field determine the MEC field, by plugging the percentile scores into the inverse CDF of $j_\nu\left(\mathbf{r}\right)\ \vert\ \rho = \rho\left(\mathbf{r}\right)$ for all $\mathbf{r}$ in the volume. %We then let the percentile random field determine the emissivities, by demanding its values become the percentile scores of the realisations of $j_\nu\left(\mathbf{r}\right)\ \vert\ \rho_\mathrm{DM} = \rho_\mathrm{DM}\left(\mathbf{r}\right)$. %The percentile random field determines the emissivities, by demanding its values are the percentile scores of the realisations of $j_\nu\left(\mathbf{r}\right)\ \vert\ \rho_\mathrm{DM} = \rho_\mathrm{DM}\left(\mathbf{r}\right)$.
%\item After having obtained a merger and accretion shock synchrotron MEC field with spatial correlation, we ray trace through this three-dimensional structure, taking into account cosmological effects.
\item After having obtained a merger and accretion shock synchrotron MEC field with spatial correlation, we ray trace through this three-dimensional structure, taking into account effects due to the Universe's expansion.
This, finally, yields a merger and accretion shock synchrotron Cosmic Web (MASSCW) specific intensity function on the sky.
\item By repeating the above steps over and over, picking different density field and GRF realisations every time, we generate a probability distribution on $\mathrm{Map}\left(\mathbb{S}^2, \mathbb{R}_{\geq 0}\right)$ for the MASSCW signal.
\item With a minor modification, this procedure naturally leads to another probability distribution on $\mathrm{Map}\left(\mathbb{S}^2, \mathbb{R}_{\geq 0}\right)$, which captures the direction-dependent mean cosmological redshift of the MASSCW.
%modern, Earth-centred --> observable --> ...
\end{enumerate}

\subsection{Modern-day total matter density field posterior}
\subsubsection{A brief history of Bayesian large-scale structure reconstruction}
A decade of research in large-scale structure reconstruction from galaxy surveys has culminated in a suite of highly principled, physics-based Bayesian inference techniques that unveil the content, dynamics and history of the local Cosmic Web.
These techniques represent the state-of-the-art of LSS reconstruction, and their applications are manifold (see e.g. \textcolor{blue}{\citet{Jasche12019}}).
The application relevant to our purposes is the reconstruction of the modern-day (i.e. $z=0$) total (i.e. baryonic and dark) matter density field of the local Universe, in the form of a probability distribution over all possible density fields.\\
%\footnote{An analogous distribution over \emph{baryonic} matter density fields might have been even more useful, but does not exist (yet) because of the complexity and computational cost of incorporating baryonic physics in the Hamiltonian Monte Carlo (HMC) formalism.}\\
%There is widespread consensus among experts that this suite of techniques represents the state-of-the-art and future of LSS reconstruction.\\
Several variations of the same general framework exist.
Inspired by the map-level CMB inference methods of \textcolor{blue}{\citet{Wandelt12004}}, some theoretical groundwork common to all these techniques has been developed by \textcolor{blue}{\citet{Kitaura12008}}, \textcolor{blue}{\citet{Jasche12010_1}} and \textcolor{blue}{\citet{Jasche12010_2}}, who also demonstrated the practical feasibility of Bayesian LSS reconstruction with the ARGO, ARES and HADES codes, respectively.
%However, the theoretical groundwork common to all these techniques has been laid by \textcolor{blue}{\citet{Kitaura12008}}, who also demonstrated the practical feasibility of Bayesian LSS reconstruction with multiple numerical methods using their ARGO code.
In \textcolor{blue}{\citet{Jasche12013}}, the authors described BORG (Bayesian Origin Reconstruction from Galaxies), the first algorithm to include physics of structure formation, and provided a proof-of-concept on simulated Sloan Digital Sky Survey Data Release 7 (SDSS DR7) data.
In \textcolor{blue}{\citet{Jasche12015}}, the \emph{actual} SDSS DR7 main galaxy sample was analysed with a refined version of BORG, yielding the publically-available BORG SDSS data release.
These data form the basis for the SCW prediction method presented in this paper.\\
The BORG SDSS has hitherto been used for Cosmic Web classification \textcolor{blue}{\citep{Leclercq12015}}, for the study of galaxy properties as a function of environment \textcolor{blue}{\citep{Leclercq12016}} and to unveil the dynamics of DM streams \textcolor{blue}{\citep{Leclercq12017}}, all in a probabilistic and time-dependent fashion.
Further refinements to BORG have been presented alongside applications to new datasets, namely 2M++ \textcolor{blue}{\citep{Lavaux12016, Jasche12019}} and SDSS3-BOSS \textcolor{blue}{\citep{Lavaux12019}}.
Algorithms related to BORG are ELUCID \textcolor{blue}{\citep{Wang12013, Wang12014}}, COSMIC BIRTH \textcolor{blue}{\citep{Kitaura12019}} and BARCODE \textcolor{blue}{\citep{Bos12019}}.
%An alternative implementation, COSMIC BIRTH, has been presented in \textcolor{blue}{\citet{Kitaura12019}}. BARCODE \textcolor{blue}{\citep{Bos12019}}, which self-consistently treats redshift space distortions, is yet another variation, but has hitherto not been applied to actual observations.
%\subsubsection{Applications}
%Because these Bayesian LSS reconstruction techniques recover the full formation history of the local, modern Universe, their applications are manifold (see e.g. \textcolor{blue}{\citet{Jasche12019}}) - and include the determination of galaxy cluster masses as a function of time (e.g. that of the Coma Cluster), the determination of the three-dimensional velocity field (which reveals regions with a low radial peculiar velocity, that are ideal for unbiased Hubble flow measurements), and the determination of the primordial density fluctuations of our part of the Universe (in contrast to CMB measurements, which reveal density fluctuations on a sphere that is currently $\sim45\ \mathrm{Gly}$ away from us (in comoving radial distance)).
%density fluctuations of a spherical shell
%in terms of comoving radial distance
%(to measure the parameters of the $\Lambda$CDM cosmological model and its extensions).
%The application most relevant to our purposes, however, is the reconstruction of the late-time DM density field of the local Universe, that takes the form of a probability distribution over all possible DM density fields.%One goal of these methods (and the most relevant for our purposes) is to
\subsubsection{Main principles}
Only a brief description of the main ideas underlying these approaches falls within the scope of this paper.
The methods all use the crucial insight that the statistical properties of the total density field in the Early Universe are well understood (i.e. Gaussian, with a theoretically predicted and observationally verified covariance function), and that the modern-day total density field relates to the initial field deterministically by means of (approximately Newtonian) gravity.\footnote{On super-Mpc scales, baryonic effects such as gas and radiation pressure play a subdominant role in structure formation, and can therefore be safely ignored.}
Thus, realisations from today's highly non-Gaussian total density field can be generated by forward modelling the effect of gravity on a collisionless\footnote{The assumption of a collisionless fluid is apt for DM, but only approximate for BM.} fluid, which is initialised as a Gaussian random field at the time of the CMB.\footnote{Because simulating gravity with an $N$-body simulation from the CMB to the present day is computationally expensive, and the method requires this process to be repeated thousands of times, all authors resort to methods that approximate Newtonian gravitational evolution.
In order of increasing accuracy (and numerical cost), they invoke either first-order Lagrangian perturbation theory (i.e. the Zel'dovich approximation (ZA) \textcolor{blue}{\citep{Zeldovich11970}}), second-order Lagrangian perturbation theory (2LPT) or particle mesh (PM) models, which approach the $N$-body solution.}
%In order of increasing accuracy (and numerical cost), these methods invoke either first-order Lagrangian perturbation theory (i.e. the Zel'dovich approximation (ZA) \textcolor{blue}{\citep{Zeldovich11970}}), second-order Lagrangian perturbation theory (2LPT) or particle mesh (PM) models, which - at least in spirit - emulate the exact solution.}
After a CMB-epoch total density GRF has been generated, and evolved into a modern total density field, a Poisson point process --- with an intensity function that attempts to capture the relation between galaxy locations and the surrounding matter distribution --- is used to calculate the likelihood of finding galaxies at their measured locations assuming the true underlying total density field is the one currently considered.
As suggested before, spectroscopic galaxy surveys, with hundreds of thousands of galaxies pinpointed in three-dimensional comoving space, form the typical input data.
Most notably, the likelihood for the modern total density field also fixes the likelihood for the initial total density field from which the modern one was evolved.
Because the prior on the initial total density fields is Gaussian, the initial total density field can not only be assigned a likelihood, but also a posterior probability.
Hamiltonian Monte Carlo (HMC) Markov Chains provide the means to explore the high-dimensional posterior distribution of \emph{initial} total density fields consistent with the galaxy data.
A posterior distribution over \emph{modern-day} total density fields, which is used in this article, is generated as a by-product of this process.
%To achieve this, data from spectroscopic galaxy surveys is used to locate millions of galaxies in three-dimensional comoving space.
%As galaxies are tracers of the DM density field, and cosmological simulations reveal
\subsubsection{BORG SDSS}
The BORG SDSS modern-day total density field posterior used in this article is based on optical galaxy data from the Sloan Digital Sky Survey (SDSS) \textcolor{blue}{\citep{York12000, Strauss12002}} Legacy Survey \textcolor{blue}{\citep{Abazajian12009}}.
%The late-time total density field posterior used in this article is the BORG SDSS \textcolor{blue}{\citep{Jasche12015}}, which is based on optical galaxy data from the Sloan Digital Sky Survey (SDSS) \textcolor{blue}{\citep{York12000, Strauss12002}} Legacy Survey \textcolor{blue}{\citep{Abazajian12009}}.
%The late-time total density field posterior used in this article is the BORG SDSS \textcolor{blue}{\citep{Jasche12015}}, which is based on optical galaxy data from the Sloan Digital Sky Survey (SDSS) \textcolor{blue}{\citep{York12000, Strauss12002}}.
%In particular, SDSS DR7 marked the end of the SDSS-II phase, and saw the public release of the SDSS Legacy Survey data \textcolor{blue}{\citep{Abazajian12009}}.
This catalogue provides right ascensions $\phi$, declinations $\theta$, spectroscopic redshifts and Petrosian r-band apparent magnitudes $m_r$ of about a million galaxies in the Northern Galactic Cap (NGC), which covers a region of $7646\ \mathrm{deg}^2$ of Northern Sky, and three stripes of the Southern Galactic Cap (SGC), which together cover $386\ \mathrm{deg}^2$.
In these regions, the SDSS Legacy Survey is spectroscopically complete for galaxies with $m_r < 17.77$.\\
%In total, the SDSS Legacy Survey thus features $8032\ \mathrm{deg}^2$ of sky coverage, with uniform selection criteria: within the survey footprint, spectroscopic detections are virtually complete for galaxies with $m_r < 17.77$.\\
The BORG SDSS, which is based on 372,198 NGC galaxies (see \textcolor{blue}{\citet{Jasche12015}} for selection details), infers structures in the density field of ${\sim}3\ \mathrm{Mpc}\ \mathrm{h}^{-1}$ and larger.
%The BORG SDSS, which is based on these data, accurately reconstructs dark matter density structures of ${\sim}3\ \mathrm{Mpc}\ \mathrm{h}^{-1}$ and larger.
Each realisation of the posterior is a cube containing $256^3$ voxels, that represents a region $\mathcal{R} \subset \mathbb{R}^3$ with comoving volume $(750\ \mathrm{Mpc}\ \mathrm{h}^{-1})^3$.
Naturally, $\mathcal{R}$ is chosen to correspond to the half of the Northern Sky that contains the NGC (up to a cosmological redshift $z \sim 0.2$).%, which is the most well-constrained part of the SDSS Legacy Survey footprint.
%As the NGC is by far the most well-constrained part of the SDSS Legacy Survey footprint, this volume corresponds to half of the Northern Sky, up to a cosmological redshift of $\sim 0.2$.

\subsection{Relating total matter density to synchrotron monochromatic emission coefficient}
Next, to convert total matter density fields $\rho$ to proper synchrotron monochromatic emission coefficient (MEC) fields $j_\nu$, we establish the relation between $\rho$ and $j_\nu$ (including its variability).
In radiative transfer theory, the MEC quantifies the amount of radiative energy released per unit of time, volume, frequency and solid angle \textcolor{blue}{\citep{Rybicki11986}}.
\subsubsection{The accretion shock Ansatz}
Cosmological simulations demonstrate that accretion shocks during large-scale structure formation are ubiquitous, and dominate the thermalisation of kinetic energy of baryons falling onto filaments.
Given the suite of known particle acceleration processes, and the fact that such shocks are almost certain to exist, shock acceleration is expected to be the biggest contributor to the synchrotron emission from filaments. 
%It is a widely held view, but not observationally demonstrated, that these accretion shocks are also the main source of synchrotron emission from the WHIM.
%Although it has not been demonstrated observationally that accretion shocks are the main source of synchrotron emission from the WHIM, it is widely thought that accretion shocks provide the dominant contribution.
This paper follows this hypothesis, by only considering the merger and accretion shock contribution to the SCW.\\
Even under the Ansatz that merger and accretion shocks drive the SCW signal, considerable uncertainty surrounding the correct physical description remains.
For example, it is an open question whether the electrons that eventually radiate in synchrotron light originate from the high-energy tail of the thermal Maxwell--Boltzmann distribution, or are rather accelerated to cosmic ray (CR) energies by SMBHs in the centres of galaxies, and then flung out via jets into the IGM.
Apart from uncertainty in the source of energetic electrons, the complex nature of plasma physics also makes it hard to establish how already-energetic electrons attain ultra-relativistic energies in magnetised shock fronts, with the theory of diffusive shock acceleration (DSA) being just one of multiple scenarios.
In DSA, CRs are accelerated by repeated crossings of the shock front, which acts as a magnetic mirror  \textcolor{blue}{\citep{Malkov12001, Xu12020}}.
Thus, we stress that the functional form of the SCW MEC, which we require to establish a connection with the total matter density field, depends on ill-constrained assumptions surrounding the exact radiation mechanism initiated by the accretion shocks.

\subsubsection{The \citet{Hoeft12007} model}
\label{sec:HoeftBruggenModel}
\textcolor{blue}{\citet{Hoeft12007}} have derived an analytic expression for the synchrotron power density of cosmological shock waves, assuming that the radiating electrons exclusively originate from the high-energy tail of the thermal Maxwell--Boltzmann distribution, and that the electron energy spectrum at the shock front is well-described by DSA.
As the DSA-based formulae of \textcolor{blue}{\citet{Hoeft12007}} have been partially successful in explaining observations of shocks in the ICM (e.g. \textcolor{blue}{\citet{vanWeeren12019, Locatelli12020}}), we postulate that the same formulae can describe synchrotron emission due to accretion shocks onto filaments.\footnote{\textcolor{blue}{\citet{Araya-Melo12012}} also took this approach and used the MareNostrum simulation to establish that, under these assumptions, filaments at a redshift of $0.15$ should produce a flux density of $10^{-1}\ \mathrm{\mu Jy}$ at $150\ \mathrm{MHz}$.}\\
%To answer the main question of this paper (`Where to look to maximise chances of detecting synchrotron radiation from the WHIM?'), it suffices to determine the MEC up to a constant.
%As long as the unknown constant is the same everywhere in the reconstructed volume, the specific intensity of one sky direction relative to some other remains well-determined.
We assume that the power density $P_\nu$ of a single shock obeys \textbf{Equation}~32 of \textcolor{blue}{\citet{Hoeft12007}}.
Let the shock surface area be $A$, and let the effective width of the downstream region be $\langle y \rangle$ (intuitively, this is the thickness of the shock in the direction perpendicular to the surface).
The effective shock volume is $V = A\langle y \rangle$.
If shocks radiate isotropically, the MEC $j_\nu$ is direction-independent.
However, like $\rho$, $j_\nu$ is a volume-averaged quantity, so that it depends on the scale on which the averaging occurs.
Index all shocks in the universe, so that $P_{\nu,i}\left(z, \nu\right)$ is the power density of shock $i$ at cosmological redshift $z$ and emission frequency $\nu$, and $A_i$ and $\langle y \rangle_i$ are its surface area and effective width.\footnote{Because the (cosmology-dependent) function $z\left(t\right)$ is strictly decreasing and thus invertible, we can use $z$ as a time coordinate.}
Moreover, let $\mathcal{R}_i\left(z\right) \subset \mathbb{R}^3$ be the region of space occupied by this shock at redshift $z$; the Lebesgue measure of $\mathcal{R}_i$ is $V_i = \int_{\mathcal{R}_i} \mathrm{d}\mathbf{r} = A_i \langle y \rangle_i$.
Under the approximation that the shock emission is homogeneous within $\mathcal{R}_i$, the average \emph{total} MEC within an arbitrary region $\bar{\mathcal{R}} \subset \mathbb{R}^3$ with Lebesgue measure $\bar{V} = \int_{\bar{\mathcal{R}}}\mathrm{d}\mathbf{r}$ is
\begin{align}
    j_\nu\left(z, \nu\right) = \frac{1}{4\pi\ \mathrm{sr}} \frac{1}{\bar{V}} \sum_i P_{\nu,i}\left(z,\nu\right) \frac{\int_{\mathcal{R}_i\left(z\right)\ \cap\ \bar{\mathcal{R}}}\mathrm{d}\mathbf{r}}{V_i}.
\end{align}
%Adapting the power density $P_\nu$ formula given by \textbf{Equation}~32 of \textcolor{blue}{\citet{Hoeft12007}} to one for the monochromatic emission coefficient $j_\nu$, we find that spacetime events\footnote{Because the (cosmology-dependent) function $z\left(t\right)$ is strictly decreasing and thus invertible, we can use $z$ as a time coordinate.} $\left(\mathbf{r}, z\left(t\right)\right)$ in the downstream region of a shock with volume $V$ obey
%and omitting universal constants,
The MEC of a \emph{single} shock (thus dropping indices) located at $\mathbf{r}$ at redshift $z$, and averaged over its own effective volume, is

\begin{align}
    &j_\nu\left(\mathbf{r}, z, \nu\right) = \frac{1}{4\pi\ \mathrm{sr}}\frac{P_\nu}{V} \nonumber \\
    &= \frac{445\ \mathrm{Jy}}{\mathrm{deg^2\ Mpc}}\ \left(\frac{\langle y \rangle}{\mathrm{Mpc}}\right)^{-1} \frac{n_{e,\mathrm{d}}\left(\mathbf{r}, z\right)}{10\ \mathrm{m}^{-3}} \frac{\xi_e}{10^{-2}} \left(\frac{\nu}{150\ \mathrm{MHz}}\right)^{\alpha\left(\mathbf{r}, z\right)} \nonumber \\
    &\left(\frac{T_\mathrm{d}\left(\mathbf{r}, z\right)}{10^8\ \mathrm{K}}\right)^\frac{3}{2} \frac{\left(\frac{B_\mathrm{d}}{\mathrm{\mu G}}\right)^{1-\alpha\left(\mathbf{r}, z\right)}\left(\mathbf{r}, z\right)}{\left(\frac{B_\mathrm{d}}{\mathrm{\mu G}}\right)^2\left(\mathbf{r}, z\right) + \left(\frac{B_\mathrm{CMB}}{\mathrm{\mu G}}\right)^2\left(z\right)} \left(\frac{150}{1400}\right)^{\alpha\left(\mathbf{r}, z\right)} \Psi\left(\mathbf{r},z\right).
    %&n_e\left(\mathbf{r}, z\right)T_\mathrm{d}^\frac{3}{2}\left(\mathbf{r}, z\right)\ \nu^{\alpha\left(\mathbf{r}, z\right)} \frac{B^{1-\alpha\left(\mathbf{r}, z\right)}\left(\mathbf{r}, z\right)}{B^2\left(\mathbf{r}, z\right) + B_\mathrm{CMB}^2\left(z\right)}\Psi\left(\mathbf{r},z\right).%\propto
    %\langle y \rangle \left(\mathbf{r},z\right)
\label{eq:properMonochromaticEmissionCoefficient}
\end{align}
Here, $n_{e,\mathrm{d}}$ is the downstream electron number density, $T_\mathrm{d}$ is the downstream plasma temperature, $\alpha$ is the integrated spectral index of the associated synchrotron emission, $B_\mathrm{d}$ is the downstream plasma magnetic field strength, $B_\mathrm{CMB}$ is the CMB magnetic field strength, and $\Psi$ is a dimensionless quantity with a strong dependence on the upstream Mach number $\mathcal{M}_\mathrm{u}$ and a weak dependence on $T_\mathrm{d}$ (as can be glanced from \textbf{Figure}~4 of \textcolor{blue}{\citet{Hoeft12007}}).
$\Psi$ approaches unity for high Mach numbers, and so for the WHIM, where the upstream Mach numbers are expected to be high, $\Psi\left(\mathbf{r},z\right) \approx 1$.
%Note that this approximation overestimates the 
Like $\Psi$, $\alpha$ does not depend on the spacetime coordinate $\left(\mathbf{r},z\right)$ \emph{directly,} but rather via $\mathcal{M}_\mathrm{u}$ and the adiabatic index $\gamma$.
%In the energy-balance scenario of DSA
%In concreto, for $\mathcal{M}_\mathrm{u} > 1$,
Concretely, for $\mathcal{M}_\mathrm{u} > 1$,
\begin{align}
    %\alpha = \alpha\left(\mathcal{M}\left(\mathbf{r},z\right)\right) = - \frac{\mathcal{M}^2 + 1}{\mathcal{M}^2 - 1},
    \alpha = \alpha\left(\mathcal{M}_\mathrm{u}\left(\mathbf{r},z\right), \gamma\left(\mathbf{r},z\right)\right) = \frac{\tfrac{1}{4}\left(1 - 3\gamma\right)\mathcal{M}_\mathrm{u}^2 - 1}{\mathcal{M}_\mathrm{u}^2 - 1}.
\end{align}
Note that $\alpha \to -\infty$ as $\mathcal{M}_\mathrm{u} \to 1+$, and that $\alpha \to \sfrac{1}{4}\left(1-3\gamma\right)$ as $\mathcal{M}_\mathrm{u} \to \infty$.
%Note that the spectral index diverges as $\mathcal{M}$ approaches 1 from above, and that it tends to $\frac{1}{4}\left(1-3\gamma\right)$ asymptotically as $\mathcal{M}$ grows large.
%For $\gamma = \sfrac{5}{3}$ (a non-relativistic monoatomic ideal gas), $\alpha \to -1$; for $\gamma = \sfrac{4}{3}$ (an ultra-relativistic monoatomic ideal gas), $\alpha \to \sfrac{-3}{4}$.
The top panel of \textbf{Figure}~\ref{fig:spectralIndex} illustrates this behaviour.
%, and demonstrates that, given the high upstream Mach numbers in the WHIM, it is reasonable to assume $\alpha = -1$ for all WHIM accretion shocks.
%from accretion shocks in filaments under DSA. \alpha \approx -1
%$\langle y \rangle$ denotes the effective width of the downstream region - intuitively, this is the `thickness' of the shock (in the direction perpendicular to the surface).
The electron acceleration efficiency $\xi_e$ quantifies the fraction of the shock's thermal energy that is used to accelerate suprathermal electrons.
$\xi_e$ is unknown for WHIM shocks, but a comparison to supernova remnant (SNR) shocks suggests $\xi_e = 0.05$ \textcolor{blue}{\citep{Keshet12004}}.
%Notice that we also omitted the dependence of the MEC on the electron acceleration efficiency $\xi_e$
%As it is not decisively established which emission mechanism(s), 
\subsubsection{Enzo simulations}
To capture, in a statistical sense, the relationship between the total matter density and the \textcolor{blue}{\citet{Hoeft12007}} monochromatic emission coefficient, we turn to MHD simulations.
In particular, we use snapshots (i.e. 3D spatial fields at constant time) of the largest uniform-grid cosmological MHD simulation to date \textcolor{blue}{\citep{Vazza12019}}, which is based on the Enzo code \textcolor{blue}{\citep{Bryan12014}}.
These cubic snapshots cover a comoving volume of $\left(100\ \mathrm{Mpc}\right)^3$ with $2400^3$ voxels, yielding a (comoving) resolution of $41\tfrac{2}{3}\ \mathrm{kpc}$ per voxel edge.
%, as one should to describe the actual Universe
The simulations recreate the evolution of the baryonic and dark matter density functions $\rho_\mathrm{BM}$ and $\rho_\mathrm{DM}$ --- as well as (thermo)dynamic quantities\footnote{As long as one considers a single shock, space (at a fixed time) can be classified into an upstream region and a downstream region, with associated temperatures ($T_\mathrm{u}$ and $T_\mathrm{d}$, respectively) and magnetic field strengths ($B_\mathrm{u}$ and $B_\mathrm{d}$, respectively). This distinction is less meaningful once one considers multiple shocks at the same time: a given location could then be upstream for some shocks, and downstream for others. Therefore, cosmological simulations do not evolve an upstream or downstream temperature field, but just a \emph{general} temperature field $T$. Analogously, cosmological simulations maintain only \emph{general} magnetic field component fields $B_x$, $B_y$ and $B_z$ (and thus a magnetic field strength field $B \coloneqq \sqrt{B_x^2 + B_y^2 + B_z^2}$), without upstream and downstream distinction.} such as the gas temperature $T$, magnetic field strength $B$ and gas velocity $v$ --- under Newtonian gravity.
(However, the effects of the expansion of the Universe as predicted by general relativity are still incorporated.)
%The simulations recreate the evolution of the baryonic and dark matter density functions $\rho_\mathrm{BM}$ and $\rho_\mathrm{DM}$ - as well as (thermo)dynamic quantities\footnote{Both in cosmological simulations and in the real world, there is no intrinsic distinction between the upstream and downstream gas temperature and magnetic field - just universal functions $T$ and $B$. These quantities (e.g. $T$) are best \emph{interpreted} as downstream (e.g. $T_\mathrm{d}$) for some voxels, and as upstream (e.g. $T_\mathrm{u}$) in the rest.} such as the gas temperature $T$, magnetic field strength $B$ and gas velocity $v$ - under Newtonian gravity.\footnote{The effects of the expansion of the Universe as predicted by general relativity are incorporated \textit{ad hoc}.}
No galactic physics is included.
%Physics of galaxies is not included, except for a parametric treatment of AGN outflows, but this feature is only used in the astrophysical seeding scenario of WHIM magnetic fields.
Shocks in the snapshots can be identified by searching for temperature and velocity jumps, as described in \textcolor{blue}{\citet{Vazza12009}}.\footnote{From the middle panel of \textbf{Figure}~\ref{fig:spectralIndex}, it is clear that shocks also induce a jump in gas density. However, this jump is modest, saturating at e.g. a factor $4$ for $\gamma = \sfrac{5}{3}$. By contrast, the jump in temperature, seen in the bottom panel of the same figure, can be several orders of magnitude --- in the WHIM, at least.}
For shocked voxels, we simply take $\langle y \rangle$ to be the voxel edge length, and calculate $\alpha$ and $\Psi$ after establishing $\mathcal{M}_\mathrm{u}$.
With the exception of $\xi_e$, which we will assume to be a constant throughout, the simulation thus enables us to compute all factors on the RHS of \textbf{Equation}~\ref{eq:properMonochromaticEmissionCoefficient}.\\
Taking their product for the $z = 0.025$ snapshot, we obtain $j_\nu$, and compare it to $\rho \coloneqq \rho_\mathrm{BM} + \rho_\mathrm{DM}$ to study their relationship.\footnote{By repeating our analysis for snapshots of other redshifts, one could study the time evolution of this relationship.
However, as the \textcolor{blue}{\citet{Hoeft12007}} model is likely only a rough description of the actual synchrotron emission mechanism in filaments, we currently consider such level of detail superfluous.}% - especially given the modest changes in large-scale filamentary properties expected between $z = 0.5$ and $z = 0$ \textcolor{blue}{\citep{Gheller12016}}.}

\subsubsection{Conditional probability distribution}
\begin{figure}
    \centering
    \includegraphics[width=\columnwidth]{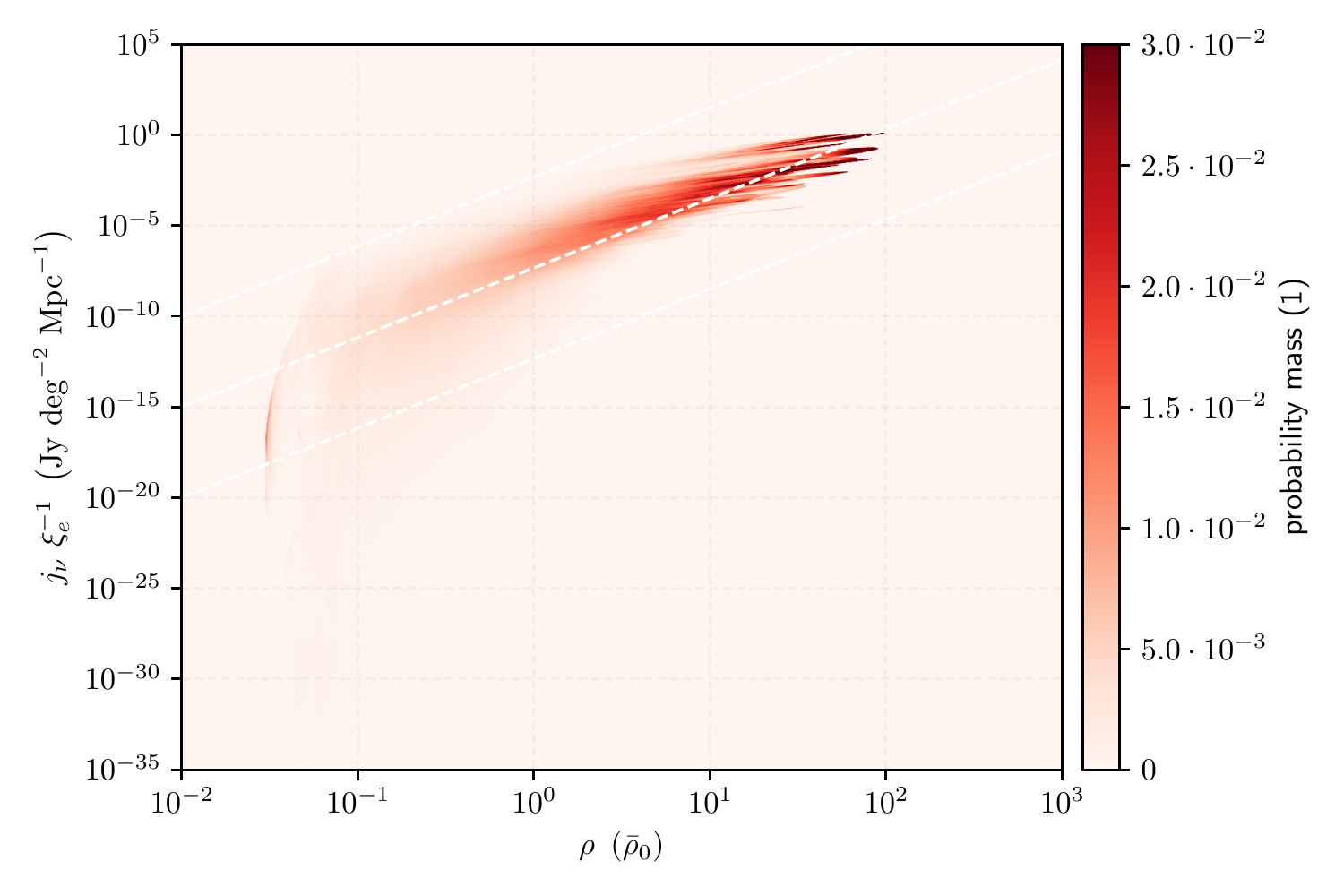}%MHDPMFConditionalConvolved2SD2Slices.pdf}
    \caption{Conditional probability distributions $P\left(j_\nu\ \vert\ \rho\right)$ of synchrotron monochromatic emission coefficient $j_\nu$ (at rest-frame (emission) frequency $\nu = 150\ \mathrm{MHz}$) given total matter density $\rho$ for the \textcolor{blue}{\citet{Hoeft12007}} formalism, as derived from $z = 0.025$ snapshots of cosmological MHD simulations by \textcolor{blue}{\citet{Vazza12019}}. The dashed lines indicate a $j_\nu \propto \rho^{\sfrac{23}{6}}$ (single shock) scaling relation for various proportionality constants. $\rho$ is shown relative to the current-day mean total matter density $\bar{\rho}_0 = \Omega_\mathrm{M,0}\ \rho_{c,0}$, while the weakly constrained electron acceleration efficiency $\xi_e$ is divided out from $j_\nu$. Each conditional is numerically approximated by a probability mass function with $1000$ bins, whose edges vary by a constant factor and span $40$ orders of magnitude in $j_\nu$.}
    %, whilst $j_\nu$ is shown relative to a value expected in typical WHIM conditions.}
    \label{fig:MHDPMFConditional}
\end{figure}

In order to convert the BORG SDSS total matter density fields into synchrotron MEC fields, we need to predict $j_\nu$ from $\rho$ --- ideally including uncertainty.
Because the BORG SDSS has a comoving resolution of $4.17\ \mathrm{Mpc}$ per voxel edge compared to Enzo's $4.17\cdot 10^{-2}\ \mathrm{Mpc}$ per voxel edge, we blur both $j_\nu$ and $\rho$ with a Gaussian kernel whose standard deviation is half of the comoving resolution ratio.
%(which, by chance, is almost precisely 100).
We then treat each $(j_\nu, \rho)$ pair of the blurred fields as a draw from the joint distribution $P\left(j_\nu, \rho\right)$.
%an independent draw
By binning the data, we perform a simplistic form of kernel density estimation (KDE).
Next, we calculate the conditional $P\left(j_\nu\ \vert\ \rho\right)$ from the joint by dividing it by the marginal $P\left(\rho\right) \coloneqq \int_0^\infty P\left(j_\nu, \rho\right) \mathrm{d}j_\nu$.\\
The result is the probability distribution shown in \textbf{Figure}~\ref{fig:MHDPMFConditional}.
We overplot the single shock scaling relation expected for the WHIM regime: $j_\nu \propto \rho^{\sfrac{23}{6}}$.
See \textbf{Appendix}~\ref{ap:scalingRelation} for a derivation.

\begin{figure*}[h!]
    \centering
    \includegraphics[width=\textwidth]{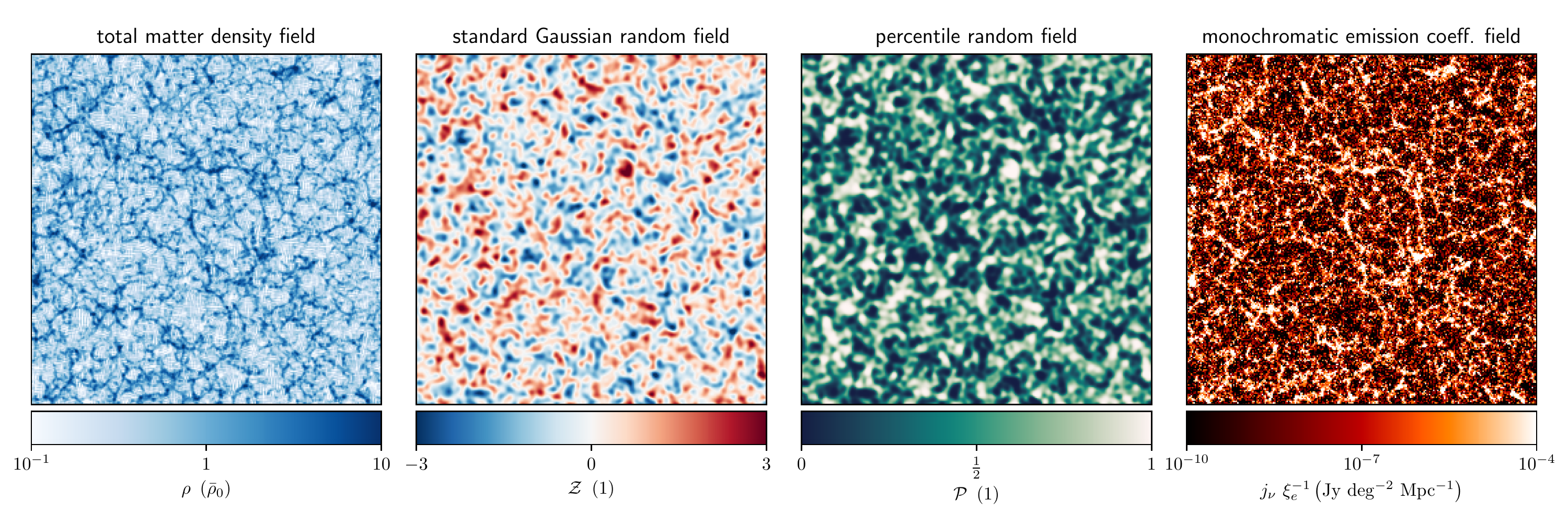}
    \caption{Overview of the procedure for converting a probability distribution over total (i.e. baryonic and dark) matter density fields into a probability distribution over monochromatic emission coefficient (MEC) fields. We show a fixed slice through $\mathcal{R}$, of $750\ \mathrm{Mpc}\ h^{-1} \times 750\ \mathrm{Mpc}\ h^{-1}$ for the density and MEC fields, and a subregion of $150\ \mathrm{Mpc}\ h^{-1} \times 150\ \mathrm{Mpc}\ h^{-1}$ for the Gaussian and percentile random fields (to more clearly illustrate the bijective mapping between the two). \textbf{Left:} BORG SDSS density $\rho$ sample, in multiples of the current-day mean total matter density $\bar{\rho}_\mathrm{0} = \Omega_\mathrm{M,0}\ \rho_\mathrm{c,0}$. \textbf{Middle left:} Gaussian random field $\mathcal{Z}$ sample using a squared exponential (SE) kernel with lengthscale $l_\mathrm{SE} = 2\ \mathrm{Mpc}$. \textbf{Middle right:} the corresponding percentile random field $\mathcal{P}$ sample. \textbf{Right:} the corresponding MEC $j_\nu$ sample at an emission frequency of $\nu = 150\ \mathrm{MHz}$, with the weakly-constrained electron acceleration efficiency $\xi_e$ divided out.}
    %, in multiples of a typical \textcolor{blue}{\citet{Hoeft12007}} monochromatic emission coefficient for filaments.}
    \label{fig:densityToMEC}
    % NOTE!!! THIS FIGURE IS NOT YET RIGHT: THE DENSITY FIELD DOES NOT CORRESPOND TO THE MEC FIELD... Now it does??
\end{figure*}
\subsection{Generating monochromatic emission coefficient fields}
\label{sec:MEC}
In principle, we could --- for each BORG SDSS sample $\rho\left(\mathbf{r}\right)$ --- convert to $j_\nu\left(\mathbf{r}\right)$ by drawing from $P\left(j_\nu\ \vert\ \rho = \rho\left(\mathbf{r}\right)\right)$ independently for every $\mathbf{r} \in \mathcal{R}$.
%In principle, using the conditional probability distribution for emissivity given dark matter density, we could convert $\rho_\mathrm{DM}\left(\mathbf{r}\right)$ to $j_\nu\left(\mathbf{r}\right)$ by drawing from $P\left(j_\nu\ \vert\ \rho_\mathrm{DM} = \rho_\mathrm{DM}\left(\mathbf{r}\right)\right)$ independently for every $\mathbf{r} \in \mathcal{R}$ of the BORG SDSS dark matter density field sample at hand.
However, this is clearly suboptimal, as accurate SCW MEC fields exhibit spatial correlation up to megaparsec scales, both under the merger and accretion shock Ansatz as well as in a turbulence scenario \textcolor{blue}{\citep{Govoni12019, Brunetti12020}}.\\
To generate spatially correlated draws from our conditional probability distribution, we use realisations of a Gaussian random field (GRF) $\mathcal{Z}$ over $\mathcal{R}$.
In general, the statistical properties of a GRF are determined by its mean, and its covariance function or \emph{kernel}.
In our case, we set the mean to 0, and choose an isotropic and stationary kernel, as suggested by the Cosmological Principle.
From the variety of remaining choices commonly used in machine learning (ML), we pick\footnote{The choice of kernel should ideally reflect the morphology of merger and accretion shocks. However, given the current spatial resolution of the BORG SDSS density fields and our requirement that the kernel be isotropic (disregarding the relation between shock morphology and local LSS orientation), the approach is already so approximate that no particular kernel is clearly preferred over the others.} the squared-exponential (SE) kernel $K_\mathrm{SE}: \mathcal{R} \times \mathcal{R} \to \mathbb{R}_{\geq 0}$ \textcolor{blue}{\citep{Rasmussen12006}}:
\begin{align}
    &K_\mathrm{SE}\left(\mathbf{r}, \mathbf{r}'\right) = \sigma_\mathrm{SE}^2 \exp{\left(-\frac{||\mathbf{r} - \mathbf{r}'||_2^2}{2 l_\mathrm{SE}^2}\right)},\ \textrm{so that} \nonumber \\
    &\mathrm{Cov}\left(\mathcal{Z}\left(\mathbf{r}\right), \mathcal{Z}\left(\mathbf{r}'\right)\right) = K_\mathrm{SE}\left(\mathbf{r}, \mathbf{r}'\right)\ \textrm{with}\ \mathbf{r}, \mathbf{r}' \in \mathcal{R}.
\end{align}
The variance $\sigma_\mathrm{SE}^2$ allows us to endow the GRF with dimensionality and scale it at will.
As our GRF will only serve to draw spatially correlated MEC samples, we set $\sigma_\mathrm{SE} = 1$.\\
The lengthscale $l_\mathrm{SE}$ is the characteristic scale of spatial correlation, and forms an important model choice.
%sets the stochasticity of the realisations.
Some of the longest coherent shocks observed thus far are the Toothbrush Relic in galaxy cluster 1RXS J0603.3+4214 \textcolor{blue}{\citep{vanWeeren12012}} and the Sausage Relic in galaxy cluster CIZA J2242.8+5301 \textcolor{blue}{\citep{diGennaro12018}}, both with a largest linear size (LLS) of ${\sim}2\ \mathrm{Mpc}$.
Although the majority of observed shocks seem smaller (suggesting $l_\mathrm{SE} < 2\ \mathrm{Mpc}$), the presence of noise in radio imagery obfuscates the full extent of shock fronts, biasing measurements towards small LLSs.
As a compromise, we choose $l_\mathrm{SE} = 2\ \mathrm{Mpc}$.\\\\
%Because the majority of observed shocks appear more compact, choosing $l_\mathrm{SE} = 2\ \mathrm{Mpc}$ seems to overestimate the typical correlation.
As is customary in cosmological simulations --- where GRFs with isotropic and stationary kernels are used to initialise the Early Universe matter density fields --- we generate our zero mean, unit covariance GRF realisations on the BORG SDSS voxel grid using Fourier analysis.
This necessitates calculating the power spectrum of the kernel, which in the case of the 3D SE kernel is
\begin{align}
    P_\mathrm{SE}\left(\mathbf{k}\right) = \left(2\pi l_\mathrm{SE}^2\right)^\frac{3}{2} \exp{\left(-2 \pi ||\mathbf{k}||_2^2 l_\mathrm{SE}^2\right)}.%,\ \textrm{where}\ k \coloneqq ||\mathbf{k}||_2.
\end{align}
Finally, Fourier-generated GRFs on a finite (numerical) grid require an appropriate normalisation; we use Parseval's theorem to find the correct factor.\\
This procedure generates a `standard normal' GRF $\mathcal{Z}$, in the sense that each location's RV is standard normal: $\mathcal{Z}\left(\mathbf{r}\right) \sim \mathcal{N}\left(0, 1\right)$ for all $\mathbf{r} \in \mathcal{R}$.
Next, we apply the CDF $\Phi$ of the standard normal to the GRF in point-wise fashion (i.e. voxel-wise, in practice), thereby creating a `percentile random field' $\mathcal{P}$:
\begin{align}
    \mathcal{P}\left(\mathbf{r}\right) \coloneqq \Phi\left(\mathcal{Z}\left(\mathbf{r}\right)\right),\ \textrm{where}\ \Phi\left(x\right) \coloneqq \frac{1}{\sqrt{2\pi}}\int_{-\infty}^x \exp{\left(\frac{-y^2}{2}\right)}\ \mathrm{d}y.
\end{align}
This field has values strictly between $0$ and $1$, and inherits the spatial correlations present in $\mathcal{Z}$.\\
Let $\Xi_\rho\left(x\right): \mathbb{R}_{\geq 0} \to [0, 1]$ be the CDF of the RV $j_\nu\ \vert\ \rho$, and let $\Xi_\rho^{-1}\left(y\right): [0, 1] \to \mathbb{R}_{\geq 0}$ be its inverse.
Then the final MEC field is given by
\begin{align}
    j_\nu\left(\mathbf{r}\right) \coloneqq \Xi_{\rho\left(\mathbf{r}\right)}^{-1}\left(\mathcal{P}\left(\mathbf{r}\right)\right),\ \textrm{for all}\ \mathbf{r} \in \mathcal{R}.
\label{eq:densityToMEC}
\end{align}
Because $\rho$ and $\mathcal{P}$ are random fields on $\mathcal{R}$, so is $j_\nu$.
A graphical summary of the procedure described in this subsection, and defined in \textbf{Equation}~\ref{eq:densityToMEC}, is shown in \textbf{Figure}~\ref{fig:densityToMEC}.

\subsection{Generating specific intensity functions}
To find the specific intensity of the SCW \emph{on the sky} ($I_\nu$), we simulate the passage of light rays through our 3D MEC fields via ray tracing.
Conveniently, projecting a ray's 4D null geodesic in a pure Friedmann--Lema\^itre--Robertson--Walker (FLRW) metric onto 3D comoving space results in a straight line.
We provide a brief derivation in \textbf{Appendix}~\ref{ap:rayTracing}.
By assuming an exact FLRW metric, we neglect all spacetime deformations due to \emph{local} (large-scale structure) energy density fluctuations.
Our results thus do not feature gravitational lens effects around massive clusters.\\
We estimate $I_\nu$ in a sky patch by sampling many directions $\hat{r}_i$ within it, and simulating the passage of light through our reconstructed SCW in comoving space for each such direction (or `ray') $\hat{r}_i$.
%Ignoring numerical considerations, ray tracing is arguably the best method to project a three-dimensional SCW onto the two-dimensional sky:
%\begin{itemize}
%    \item Ray tracing takes the full 3D nature of the SCW into account, correctly handling emission from objects with complex shapes, or from partly overlapping objects;
%    \item Ray tracing naturally causes objects that are further away to become smaller on the sky. Note that we ignore the effect of the expansion of the Universe on the angular size redshift relation; however, this is not deemed problematic as we ray trace through LSS at $z \ll 1.5$, the approximate redshift of peak angular diameter distance.
%    \item Ray tracing avoids assigning single redshifts to objects of cosmological spatial extent, and other discretisation effects;
%    \item Ray tracing generates samples from the specific intensity function, that respect sky curvature when mirrored or translated (or, equivalently, rotated) over $S^2$;
    % Parallellisability is a property of ray tracing, but not exclusively for ray tracing; e.g. Vernstrom's stacking of Gaussians could also be done in a parallel way. Remove?
%    \item Ray tracing is parallellisable and extendable (e.g. to include absorption).
%\end{itemize}
\subsubsection{From proper monochromatic emission coefficient to observer's specific intensity}
\label{sec:observersSCWSpecificIntensity}
Next, we establish the relation between the observed specific intensity $I_\nu$ of a ray that has travelled through the SCW to Earth from direction $\hat{r}$, and the MEC $j_\nu$ along its path.\\
%$\mathcal{L}$.\\%\left(\hat{r}\right)
%\subsection{Observer's SCW specific intensity}
%We derive the specific intensity  of a light ray  .
%First, we note that the Universe is mostly optically thin for electromagnetic (EM) waves with frequencies around 150 MHz; we assume this holds perfectly.
%Thus, absorption is neglected and the specific intensity of the ray only accumulates as the ray travels through LSS to the observer.\\
%Be $j_\nu, j_{\nu, \mathrm{obs}}: \mathbb{R}^3 \times \mathbb{R}_{\geq 0} \times \mathbb{R}_{> 0} \to \mathbb{R}_{\geq 0}$ the SCW \emph{proper} and \emph{observer's} monochromatic emission coefficients, respectively (with SI units $\mathrm{W\ m^{-3}\ Hz^{-1}\ sr^{-1}}$): these function assign their corresponding light-emission properties to each 3-tuple $\left(\mathbf{r}, z, \nu\right)$ containing a comoving location $\mathbf{r}$, a redshift $z$ and an EM wave frequency $\nu$.
Be $j_\nu: \mathbb{R}^3 \times \mathbb{R}_{\geq 0} \times \mathbb{R}_{> 0} \to \mathbb{R}_{\geq 0}$ the function that assigns to each 3-tuple $\left(\mathbf{r}, z, \nu\right)$ containing a comoving locus $\mathbf{r}$, a cosmological redshift $z$ (which represents time) and an EM wave frequency $\nu$, the \emph{proper} (rather than comoving) MEC $j_\nu\left(\mathbf{r}, z, \nu\right)$ (with SI units $\mathrm{W\ m^{-3}\ Hz^{-1}\ sr^{-1}}$).
Conceptually, we must differentiate between the specific intensity of large-scale structure at the time of emission (`there and then'), and at today's observing epoch (`here and now').
%$I_{\nu, \mathrm{prop}}$
%We compute this last quantity by summing up the contributions for each infinitesimal proper length $\mathrm{d}l = c\ \mathrm{d}t$, yielding
\emph{Today's} quantity $I_\nu$ for direction $\hat{r}$ and observing frequency $\nu_\mathrm{obs}$ is
\begin{align}
%\boxed{
%&I_\nu\left(\hat{r}, \nu_\mathrm{obs}\right) = \int_0^\infty \frac{\mathrm{d}I_\nu}{\mathrm{d}z}\left(\hat{r},\nu_\mathrm{obs},z\right)\ \mathrm{d}z \nonumber \\
I_\nu\left(\hat{r},\nu_\mathrm{obs}\right) = \frac{c}{H_0} \int_0^\infty \frac{j_\nu\left(r\left(z\right) \hat{r},z,\nu_\mathrm{obs}\left(1+z\right)\right)}{\left(1+z\right)^4 E\left(z\right)} \mathrm{d}z.
%&= \int_0^\infty \frac{c\ H_0^{-1}\ j_\nu \left(r\left(z\right)\hat{r}, z, \nu_\mathrm{obs}\left(1 + z\right)\right)}{(1 + z)^4 \sqrt{\left(1+z\right)^2\left(1 + \Omega_{\mathrm{M},0} z\right) - \Omega_{\Lambda,0} z \left(2 + z\right)}}\ \mathrm{d}z.
%}
%\boxed{
%&I_{\nu, \mathrm{obs}}(\hat{r}, \nu) =\\
%& \frac{c}{H_0} \int_0^\infty \frac{j_\nu\left(\mu\left(z\right)\hat{r}, z, \nu\left(1 + z\right)\right)}{(1 + z)^4 \sqrt{\left(1+z\right)^2\left(1 + \Omega_\mathrm{M} z\right) - \Omega_\Lambda z \left(2 + z\right)}}\ \mathrm{d}z%}\ .
\label{eq:specificIntensity}
\end{align}
We provide a brief derivation, alongside explicit expressions for $r\left(z\right)$ and $E\left(z\right)$, in \textbf{Appendix}~\ref{ap:SIFormula}.\\
%where $I_{\nu,\mathrm{obs}}$ is shorthand for $I_{\nu,\mathrm{obs}}\left(\hat{r},\nu\right)$.\\\\
For each specific intensity function that we wish to generate, we now draw $M = 10^6$ ray directions uniformly from the $\pi\ \mathrm{sr}$ lune that covers the SDSS-constrained half of the Northern Sky.
For each ray, we calculate the corresponding specific intensity by combining \textbf{Equations}~\ref{eq:properMonochromaticEmissionCoefficient} and \ref{eq:specificIntensity}.\\
Note that this requires evaluating the MEC field at a range of emission frequencies.
Numerically, we realise this by building both the Enzo-derived conditional probability distribution $P\left(j_\nu\ \vert\ \rho\right)$ and the MEC field sample (using the same density field sample and GRF sample) at two emission frequencies $\nu$: $\nu = \nu_\mathrm{obs}$ and $\nu = \nu_\mathrm{obs}\left(1+z_\mathrm{max}\right)$.
Next, for each voxel, we calculate a spectral index from the two MEC fields by assuming a power-law spectrum, and use it to find the MEC at the emission frequency $\nu = \nu_\mathrm{obs}\left(1 + z\right)$ needed given the voxel's cosmological redshift $z$.\\\\
We repeat this process many times by selecting $1000$ density samples from the BORG SDSS posterior (we discard the first $2110$ samples due to burn-in, and thin the Monte Carlo Markov chain by a factor $10$) and generating $1$ independent and identically distributed (IID) GRF sample for each.
(There is no compelling reason to thin the chain and we could use multiple GRF samples per density sample; the current approach is merely to limit data storage, compute time and energy usage.)
Thus, we numerically realise $1000 \cdot 1 = 1000$ specific intensity functions.
Together, they form a probability distribution over MASSCW specific intensity functions over a quarter of the sky.
%This approach is only mathematically valid if: 1) spectral index is a non-decreasing function of MEC, and the percentile random field corresponds to \nu = \nu_\mathrm{obs}. 2) spectral is a non-increasing function of MEC, and the percentile random field corresponds to \nu = \nu_\mathrm{obs} \left(1 + z_\mathrm{max}\right). 3) there is a constant and universal spectral index, and the percentile random field corresponds to all frequencies.
% The mathematically correct way of doing things is P\left(j_\nu, \alpha $\vert$ \rho\right) = P\left(\alpha $\vert$ j_\nu, \rho\right) P\left(j_\nu $\vert$ \rho\right).

\subsection{Generating redshift functions}
Our methodology can also be used to characterise the redshift properties of the MASSCW.
%Since we generate 3D radiating structures at known redshifts, we can readily calculate the specific-intensity-weighted mean redshift 
%Because we calculate a ray's total specific intensity by summing up the specific intensity contributions from all voxels along its path, each of which has a known redshift, we can also cheaply calculate the specific-intensity-weighted mean redshift function for a given MASSCW prior realisation.
%As with specific intensity, iteration over DM density and GRF samples leads to a probability distribution over MASSCW specific-intensity-weighted mean redshift functions over a quarter of the sky.
%We thus obtain a second valuable probability distribution over the same sky region.
%Because our goal is to 
%the 
%and $\mathcal{M}$, and thus allow for the calculation of $j_\nu$, whilst evidently al
%These between $\rho_\mathrm{DM}$ and , we turn to cosmological simulations, w
%\textcolor{blue}{\citet{Vazza12019}} describes the largest uniform-grid cosmological MHD simulation hitherto performed, realised using the Enzo code \textcolor{blue}{\citep{Bryan12014}}.
\subsubsection{Specific-intensity-weighted mean redshift}
%One might wonder, \emph{If I were to pick a random direction on the sky, from what redshift would its MASSCW signal come predominantly?}
%Our methodology can address this question, since we generate 3D radiating structures at known redshifts.
A ray's total specific intensity is found by summing up the specific intensity contributions from all voxels along its path, each of which has a known redshift.
Therefore, a ray's mean redshift is found by weighting each voxel's redshift by the corresponding specific intensity contribution, before dividing by the sum of such contributions.
Concretely, we define the specific-intensity-weighted mean redshift $\bar{z}$ at observing frequency $\nu_\mathrm{obs}$ of a ray $\hat{r}$ passing through MEC field $j_\nu$ to be
%for each ray passing through a given MEC field, we calculate a specific-intensity-weighted mean redshift
\begin{align}
%    \bar{z}\left(\hat{r},\nu\right) = \frac{\displaystyle \int_0^\infty z\ \frac{\mathrm{d}I_{\nu,\mathrm{obs}}}{\mathrm{d}z}\ \mathrm{d}z}{I_{\nu,\mathrm{obs}}\left(\hat{r},\nu\right)}.
    \bar{z}\left(\hat{r},\nu_\mathrm{obs}\right) \coloneqq I_\nu^{-1}\left(\hat{r},\nu_\mathrm{obs}\right)\ \int_0^\infty z\ \frac{\mathrm{d}I_\nu}{\mathrm{d}z}\left(\hat{r},\nu_\mathrm{obs},z\right)\ \mathrm{d}z.
\end{align}
%Doing so for many rays leads to a MASSCW specific-intensity-weighted mean redshift function.
By iterating over total matter density and GRF samples, one generates a probability distribution over the function $\bar{z}\left(\hat{r},\nu_\mathrm{obs}\right)$, in exact analogy to the generation of the distribution over specific intensity functions $I_\nu\left(\hat{r},\nu_\mathrm{obs}\right)$.\\
%where $\bar{z}$ is shorthand for $\bar{z}\left(\hat{r},\nu\right)$.\\
One might wonder, \emph{If I were to pick a random direction on the sky, what would be the specific-intensity-weighted mean redshift of its MASSCW signal?}
The set $\{\bar{z}\left(\hat{r}_i,\nu_\mathrm{obs}\right)\ \vert\ i \in \{1, 2, ..., M\}\}$, with rays $\hat{r}_i$ drawn from a uniform distribution over the sky, can be viewed as a random sample from a specific-intensity-weighted mean redshift \emph{random variable} $\bar{Z}$. %, which we characterise by its CDF.
An RV is fully characterised by its CDF.
%By viewing $\bar{z}\left(\hat{r}_i,\nu\right)$ for many rays $\hat{r}_i$ as samples from a specific-intensity-weighted mean redshift \emph{random variable} $\bar{Z}$
%By simply aggregating $\bar{z}\left(\hat{r}_i,\nu\right)$ for many rays $\hat{r}_i$, which are drawn from a uniform distribution over the sky, we approximate the distribution of the specific-intensity-weighted mean redshift \emph{random variable} $\bar{Z}$.
%which can be summarised by its CDF.
Each MASSCW prior realisation generates another empirical CDF (ECDF) for $\bar{Z}$.\footnote{The probability distribution over ECDFs of $\bar{Z}$ thus obtained is less informative than the probability distribution over functions $\bar{z}$, because it disposes of the directional correlations in specific-intensity-weighted mean redshift $\bar{z}\left(\hat{r},\nu_\mathrm{obs}\right)$.}
%Ignoring the correlation between MEC-weighted redshifts of adjacent directions (or rays), we obtain an estimate of the MEC-weighted mean redshift distribution
%by simply aggregating many rays, we obtain 
%directional correlations between 
%Aggregating many rays (ignoring the directional correlations between their MEC-weighted redshifts), provide a sample 
%Many rays together (ignoring the directional correlations provide a sample from the MEC-weighted redshift distribution of the MASSCW signal over the sky, as in \textbf{Figure}~\ref{fig:realisationSphereRedshift}.
%As we can generate many such realisations, we also obtain an MASSCW \emph{redshift} prior.
%Ignoring the directional correlations of the MEC-weighted redshifts by simply aggregating them, we can calculate \emph{single}-direction redshift distributions to answer this paragraph's opening question.

\subsubsection{Flux-density-weighted mean redshift}
%(all-sky)
%One might also wonder, \emph{What redshift has the largest contribution to the \emph{overall} MASSCW signal (in a specific-intensity sense) arriving at Earth?}
%One might also wonder, \emph{What is the typical redshift of the \emph{overall} MASSCW signal (in a flux-density sense)?}
One might also wonder, \emph{What is the mean redshift of the MASSCW signal --- not for a single direction, but `overall'?}
Due to strong attenuation with redshift, the MASSCW specific intensity is usually `high' for directions with a `low' specific-intensity-weighted mean redshift, and vice versa.
This means that, although the signal could originate from far away for most of the sky (given sufficiently small cluster and filament volume-filling fractions), the \emph{sky-averaged} redshift can still be low.
%sky-averaged
%This means that, although most of the sky could have its signal come from far away (given sufficiently small cluster and filament volume filling fractions), the sky-averaged, specific-intensity-weighted redshift can still be low.
%Therefore, we also calculate the typical \emph{overall} redshift of the MASSCW signal, by integrating over the sky and weighing each direction's specific-intensity-weighted mean redshift by the corresponding total specific intensity
We define the flux-density-weighted mean redshift of the MASSCW signal at observing frequency $\nu_\mathrm{obs}$ to be
\begin{align}
    \bar{\bar{z}} \coloneqq \left(\int_{\mathbb{S}^2} I_\nu\left(\hat{r},\nu_\mathrm{obs}\right) \mathrm{d}\Omega\right)^{-1} \int_{\mathbb{S}^2} \int_0^\infty z\ \frac{\mathrm{d}I_\nu}{\mathrm{d}z}\left(\hat{r},\nu_\mathrm{obs},z\right)\ \mathrm{d}z\ \mathrm{d}\Omega,
\end{align}
where $\bar{\bar{z}}$ is shorthand for $\bar{\bar{z}}\left(\nu_\mathrm{obs}\right)$.

\section{Redshift predictions from geometric cosmic web model}
\label{sec:model}
\subsection{Overcoming the redshift limitation}
\def\filamentWidth{.8}
\def\filamentLength{6}
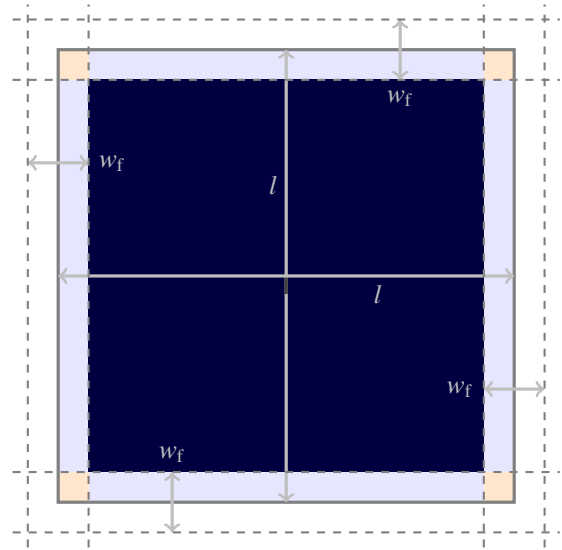
\begin{figure}[h!]
\centering
\begin{tikzpicture}
\filldraw [fill = blue!10!white, draw = none, very thick] (0, 0) rectangle (\filamentLength, \filamentLength);

\filldraw [fill = orange!20!white,  draw = none, very thick] (0, 0) rectangle (\filamentWidth/2, \filamentWidth/2);
\filldraw [fill = orange!20!white,  draw = none, very thick] (0, \filamentLength - \filamentWidth/2) rectangle (\filamentWidth/2, \filamentLength);
\filldraw [fill = orange!20!white,  draw = none, very thick] (\filamentLength - \filamentWidth/2, \filamentLength - \filamentWidth/2) rectangle (\filamentLength, \filamentLength);
\filldraw [fill = orange!20!white,  draw = none, very thick] (\filamentLength - \filamentWidth/2, 0) rectangle (\filamentLength, \filamentWidth/2);

\filldraw [fill = none, draw = gray, very thick] (0, 0) rectangle (\filamentLength, \filamentLength);
\filldraw [fill = blue!25!black, draw = none, very thick] (\filamentWidth/2, \filamentWidth/2) rectangle (\filamentLength - \filamentWidth/2, \filamentLength - \filamentWidth/2);

\draw [lightgray, <->, very thick] (\filamentLength/2, 0) -- (\filamentLength/2, \filamentLength);

\draw [darkgray, very thick] (\filamentLength/2, \filamentLength/2) -- (\filamentLength/2, \filamentLength*.46);

\draw [lightgray, <->, very thick] (0,\filamentLength/2) -- (\filamentLength, \filamentLength/2);
%\filldraw [fill=blue!25!black, draw = none] (\filamentLength/2, \filamentLength/2) circle (7pt) node[lightgray, anchor=center] {$l$};

\draw [lightgray] (\filamentLength*.7, \filamentLength/2) circle (0pt) node[anchor=north] {$l$};
\draw [lightgray] (\filamentLength/2, \filamentLength*.7) circle (0pt) node[anchor=east] {$l$};

\draw [lightgray, <->, very thick] (-\filamentWidth/2, \filamentLength/4*3) -- (\filamentWidth/2, \filamentLength/4*3) node[anchor=west] {$w_\mathrm{f}$};

\draw [lightgray, <->, very thick] (\filamentLength/4*3, \filamentLength + \filamentWidth/2) -- (\filamentLength/4*3, \filamentLength - \filamentWidth/2) node[anchor=north] {$w_\mathrm{f}$};

\draw [lightgray, <->, very thick] (\filamentLength + \filamentWidth/2, \filamentLength/4) -- (\filamentLength - \filamentWidth/2, \filamentLength/4) node[anchor=east] {$w_\mathrm{f}$};

\draw [lightgray, <->, very thick] (\filamentLength/4, -\filamentWidth/2) -- (\filamentLength/4, \filamentWidth/2) node[anchor=south] {$w_\mathrm{f}$};
%\draw [black] (0, \filamentLength/4*3) circle (0pt) node[anchor=north] {$w_\mathrm{f}$};

\draw[gray, thick, dashed] (-\filamentWidth/2, -\filamentWidth/4*3) -- (-\filamentWidth/2, \filamentLength + \filamentWidth/4*3);
\draw[gray, thick, dashed] (\filamentWidth/2, -\filamentWidth/4*3) -- (\filamentWidth/2, \filamentLength + \filamentWidth/4*3);

\draw[gray, thick, dashed] (\filamentLength - \filamentWidth/2, -\filamentWidth/4*3) -- (\filamentLength - \filamentWidth/2, \filamentLength + \filamentWidth/4*3);
\draw[gray, thick, dashed] (\filamentLength + \filamentWidth/2, -\filamentWidth/4*3) -- (\filamentLength + \filamentWidth/2, \filamentLength + \filamentWidth/4*3);

\draw[gray, thick, dashed] (-\filamentWidth/4*3, -\filamentWidth/2) -- (\filamentLength + \filamentWidth/4*3, -\filamentWidth/2);
\draw[gray, thick, dashed] (-\filamentWidth/4*3, \filamentWidth/2) -- (\filamentLength + \filamentWidth/4*3, \filamentWidth/2);

\draw[gray, thick, dashed] (-\filamentWidth/4*3, \filamentLength - \filamentWidth/2) -- (\filamentLength + \filamentWidth/4*3, \filamentLength - \filamentWidth/2);
\draw[gray, thick, dashed] (-\filamentWidth/4*3, \filamentLength + \filamentWidth/2) -- (\filamentLength + \filamentWidth/4*3, \filamentLength + \filamentWidth/2);

%\filldraw[black] (0,0) circle (2pt) node[anchor=west] {Intersection point};
\end{tikzpicture}
\caption{
%Geometry of the simple model, in which we treat the cosmic web as an amorphous solid with randomly-displaced cubic unit cells.
Geometry of the simple model, in which we treat the cosmic web as a collection of randomly-displaced cubic unit cells.
We show one face (grey square) of a unit cell, with edges of comoving length $l$.
Filaments of typical comoving width $w_\mathrm{f}$ and length $l_\mathrm{f} \coloneqq l - w_\mathrm{f}$ cover the light-blue-shaded region of the face, and extend to the faces of neighbouring unit cells (grey dashed lines).
Where filaments meet, galaxy clusters reside (orange-shaded regions), each of which is also connected to a filament oriented away from the observer.
The central part of the face represents a sheet (dark blue), behind which a large void looms (not depicted).
The figure is not to scale: filament widths are exaggerated compared to realistic filament lengths.}
\label{fig:modelGeometry}
\end{figure}\noindent
Although we can calculate the specific intensity function $I_\nu\left(\hat{r},\nu_\mathrm{obs}\right)$, the specific-intensity-weighted mean redshift function $\bar{z}\left(\hat{r},\nu_\mathrm{obs}\right)$, the single-direction redshift RV $\bar{Z}$ ECDF and the flux-density-weighted mean redshift $\bar{\bar{z}}\left(\nu_\mathrm{obs}\right)$ for each realisation of the MASSCW prior, all four quantities suffer from the fact that the BORG SDSS reconstructions stop at a redshift $z_\mathrm{max} = 0.2$.
It is of prime interest to know to what extent such redshift limitations, which we anticipate will become less stringent in the future, affect our inferences.
For example, if most of the real MASSCW signal originates from $z > 0.2$, the predictive power of our specific intensity function distribution would be limited.\\
Here we introduce a simple analytical model that allows us to calculate both the distribution of $\bar{Z}$, and $\bar{\bar{z}}$, for an arbitrary value of $z_\mathrm{max}$.
The model reproduces the results calculated from the MASSCW prior, if it is equally limited to $z_\mathrm{max} = 0.2$.
This provides tentative evidence that the model captures the essential elements of the MASSCW signal, so that it might be used for extrapolation.
We thus use the model to predict the distribution of $\bar{Z}$ and $\bar{\bar{z}}$, for the case that describes the actual Universe: $z_\mathrm{max} \to \infty$.
%We thus use the model to predict the single-direction mean redshift distribution and the all-sky mean redshift in the real-world case $z_\mathrm{max} \to \infty$.

\subsection{Model formulation}
\begin{figure}
    \centering
    \includegraphics[width=\columnwidth]{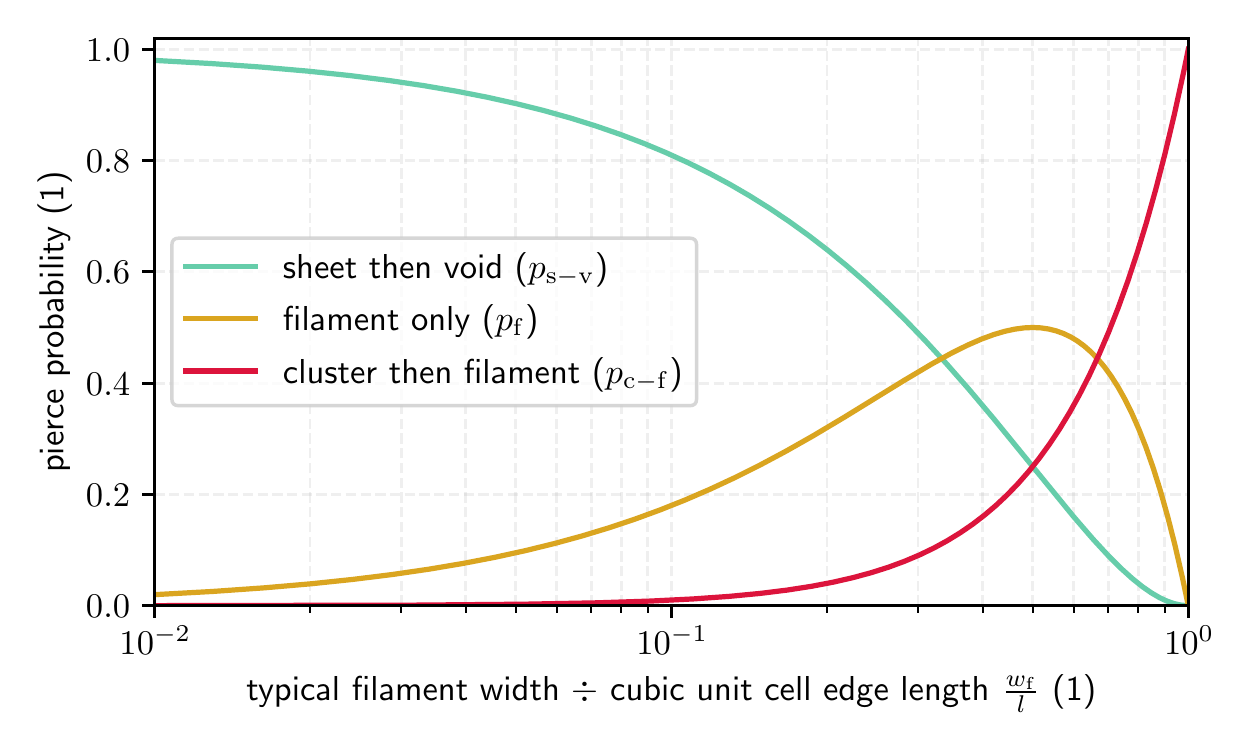}
    \caption{
    When a sightline enters a cubic unit cell of the cosmic web perpendicularly to a face (as in \textbf{Figure}~\ref{fig:modelGeometry}) and at a random position on it, one of three different LSS-piercing events occurs.
    %When a sightline enters a cubic unit cell of the cosmic web perpendicularly at a random position on a face as in \textbf{Figure}~\ref{fig:modelGeometry}, one of three different LSS-piercing events occurs.
    We show the probabilities for each event as a function of the ratio of the typical filament width and cubic unit cell edge length (see \textbf{Equations}~\ref{eq:pierceProbabilitySV}, \ref{eq:pierceProbabilityF} and \ref{eq:pierceProbabilityCF}).}
    %Probabilities for a sightline to pierce through a filament upon entering a cubic unit cell of the Cosmic Web at a random position on a face as in \textbf{Figure}~\ref{fig:modelGeometry}, given as a function of the ratio of the typical filament width and length. See \textbf{Equation}~\ref{eq:pierceProbability}.}
    \label{fig:modelProbability}
\end{figure}

%To retrieve general redshift properties of the MASSCW, we propose a simple geometric model.
To retrieve MASSCW redshift properties, we propose a simple geometric model.
The key idea is that the dominant MASSCW redshift for an arbitrary sightline can be calculated via a weighted sum of the redshifts of the clusters and filaments it passes through, where the weights are set by the geometry and MECs of the LSS pierced.
%, and cosmological attenuation factors. redshift-dependent MECs
The redshifts of the structures a sightline pierces through depend on the particular LSS realisation the observer is immersed in.
However, the cosmological principle dictates that LSS is statistically similar everywhere, and thus typical geometric cosmic web parameters must exist (and can be retrieved from numerical simulations; see e.g. \textcolor{blue}{\citet{Gheller12015}}).\\
Here we assume that sightlines pass through an arrangement of identical cubic unit cells --- ignoring morphological variation among clusters, filaments, sheets and voids --- with comoving edge length $l$.
The edges are surrounded by square cuboids, which represent filaments, of typical comoving width $w_\mathrm{f}$ and length $l_\mathrm{f} \coloneqq l - w_\mathrm{f}$ (and thus volume $w_\mathrm{f}^2 l_\mathrm{f}$).\footnote{Often, filaments are modelled as cylinders. In such cases, the length of the path of a sightline through a filament depends on the exact point of incidence on the unit cell boundary. However, in our simplistic geometric model, this effect does not arise.}
%The non-overlapping edges of these unit cells represent filaments, so that $l_\mathrm{f} \coloneqq l - w_\mathrm{f}$ is the typical filament length, and $w_\mathrm{f}$ is the typical filament width.
Galaxy clusters reside where filaments meet, and --- in this simplistic model --- are cubes with comoving edge length $w_\mathrm{f}$ (and thus volume $w_\mathrm{f}^3$).
By far the largest component of a unit cell is its central void, a cubical region of comoving edge length $l_\mathrm{f}$ (and thus volume $l_\mathrm{f}^3$).
Finally, unit cells contain sheets, which fill the regions bounded by filaments and voids.
They are of typical comoving length $l_\mathrm{f}$ and thickness $w_\mathrm{f}$ (and thus volume $l_\mathrm{f}^2 w_\mathrm{f}$).
\textbf{Figure}~\ref{fig:modelGeometry} depicts a face of a unit cell.
If we were to assume a perfect crystal structure for the cosmic web, some sightlines would never encounter clusters or filaments whilst others would consistently do so at regular (comoving) intervals.
To avoid this unphysical scenario, we assume that our sightline-of-interest enters every unit cell it encounters at a random position on the face of incidence.
For additional simplicity, we assume that the sightline always hits such faces perpendicularly.\\
If one would choose a point-of-incidence on the face shown in \textbf{Figure}~\ref{fig:modelGeometry} in uniform fashion, the sightline would hit a sheet and then a void (dark-blue-shaded area) with probability
\begin{align}
    p_\mathrm{s-v}\left(\frac{w_\mathrm{f}}{l}\right) = \frac{\left(l - w_\mathrm{f}\right)^2}{l^2} = \left(1 - \frac{w_\mathrm{f}}{l}\right)^2.
\label{eq:pierceProbabilitySV}
\end{align}
Note that $p_\mathrm{s-v} \to 1$ when $\frac{w_\mathrm{f}}{l} \to 0$, and $p_\mathrm{s-v} \to 0$ when $\frac{w_\mathrm{f}}{l} \to 1$, as required.
Likewise, the probability that the sightline hits a filament only (light-blue-shaded area) is
\begin{align}
    p_\mathrm{f}\left(\frac{w_\mathrm{f}}{l}\right) = \frac{2 w_\mathrm{f} \left(l - w_\mathrm{f}\right)}{l^2} = 2 \frac{w_\mathrm{f}}{l}\left(1-\frac{w_\mathrm{f}}{l}\right).
\label{eq:pierceProbabilityF}
\end{align}
%Finally, the probability that the sightline hits a cluster first, and then a filament with which it is parallel (orange-shaded area) is
Finally, the probability that the sightline hits a cluster first, and then the filament behind it (orange-shaded area) is
\begin{align}
    p_\mathrm{c-f}\left(\frac{w_\mathrm{f}}{l}\right) = \left(\frac{w_\mathrm{f}}{l}\right)^2.
\label{eq:pierceProbabilityCF}
\end{align}
\textbf{Figure}~\ref{fig:modelProbability} depicts these pierce probabilities as a function of the ratio between filament width and unit cell edge length $\frac{w_\mathrm{f}}{l}$.
For a comparison between the LSS volume-filling factors (VFFs) of this simple model and those from cosmological simulations, see \textbf{Appendix}~\ref{ap:VFFs}.
%If one would choose a point-of-incidence on the face shown in \textbf{Figure}~\ref{fig:modelGeometry} in uniform fashion, the sightline would hit a filament (light-blue-shaded area) with probability
%\begin{align}
%    p = 1 - \frac{\left(l - w_\mathrm{f}\right)^2}{l^2} = \frac{w_\mathrm{f}}{l}\left(2-\frac{w_\mathrm{f}}{l}\right).
%\label{eq:pierceProbability}
%\end{align}
%Note that $p \to 0$ when $\frac{w_\mathrm{f}}{l} \to 0$, and $p \to 1$ when $\frac{w_\mathrm{f}}{l} \to 1$, as required. See \textbf{Figure}~\ref{fig:modelProbability}.\\
\\
Pick a sightline, and successively label each unit cell boundary this sightline crosses with a natural number $n \in \mathbb{N}_{\geq 1}$.
%Let $X_n$ be the RV denoting whether or not the sightline hits a filament (thereby picking up MASSCW signal) at the $n$-th crossing (i.e. upon entering the $n$-th new unit cell from the observer), so that $X_n \sim \mathrm{Bernoulli}\left(p\right)$ for all $n$.
%, with the contribution obtained by piercing through a filament only normalised to 1.
If the first boundary crossing occurs at a comoving distance $d_1$, then the $n$-th crossing happens at comoving distance $d_n = d_1 + \left(n - 1\right) l$.
Let $\mathfrak{M}$ be the cosmological model of preference, and let $z_\mathfrak{M}\left(d\right)$ denote the function that converts comoving distance to cosmological redshift under this model.
The unit cell boundary crossings occur at redshifts $z_n \coloneqq z_\mathfrak{M}\left(d_n\right)$ for all $n \in \mathbb{N}_{\geq 1}$.\\
Because we want to be able to calculate the dominant redshift of the MASSCW signal produced by filaments and clusters up to a given cosmological redshift $z_\mathrm{max}$ only, we introduce a parameter $N$.
Intuitively, $N \in \mathbb{N}_{\geq 1}$ is the label of the last boundary crossing within the LSS considered.
More formally, $N \coloneqq \max\ \{n \in \mathbb{N}_{\geq 1}\ |\ z_n < z_\mathrm{max} \}$.\\
Let $X_n\left(\nu_\mathrm{obs}\right)$ be the RV denoting the contribution to the MASSCW specific intensity at observing frequency $\nu_\mathrm{obs}$ picked up by the sightline during the $n$-th full unit cell crossing (i.e. whilst travelling through the $n$-th \emph{newly entered} unit cell; the unit cell the observer resides in does not contribute).
To retain low complexity, we assign the complete specific intensity contribution from the $n$-th crossing to the redshift of incidence $z_n$.
It is useful to regard $X_n$ as an instance of a more general random variable: $X_n\left(\nu_\mathrm{obs}\right) \coloneqq X\left(\nu_\mathrm{obs}, z_n\right)$.
$X\left(\nu_\mathrm{obs}, z\right)$ is the RV with support $\left(0, \Delta I_{\nu,\mathrm{f}}\left(\nu_\mathrm{obs}, z\right), \Delta I_{\nu,\mathrm{c-f}}\left(\nu_\mathrm{obs}, z\right)\right)$, and corresponding probabilities $\left(p_\mathrm{s-v}, p_\mathrm{f}, p_\mathrm{c-f}\right)$.
Note that we assume that sheets and voids have a vanishing contribution to the MASSCW signal.
Here,
\begin{align}
    \Delta I_{\nu,\mathrm{f}}\left(\nu_\mathrm{obs}, z\right) &\coloneqq w_\mathrm{f} j_{\nu,\mathrm{f}}\left(\nu_\mathrm{obs}\left(1+z\right), z\right)\left(1+z\right)^{-4}\\
    \Delta I_{\nu,\mathrm{c-f}}\left(\nu_\mathrm{obs}, z\right) &\coloneqq (w_\mathrm{f}j_{\nu,\mathrm{c}}\left(\nu_\mathrm{obs}\left(1+z\right), z\right) \nonumber \\
    &+ l_\mathrm{f}j_{\nu,\mathrm{f}}\left(\nu_\mathrm{obs}\left(1+z\right), z\right))\left(1+z\right)^{-4}.
\end{align}
%where we used the abbreviations $\Delta I_{\nu,\mathrm{f}}\left(\nu_\mathrm{obs}, z\right)$ and $\Delta I_{\nu,\mathrm{c-f}}\left(\nu_\mathrm{obs}, z\right)$.
%\textcolor{red}{The factor $\left(1+z\right)^{-4}$ follows from combining \textbf{Equations}~\ref{eq:timeAndComovingRadialDistance} and \ref{eq:specificIntensityTime} to obtain an expression for specific intensity as an integral over comoving radial distance.}
The factor $\left(1+z\right)^{-4}$ follows from \textbf{Equation}~\ref{eq:specificIntensityComovingRadialDistance}: the specific intensity as an integral over comoving radial distance.
As we assume LSS to occur at regularly-spaced radial comoving distances $d_n$, this integral can be approximated by a Riemann sum, where each term equals the integrand multiplied by $\Delta r = l$.\\
Although redshift-dependent, we assume the MEC to be constant (at a given emission frequency) within --- but different between --- clusters and filaments of the same unit cell.
Respectively, $j_{\nu,\mathrm{f}}\left(\nu, z\right)$ and $j_{\nu,\mathrm{c}}\left(\nu, z\right)$ represent the typical proper filament and cluster SCW MECs at emission frequency $\nu$ and cosmological redshift $z$.
We use the dimensionless parameter $\mathcal{C}$ to denote the typical cluster-to-filament-SCW MEC ratio at some reference emission frequency $\nu = \nu_\mathrm{ref}$ and $z = 0$: $j_{\nu,\mathrm{c}}\left(\nu_\mathrm{ref}, 0\right) \coloneqq \mathcal{C} j_{\nu,\mathrm{f}}\left(\nu_\mathrm{ref}, 0\right)$.
To retain minimal complexity, we propose to describe the spectral and temporal dependencies of both $j_{\nu,\mathrm{f}}\left(z\right)$ and $j_{\nu,\mathrm{c}}\left(z\right)$ as power laws in $\frac{\nu}{\nu_\mathrm{ref}}$ and $\left(1 + z\right)$:
\begin{align}
    %\begin{cases}
    j_{\nu,\mathrm{f}}\left(\nu, z\right) &= j_{\nu,\mathrm{f}}\left(\nu_\mathrm{ref}, 0\right) \left(\frac{\nu}{\nu_\mathrm{ref}}\right)^{\alpha_\mathrm{f}} \left(1+z\right)^{\beta_\mathrm{f}}\\
    j_{\nu,\mathrm{c}}\left(\nu, z\right) &= j_{\nu,\mathrm{c}}\left(\nu_\mathrm{ref}, 0\right) \left(\frac{\nu}{\nu_\mathrm{ref}}\right)^{\alpha_\mathrm{c}} \left(1+z\right)^{\beta_\mathrm{c}},%, j_{\nu,\mathrm{c}}\left(\nu_\mathrm{ref}, 0\right) \coloneqq \mathcal{C} j_{\nu,\mathrm{f}}\left(\nu_\mathrm{ref}, 0\right)
    %\end{cases}
\end{align}
with $\alpha_\mathrm{f}$, $\beta_\mathrm{f}$, $\alpha_\mathrm{c}$ and $\beta_\mathrm{c}$ constants.
%Here, $X\left(z\right)$ is the RV with support $\left(0, w_\mathrm{f} j_{\nu,\mathrm{f}}\left(z\right)\left(1+z\right)^{-4}, \left(w_\mathrm{f}j_{\nu,\mathrm{c}}\left(z\right) + l_\mathrm{f}j_{\nu,\mathrm{f}}\left(z\right)\right)\left(1+z\right)^{-4}\right)$, and probability distribution $\left(p_\mathrm{s-v}, p_\mathrm{f}, p_\mathrm{c-f}\right)$.
%Under these conditions, the $X_n$ are IID discrete RVs with support $\left(0, 1, \frac{l}{w_\mathrm{f}} + \mathcal{C} - 1\right)$, and probability distribution $\left(p_\mathrm{s-v}, p_\mathrm{f}, p_\mathrm{c-f}\right)$.
%The expectation value of $X_n$ is
Under this choice of MEC function parametrisation, the expectation value of $X\left(\nu_\mathrm{obs}, z\right)$ becomes
\begin{align}
    &\mathbb{E}\left(X\left(\nu_\mathrm{obs}, z\right)\right) = p_\mathrm{f} \Delta I_{\nu,\mathrm{f}}\left(\nu_\mathrm{obs}, z\right) + p_\mathrm{c-f} \Delta I_{\nu,\mathrm{c-f}}\left(\nu_\mathrm{obs}, z\right)\\
    &= \left(1+z\right)^{-4}\left(\frac{w_\mathrm{f}}{l}\right)^2 j_{\nu,\mathrm{f}}\left(\nu_\mathrm{ref}, 0\right) \cdot \nonumber \\
    &\left(3 l_\mathrm{f}\left(\frac{\nu_\mathrm{obs}}{\nu_\mathrm{ref}}\right)^{\alpha_\mathrm{f}}\left(1+z\right)^{\alpha_\mathrm{f} + \beta_\mathrm{f}} + \mathcal{C} w_\mathrm{f} \left(\frac{\nu_\mathrm{obs}}{\nu_\mathrm{ref}}\right)^{\alpha_\mathrm{c}}\left(1+z\right)^{\alpha_\mathrm{c} + \beta_\mathrm{c}}\right).
\label{eq:SIContributionExpectationValue}
\end{align}

\subsection{Specific-intensity-weighted mean redshift}
%Usually, filaments are thought of as cylindrical (at least locally), so that the within-filament length of a sightline piercing a filament at some unit cell boundary crossing depends on the exact location \emph{where} the sightline is incident.
%For simplicity, we ignore this complication, instead assigning an equal specific intensity contribution to all sightlines that pierce filaments at a given redshift $z_n$.
%This specific intensity contribution, including cosmological attenuation effects, is proportional to $\left(1 + z_n\right)^{-9}$.
%This is seen by first combining \textbf{Equations}~\ref{eq:timeAndComovingRadialDistance} and \ref{eq:specificIntensityTime} to obtain an expression for specific intensity as an integral over comoving radial distance.
%As we assume LSS to occur at regularly-spaced radial comoving distances $d_n$, this integral can be approximated by a Riemann sum, where each term equals the integrand multiplied by $\Delta r = l$.
%\footnote{It seems reasonable, however, to assume at least some redshift evolution of the typical cluster and filament magnetic field strength, with stronger fields at lower redshifts. This would further steepen the redshift scaling.}\\
Consider the random vector $\left[X_1, X_2, ..., X_N\right]^T$.
Whenever $\max\{X_1, X_2, ..., X_N\} > 0$, we can define the MASSCW specific-intensity-weighted mean redshift RV
%Whenever $1 \in \{X_1, X_2, ..., X_N\}$, we can define the dominant MASSCW redshift
%After all, although filaments in the actual world have varying lengths and thicknesses, \emph{typical} parameter values are known from numerical simulations.
%Besides, the 
%that this average redshift is high for most of the sky, but contribute 
%Another interesting question is to ask what the dominant redshift is of the MASSCW signal as a whole, integrating over the sky and weighing each sky direction's emissivity-weighted redshift by 
\begin{align}
    \bar{Z}\left(\nu_\mathrm{obs}\right) \coloneqq \frac{\sum_{n=1}^N z_n X_n\left(\nu_\mathrm{obs}\right)}{\sum_{n=1}^N X_n\left(\nu_\mathrm{obs}\right)}.
    %\bar{Z} = \frac{\sum_{n=1}^N X_n z_n \left(1 + z_n\right)^{-9}}{\sum_{n=1}^N X_n \left(1 + z_n\right)^{-9}}.
\label{eq:redshiftWeightedSpecificIntensityRV}
\end{align}
We stress that $\bar{Z}$, being a deterministic function of the RVs $\{X_1, X_2, ..., X_N\}$, is itself an RV.
It quantifies the variety of specific-intensity-weighted mean redshifts that could occur in \emph{different} LSS realisations (with the same model parameters) for a \emph{fixed} direction (sightline) of the sky.
Conversely, it could be interpreted as representing the redshift variety that occurs in a \emph{fixed} LSS realisation for \emph{different} directions of the sky, provided that the sky contains many (almost) independent patches of LSS.
%Conversely, it could be interpreted as representing the redshift variety that occurs in a \emph{fixed} LSS realisation for \emph{different} directions of the observer's sky, provided that filaments are small on the sky - in other words, that the sky contains many (almost) independent patches of LSS.
This second interpretation invites for a comparison between the specific-intensity-weighted mean redshift distribution of our MASSCW prior skies, and the distribution of $\bar{Z}$.\\
Because $X_n\left(\nu_\mathrm{obs}\right)$ features in both numerator and denominator of the fraction in \textbf{Equation}~\ref{eq:redshiftWeightedSpecificIntensityRV}, redshift-independent multiplicative factors cancel out.
Therefore, $j_{\nu,\mathrm{f}}\left(\nu_\mathrm{ref}, 0\right)$ need not be specified; also, $\bar{Z}$ becomes $\nu_\mathrm{obs}$-independent if $\alpha_\mathrm{f} = \alpha_\mathrm{c}$.
In general, the distribution of $\bar{Z}$ is determined by the cosmological model $\mathfrak{M}$ and 9 additional parameters ($w_\mathrm{f}$, $l$, $d_1$, $z_\mathrm{max}$, $\alpha_\mathrm{f}$, $\beta_\mathrm{f}$, $\alpha_\mathrm{c}$, $\beta_\mathrm{c}$ and $\mathcal{C}$).
Using $d_1$, we can force the filaments and clusters nearest to the observer to occur at a fixed comoving distance.
Alternatively, one could randomise the observer's position with respect to the unit cells by choosing $d_1 \sim \mathrm{Uniform}\left(0, l\right)$, making it an RV.
%, where the RV $U \sim \mathrm{Uniform}\left(0, 1\right)$.
In such case, 8 parameters remain that determine the distribution of $\bar{Z}$.

\subsection{Flux-density-weighted mean redshift}
The geometric model also allows to calculate the flux-density-weighted mean (i.e. sky-averaged) MASSCW redshift $\bar{\bar{z}}\left(\nu_\mathrm{obs}\right)$.
We do so by considering $M \in \mathbb{N}_{\geq 1}$ sightlines (instead of one) and their associated specific-intensity-weighted mean redshift RVs $\{\bar{Z}_1, \bar{Z}_2, ..., \bar{Z}_M\}$ in the sense of \textbf{Equation}~\ref{eq:redshiftWeightedSpecificIntensityRV}.
In total, these RVs depend on a set of $N \cdot M$ discrete RVs $X_{nm}$, which can be partitioned into $N$ subsets of $M$ IID RVs: $X_{nm} \sim X_n$.
% \sim \mathrm{Bernoulli}\left(p\right)$.
We find $\bar{\bar{z}}$ by summing the elements of $\{\bar{Z}_1, \bar{Z}_2, ..., \bar{Z}_M\}$, each weighted by the corresponding specific intensity, and dividing the result by the sum of these weights; all whilst we let $M \to \infty$:
\begin{align}
    \bar{\bar{z}}\left(\nu_\mathrm{obs}\right) &\coloneqq \lim_{M \to \infty} \frac{\sum_{m=1}^M \sum_{n=1}^N z_n X_{nm}\left(\nu_\mathrm{obs}\right)}{\sum_{m=1}^M \sum_{n=1}^N X_{nm}\left(\nu_\mathrm{obs}\right)}\\
    %\bar{\bar{z}} &\coloneqq \lim_{M \to \infty} \frac{\sum_{m=1}^M \sum_{n=1}^N X_{nm} z_n \left(1 + z_n\right)^{-9}}{\sum_{m=1}^M \sum_{n=1}^N X_{nm} \left(1 + z_n\right)^{-9}}\\
    &= \lim_{M \to \infty} \frac{\sum_{n=1}^N \left(z_n M^{-1} \sum_{m=1}^M X_{nm}\left(\nu_\mathrm{obs}\right)\right)}{\sum_{n=1}^N \left(M^{-1} \sum_{m=1}^M X_{nm}\left(\nu_\mathrm{obs}\right)\right)}\\
    %&= \lim_{M \to \infty} \frac{\sum_{n=1}^N \left(z_n \left(1 + z_n\right)^{-9} M^{-1} \sum_{m=1}^M X_{nm}\right)}{\sum_{n=1}^N \left(\left(1 + z_n\right)^{-9} M^{-1} \sum_{m=1}^M X_{nm}\right)}\\
    &= \frac{\sum_{n=1}^N z_n \mathbb{E}\left(X_n\left(\nu_\mathrm{obs}\right)\right)}{\sum_{n=1}^N \mathbb{E}\left(X_n\left(\nu_\mathrm{obs}\right)\right)}.
    %&= \frac{\sum_{n=1}^N z_n \left(1 + z_n\right)^{-9} p}{\sum_{n=1}^N \left(1 + z_n\right)^{-9} p}\\
    %&= \frac{\sum_{n=1}^N z_n \left(1 + z_n\right)^{-9}}{\sum_{n=1}^N \left(1 + z_n\right)^{-9}}.
\label{eq:redshiftWeightedFluxDensity}
\end{align}
Here we use the fact that the limit of a ratio of two sequences equals the ratio of the sequences' limits (provided that the denominator sequence does not converge to $0$), and that the sample mean $M^{-1} \sum_{m=1}^M X_{nm} \to \mathbb{E}\left(X_n\right)$ as $M \to \infty$, for all $n \in \{1, 2, ..., N\}$.
Like before, as $\mathbb{E}\left(X_n\left(\nu_\mathrm{obs}\right)\right)$ features in both numerator and denominator of the fraction in \textbf{Equation}~\ref{eq:redshiftWeightedFluxDensity}, redshift-independent multiplicative factors cancel out.
Thus, $\bar{\bar{z}}$ is $\nu_\mathrm{obs}$-independent when $\alpha_\mathrm{f} = \alpha_\mathrm{c}$.\\
%Here we use that the sample mean $M^{-1} \sum_{m=1}^M X_{nm}$ has expectation value $p$ and vanishing variance as $M \to \infty$.
\textbf{Equation}~\ref{eq:SIContributionExpectationValue} makes clear that clusters dominate over filaments at cosmological redshift $z$ when
\begin{align}
    \mathcal{C} \gg 3 \frac{l_\mathrm{f}}{w_\mathrm{f}}\left(\frac{\nu_\mathrm{obs}}{\nu_\mathrm{ref}}\right)^{\alpha_\mathrm{f}-\alpha_\mathrm{c}}\left(1+z\right)^{\alpha_\mathrm{f} - \alpha_\mathrm{c} + \beta_\mathrm{f} - \beta_\mathrm{c}}.
\label{eq:clusterDomination}
\end{align}
At $\nu_\mathrm{obs} = \nu_\mathrm{ref}$ and $z = 0$, this inequality reduces to
\begin{align}
    \mathcal{C} \gg 3 \left(\frac{l}{w_\mathrm{f}}-1\right),
\end{align}
which is amply satisfied (see \textbf{Section}~\ref{sec:parameterEstimates} for parameter estimates).
Near $\nu_\mathrm{obs} = \nu_\mathrm{ref}$ and for $z$ of order unity, the inequality continues to hold as long as $\alpha_\mathrm{f} - \alpha_\mathrm{c} + \beta_\mathrm{f} - \beta_\mathrm{c}$ is a number of order unity at most.
In addition, most redshifts high enough that the inequality is violated, do not contribute meaningfully to $\bar{\bar{z}}$.
Therefore, $\bar{\bar{z}}$ around $\nu_\mathrm{obs} = \nu_\mathrm{ref}$ is typically dominated by clusters, and $\bar{\bar{z}}\left(\nu_\mathrm{obs}\right) \approx \bar{\bar{z}}$.
As long as \textbf{Equation}~\ref{eq:clusterDomination} holds, $\bar{\bar{z}}$ becomes insensitive to variations in $\mathcal{C}$, $w_\mathrm{f}$, $\alpha_\mathrm{f}$ and $\beta_\mathrm{f}$; setting $z_\mathrm{max} = \infty$, just \emph{3} significant parameters remain: $l$, $d_1$ and $\alpha_\mathrm{c}+\beta_\mathrm{c}$.\\
Finally, it is of significant interest to calculate $\bar{\bar{z}}$ for the filaments' SCW signal \emph{only}, discarding the dominant SCW signal from merger and accretion shocks in galaxy clusters.
To find a filament-only $\bar{\bar{z}}$ with the geometric model, we just set $\mathcal{C} = 0$.
This simplifies the expression for $\mathbb{E}\left(X\left(\nu_\mathrm{obs}, z\right)\right)$, and reduces the number of relevant parameters considerably.
When $\mathcal{C} = 0$, $\bar{\bar{z}}$ becomes not only independent of $\alpha_\mathrm{c}$ and $\beta_\mathrm{c}$, but also of $w_\mathrm{f}$ and $\nu_\mathrm{obs}$:
\begin{align}
    \bar{\bar{z}}\left(\nu_\mathrm{obs}\right) = \bar{\bar{z}} = \frac{\sum_{n=1}^N z_n \left(1+z_n\right)^{-4 + \alpha_\mathrm{f} + \beta_\mathrm{f}}}{\sum_{n=1}^N \left(1+z_n\right)^{-4 + \alpha_\mathrm{f} + \beta_\mathrm{f}}}.
\end{align}
As before, if we set $z_\mathrm{max} = \infty$, $\bar{\bar{z}}$ is a function of just \emph{3} parameters: $l$, $d_1$ and $\alpha_\mathrm{f} + \beta_\mathrm{f}$, in this case.
%Note that $\bar{\bar{z}}$ is a function of $l$, $d_1$, $z_\mathrm{max}$ and $\mathfrak{M}$ only - it does \emph{not} depend on $w_\mathrm{f}$ or $\mathcal{C}$!

\subsection{Parameter estimates}
\label{sec:parameterEstimates}
\begin{figure}
    \centering
    \includegraphics[width=\columnwidth]{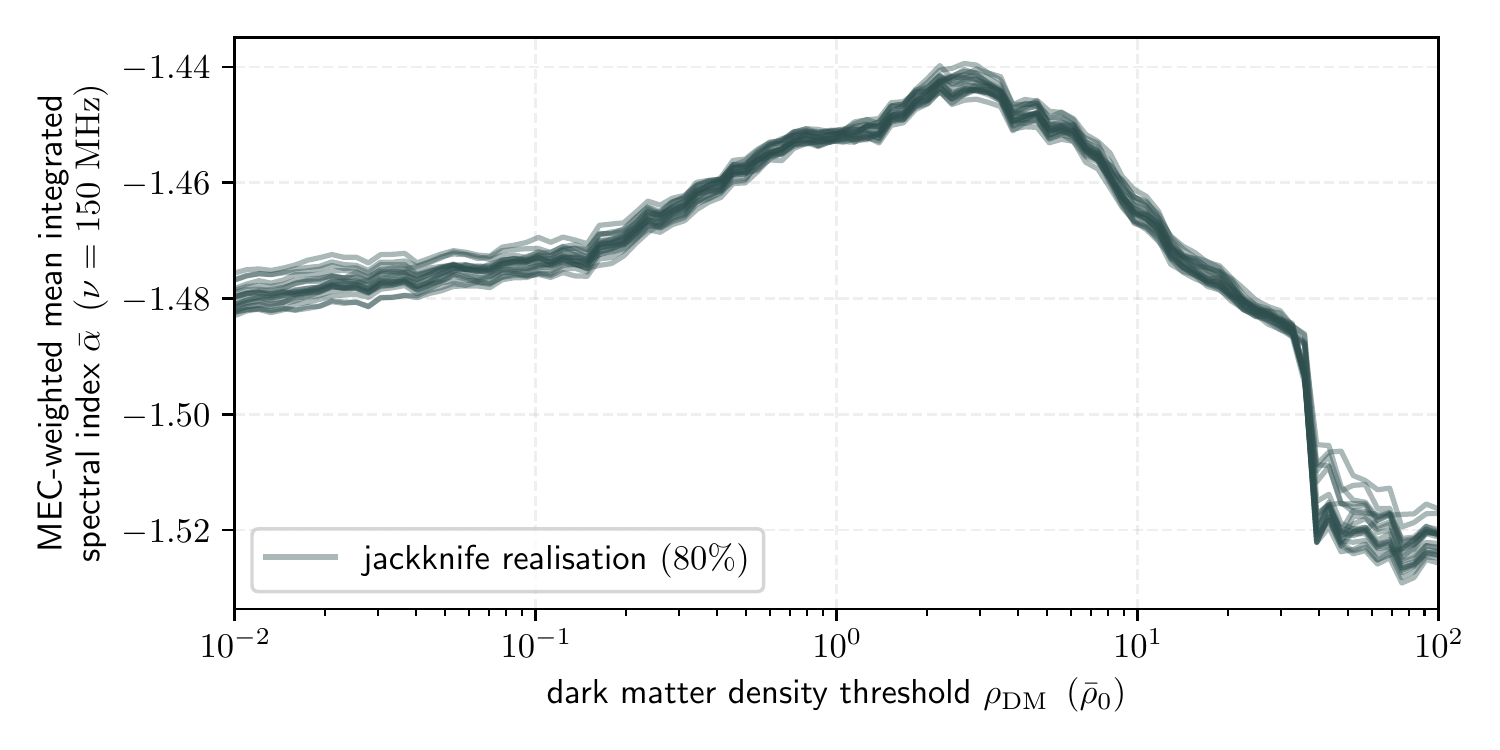}
    \caption{The monochromatic-emission-coefficient-weighted mean integrated spectral index $\bar{\alpha}$ at an emission frequency of $\nu = 150\ \mathrm{MHz}$, including all shocks that occur at locations where the dark matter (DM) density $\rho_\mathrm{DM}$ does not exceed some threshold. Larger thresholds incorporate shocks from a wider range in $\rho_\mathrm{DM}$, eventually including shocks from both filament and cluster environments. To provide an idea of the uncertainty in $\bar{\alpha}$, we use jackknife realisations that each contain $80\%$ of the shocked voxels present in our Enzo snapshot at $z = 0.025$.}
    \label{fig:EnzoSpectralIndexMean150MHz}
\end{figure}
This subsection provides concrete parameter estimates, which are necessary to run the model in practice.\\
Cosmological simulations indicate $w_\mathrm{f} \sim 10^0 - 10^1\ \mathrm{Mpc}$, while $l \sim 10^1 - 10^2\ \mathrm{Mpc}$.
The ratio of these parameters is constrained too: \textbf{Table}~\ref{tab:VFFs} shows that a VFF comparison between the geometric model and cosmological simulations favours $\frac{w_\mathrm{f}}{l} \sim 10^{-1}$.\\
From \textbf{Figure}~\ref{fig:MHDPMFConditional}, we estimate $\mathcal{C} \sim 10^5 - 10^7$ for $\nu_\mathrm{ref} = 150\ \mathrm{MHz}$, assuming filament environments are characterised by $\rho \sim \bar{\rho}_0$, and cluster environments are characterised by $\rho \sim 10^2\ \bar{\rho}_0$ (at the ${\sim}3\ \mathrm{Mpc}\ h^{-1}$ resolution, at least).
Technically, \textbf{Figure}~\ref{fig:MHDPMFConditional} does not show $P\left(j_\nu\ \vert\ \rho\right)$ for $z = 0$ (the redshift at which $\mathcal{C}$ is defined to be the cluster-to-filament SCW MEC ratio), but for $z = 0.025$.
However, for our purposes, the evolution of this relationship is likely of negligible importance.\\
To find appropriate values for $\alpha_\mathrm{f}$ and $\alpha_\mathrm{c}$, we calculate the MEC-weighted mean integrated spectral index $\bar{\alpha}$ from our Enzo simulation snapshot as a function of shock environment.
\textbf{Figure}~\ref{fig:EnzoSpectralIndexMean150MHz} shows the result when weighing by MECs at emission frequency $\nu = 150\ \mathrm{MHz}$; in general, $\bar{\alpha} = \bar{\alpha}\left(\nu\right)$.
Nevertheless, it seems that $\alpha_\mathrm{f} = \alpha_\mathrm{c} = -\sfrac{3}{2}$ are reasonable choices.\\
Next, to find $\beta_\mathrm{f}$ and $\beta_\mathrm{c}$, we revisit the single-shock MEC expression of \textbf{Equation}~\ref{eq:properMonochromaticEmissionCoefficient}.
Inverse Compton (IC) scattering to the CMB contributes a factor $\left(1+z\right)^{-4}$ to the single-shock MEC as $j_\nu \propto B_\mathrm{CMB}^{-2}$ and $B_\mathrm{CMB} \propto \left(1+z\right)^2$ (see \textbf{Appendix}~\ref{ap:scalingRelation}).
Further factors of $\left(1+z\right)$ follow by considering the typical redshift evolution of the proper BM density $\rho_\mathrm{BM}$, proper magnetic field strength $B$, proper shock velocity $v$ relative to the upstream IGM and \emph{comoving} shock number density; all for both filament and cluster environments.
In the linear regime of structure formation within an Einstein--de Sitter universe, the density contrast is proportional to $\left(1+z\right)^{-1}$. As the proper \emph{mean} matter density is proportional to $\left(1+z\right)^3$, the proper BM density in filaments is expected to be proportional to $\left(1+z\right)^2$.
In clusters, the density contrast grows more strongly; numerical simulations suggest that the density contrast is roughly proportional to $\left(1+z\right)^{-2}$.
Likewise, this would imply that the proper BM density in clusters is proportional to $\left(1+z\right)$.
Assuming that the magnetic field strength only evolves due to field line compression, we have $B \propto \rho_\mathrm{BM}^{\sfrac{2}{3}}$ (see \textbf{Appendix}~\ref{ap:scalingRelation}).\footnote{Ignoring magnetogenesis by outflows from AGN and supernovae, this Ansatz likely underestimates the actual magnetic field strength growth with time.}
In filaments then, $B_\mathrm{f} \propto \left(1+z\right)^{\sfrac{4}{3}}$, whilst in clusters $B_\mathrm{c} \propto \left(1+z\right)^{\sfrac{2}{3}}$.
As the single-shock MEC is proportional to $B^{1-\alpha}$, the proper magnetic field strength approximately contributes a factor $\left(1+z\right)^{\sfrac{10}{3}}$ for filaments, and a factor $\left(1+z\right)^{\sfrac{5}{3}}$ for clusters.
Lacking further knowledge, we apply Occam's razor and assume that the proper shock velocity and comoving shock number density do not evolve with redshift.\\
Collecting factors of $\left(1+z\right)$, we find $\beta_\mathrm{f} = \sfrac{4}{3}$ and $\beta_\mathrm{c} = -\sfrac{4}{3}$.
We stress that these values are highly uncertain and should be used as indications only.
%If no cosmological redshift evolution of the typical cluster and filament baryon density, temperature and magnetic field strength is assumed, the aforementioned steep redshift scaling is obtained.\\
%The integrand features the proper MEC $j_\nu\left(r\ \hat{r}, z, \nu_\mathrm{obs}\left(1+z\right)\right)$, for which we use \textbf{Equation}~\ref{eq:properMonochromaticEmissionCoefficient}.
%Assuming $\alpha = -1$, cosmological redshifting provides a factor $\left(1+z\right)^{-1}$, whilst inverse Compton (IC) scattering to the CMB provides another factor $\left(1+z\right)^{-4}$ as $j_\nu \propto B_\mathrm{CMB}^{-2}$ and $B_\mathrm{CMB} \propto \left(1+z\right)^2$ (see \textbf{Appendix}~\ref{ap:scalingRelation}).
%\begin{align}
%    \mathbb{E}\left(X_n\right) = 3 \frac{w_\mathrm{f}}{l} + \left(\mathcal{C}-3\right)\left(\frac{w_\mathrm{f}}{l}\right)^2.
%\end{align}
%Instead of modelling the Cosmic Web as a pure crystal, in which some sightlines would never encounter filaments while others would be guaranteed to do so at regular (comoving) intervals, we assume that the ray enters every new unit cell at a random position on the face it is incident to.

\section{Results}
\label{sec:results}
\begin{figure*}
\centering
\begin{subfigure}{\textwidth}
    \includegraphics[width=\textwidth]{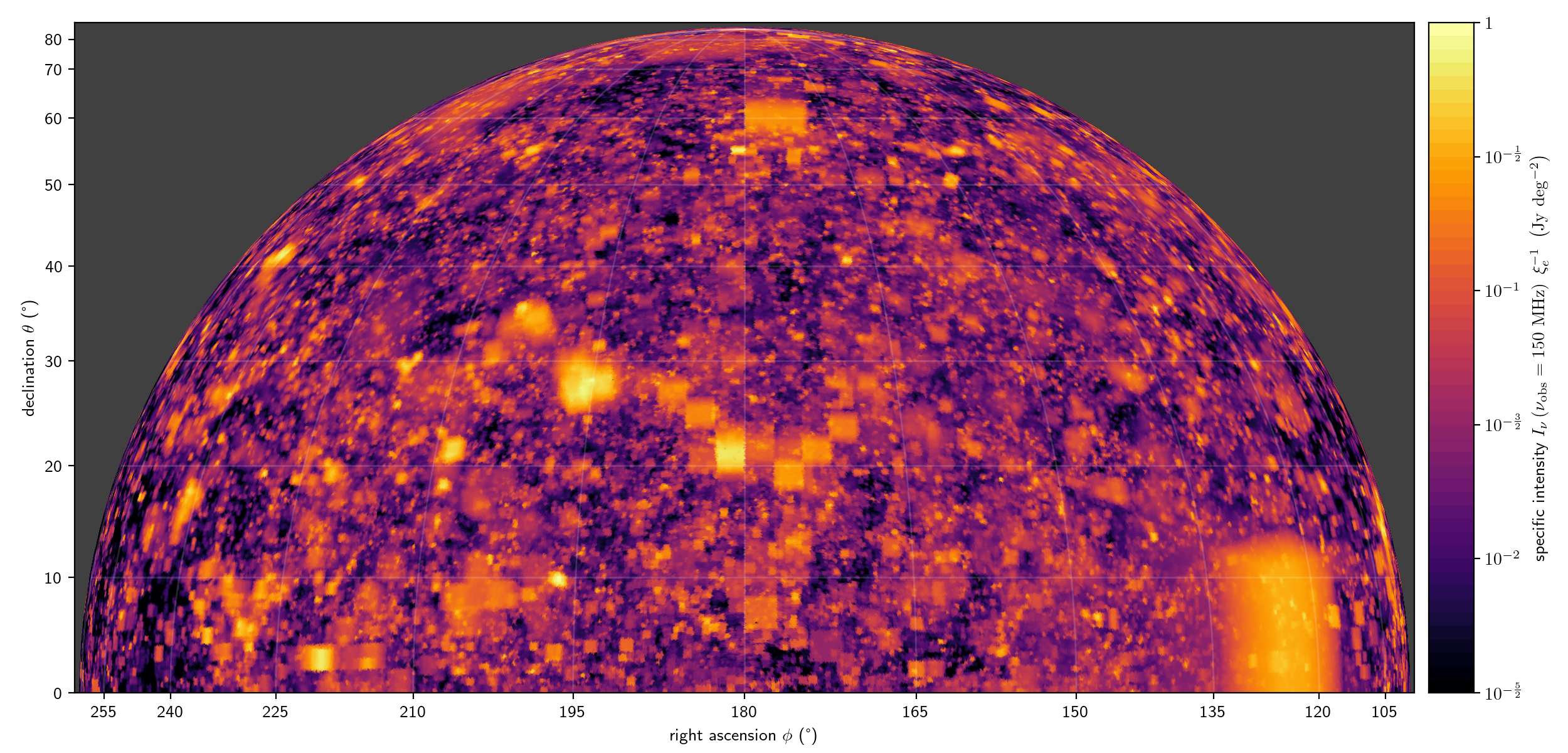}
    \subcaption{}
    \label{fig:realisationSphereSpecificIntensity}
\end{subfigure}
\begin{subfigure}{\textwidth}
    \includegraphics[width=\textwidth]{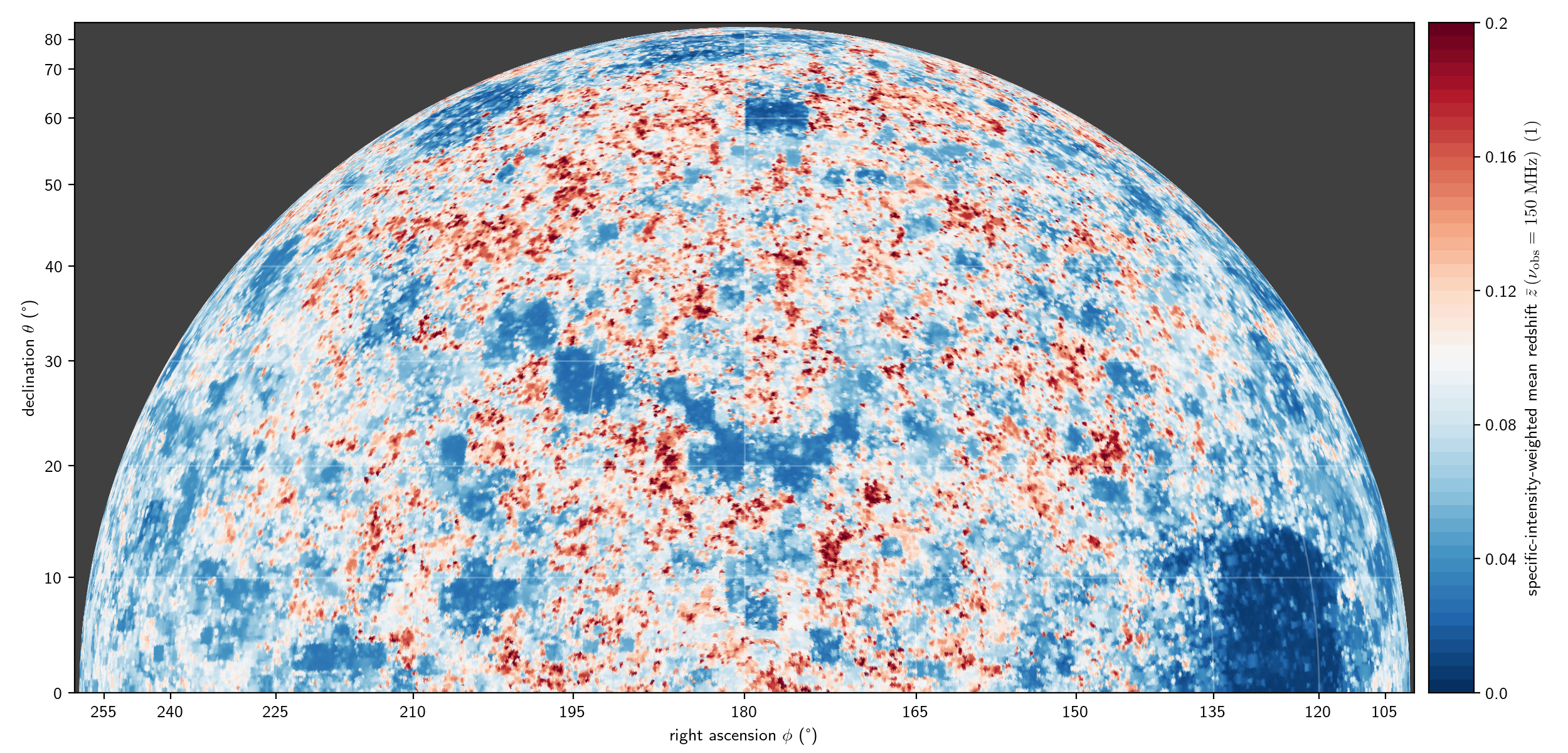}
    \subcaption{}
    \label{fig:realisationSphereRedshift}
\end{subfigure}
\caption{Realisation of the merger- and accretion-shocked synchrotron Cosmic Web (MASSCW) priors at observing frequency $\nu_\mathrm{obs} = 150\ \mathrm{MHz}$. We show exactly \sfrac{1}{4} of the total sky. \textbf{Top:} specific intensity $I_\nu$. \textbf{Bottom:} specific-intensity-weighted mean (cosmological) redshift $\bar{z}$.}
\end{figure*}

\subsection{MASSCW specific intensity function distribution: general results}
\begin{figure*}
    \centering
\begin{subfigure}{\textwidth}
    \includegraphics[width=\textwidth]{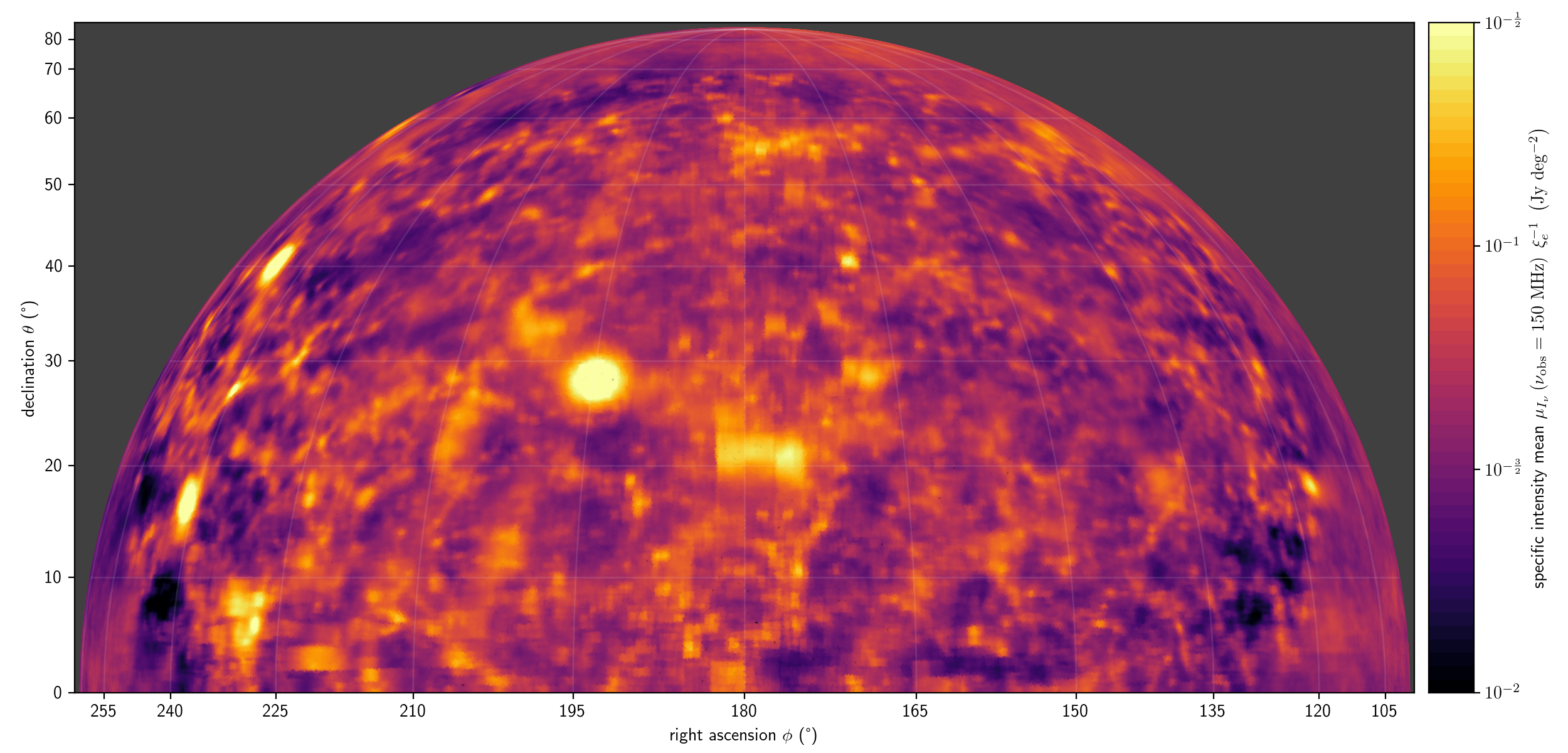}
    \subcaption{}
    \label{fig:specificIntensityMeanSphere150MHz}
\end{subfigure}
\begin{subfigure}{\textwidth}
    \includegraphics[width=\textwidth]{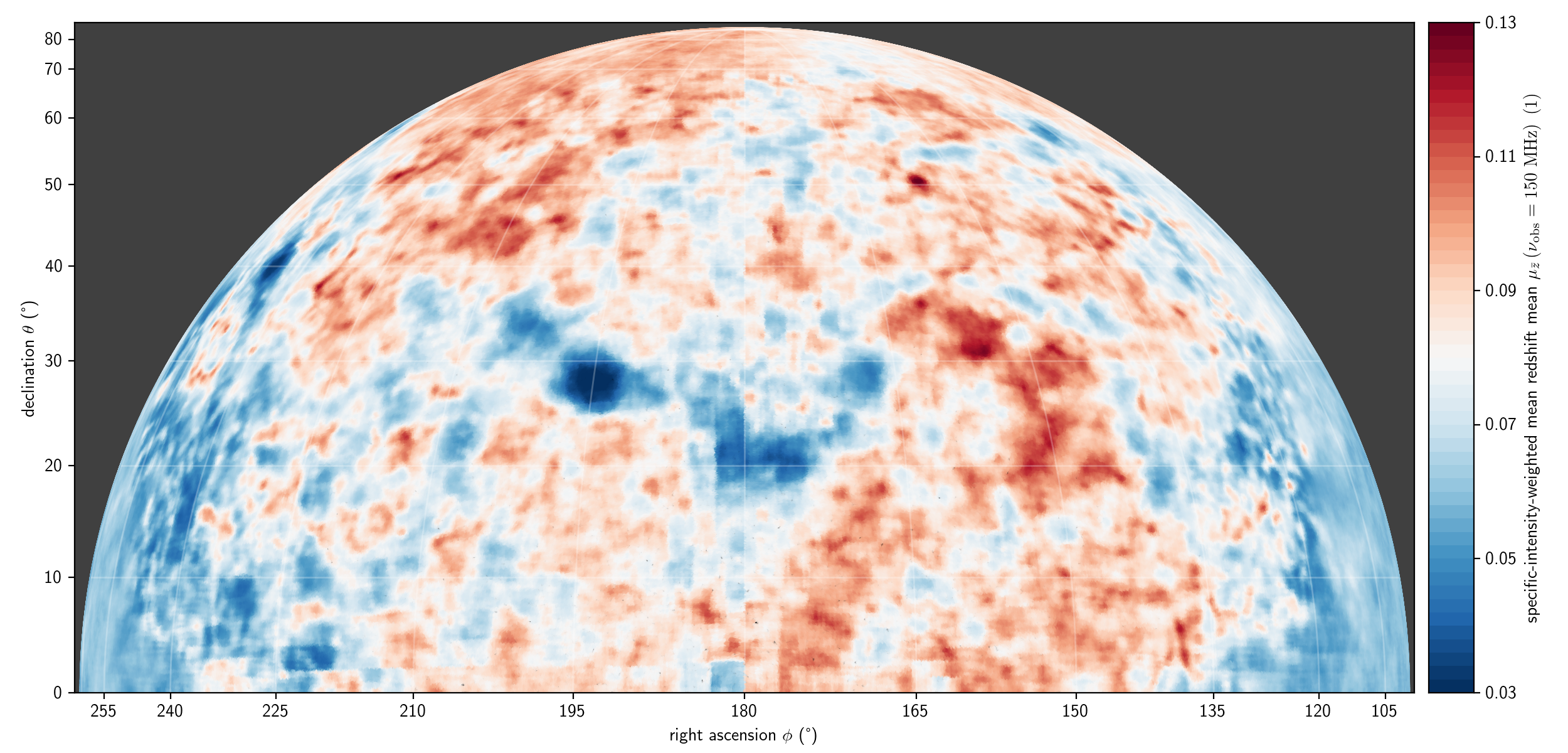}
    \subcaption{}
    \label{fig:redshiftWeightedMeanSphere150MHz}
\end{subfigure}
\caption{Per-direction mean of the MASSCW priors at observing frequency $\nu_\mathrm{obs} = 150\ \mathrm{MHz}$, exhibiting less variability than the single realisation of \textbf{Figures}~\ref{fig:realisationSphereSpecificIntensity} and \ref{fig:realisationSphereRedshift} (note the smaller colour ranges). Three famous galaxy clusters stand out; these are (in increasing order of declination) the Hercules Cluster, the Coma Cluster, and Abell 2199. \textbf{Top:} specific intensity mean. \textbf{Bottom:} specific-intensity-weighted mean (cosmological) redshift mean.}
\end{figure*}

All the results given are for $\nu_\mathrm{obs} = 150\ \mathrm{MHz}$.
At the BORG SDSS LSS reconstruction resolution, realisations from our MASSCW specific intensity function distribution exhibit a factor ${\sim}10^3$ of specific intensity variation over the sky, stemming from the highly variable power and localised nature of merger and accretion shocks in the Cosmic Web.
These variations clearly appear in \textbf{Figure}~\ref{fig:realisationSphereSpecificIntensity}, where we show a single realisation of $I_\nu\left(\hat{r},\nu_\mathrm{obs}\right)$ over the full lune.
Some sharp-edged structures are visible: these are due to the voxelised nature of the BORG SDSS density field samples.
The effect is most pronounced for high-MEC voxels close to the observer, which result in bright patches that span hundreds of square degrees.\\
In \textbf{Figure}~\ref{fig:realisationSphereRedshift}, we show the corresponding MASSCW specific-intensity-weighted mean redshift function realisation $\bar{z}\left(\hat{r},\nu_\mathrm{obs}\right)$.
All redshifts are within $[0, 0.2]$, because the BORG SDSS reconstructions are limited to $z_\mathrm{max} = 0.2$.
A comparison with \textbf{Figure}~\ref{fig:realisationSphereSpecificIntensity} reveals that generally, sky patches of high MASSCW specific intensity are due to structures at low redshift, and vice versa.\\
In \textbf{Figure}~\ref{fig:specificIntensityMeanSphere150MHz} and \ref{fig:redshiftWeightedMeanSphere150MHz}, we show the mean specific intensity function $\mu_{I_\nu}\left(\hat{r},\nu_\mathrm{obs}\right)$ and mean specific-intensity-weighted mean redshift function $\mu_{\bar{z}}\left(\hat{r},\nu_\mathrm{obs}\right)$ at $\nu_\mathrm{obs} = 150\ \mathrm{MHz}$ for each of the $10^6$ ray-traced directions, pooling all $10^3$ realisations.
\noindent
Both of these functions represent a summary statistic calculated from each ray's marginal $I_\nu$- and $\bar{z}$-distribution (i.e. the distribution obtained by marginalising out --- from the joint (prior) distribution --- the RVs of all directions but one).
We calculated the mean specific intensity after removing, for each marginal distribution separately, the lowest $1\%$ and highest $1\%$ of values.\\
Three bright spots stand out; these are (in increasing order of declination) the Hercules Cluster, the Coma Cluster, and Abell 2199.
Note that we have not included the specific intensity contribution of radio halos around galaxy clusters, which are of different origin: turbulent reacceleration \textcolor{blue}{\citep{Brunetti12001, Petrosian12001}}.
As observations suggest that the radio halo contribution usually dominates over the merger and accretion shock contribution, our results cannot be directly compared to actual galaxy cluster images.\\
The \emph{median} specific intensity function $m_{I_\nu}\left(\hat{r},\nu_\mathrm{obs}\right)$ (not shown) strongly resembles the mean specific intensity function, but is smaller over the whole lune.
This is indicative of skewed (single-direction) marginal specific intensity distributions.
%(i.e. the distributions obtained by marginalising out --- from the joint (prior) distribution --- the specific intensity RVs of all directions but one).
The $5^\mathrm{th}$ to $95^\mathrm{th}$ percentile mean-to-median ratios span the interval $(1.3, 2.1)$; the median mean-to-median ratio is $1.5$.\\
Probabilistic approaches also allow for quantification of prediction uncertainty.
In \textbf{Figure}~\ref{fig:specificIntensitySDSphere150MHz} and \textbf{Figure}~\ref{fig:specificIntensitySDOverMeanSphere150MHz}, we present both an absolute and relative measure of spread, again calculated from the MASSCW specific intensity prior marginals.
\textbf{Figure}~\ref{fig:specificIntensitySDSphere150MHz} shows the standard deviation $\sigma_{I_\nu}\left(\hat{r},\nu_\mathrm{obs}\right)$ after performing the same filtering as was done for the mean.
The resulting function is highly similar to the mean: directions that are brighter on average also tend to have larger absolute prediction uncertainties.
Of natural interest is also the relative prediction uncertainty $\sigma_{I_\nu}\mu_{I_\nu}^{-1}$, which we show in \textbf{Figure}~\ref{fig:specificIntensitySDOverMeanSphere150MHz}.
In this sense, the specific intensity function can be most accurately predicted around galaxy clusters and superclusters, reaching $\sigma_{I_\nu} \sim 60\%\ \mu_{I_\nu}$.
As expected, the relative prediction uncertainty is highest outside the SDSS DR7 coverage (i.e. near the $\phi = 90\degree$ and $\phi = 270\degree$ edges of the lune).
It is also high in regions within the SDSS DR7 coverage that have low average brightness, inverting the trend that characterises the absolute prediction uncertainty.
%However, there are some clear exceptions to this rule, where regions of several square degrees - with specific intensities comparable to those predicted near galaxy clusters - exhibit a ${\sim}0.2$ mean redshift.

\begin{figure*}
    \centering
    \includegraphics[width=.91\textwidth]{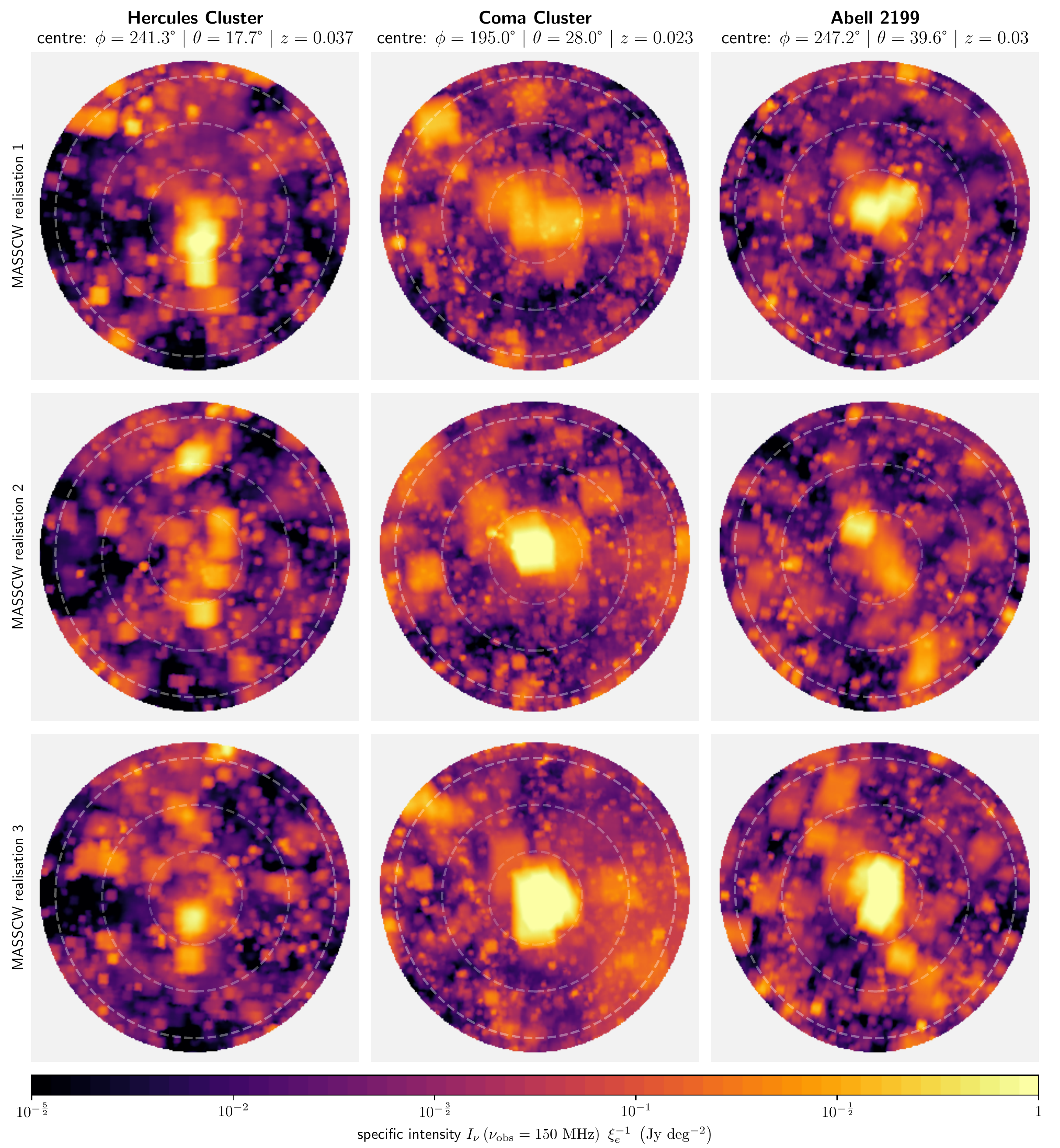}
    \caption{Three realisations (rows) of the merger- and accretion-shocked synchrotron Cosmic Web (MASSCW) specific intensity prior at observing frequency $\nu_\mathrm{obs} = 150\ \mathrm{MHz}$, showing zoom-ins around three massive Northern Sky galaxy clusters (columns). The zoom-ins show caps of the celestial sphere of $10\degree$ radius. The dashed circles are at $3\degree$, $6\degree$ and $9\degree$ from the cluster centre. Note that the usual radio halos that permeate galaxy clusters are not shown; these are caused by turbulent reacceleration, and we only show the merger and accretion shock contribution. The ${\sim}3\ \mathrm{Mpc}\ h^{-1}$ resolution of the 3D total matter density and monochromatic emission coefficient (MEC) fields limits the resolution of the 2D specific intensity fields, so that individual shocks and ${\sim}1\ \mathrm{Mpc}$ MEC - total matter density (anti)correlations are not resolved.}
    \label{fig:clusterFieldsSpecificIntensity150MHz}
\end{figure*}

\subsection{MASSCW specific intensity function distribution: outskirts of massive galaxy clusters}
Upon approaching the virial radius of a galaxy cluster from a connected filament, the WHIM transitions into the ICM, and the IGM's magnetic field grows stronger.
Especially in models such as \textcolor{blue}{\citet{Hoeft12007}}, in which the synchrotron-emitting electrons originate from the thermal pool, these cluster outskirts constitute the most promising targets to find synchrotron emission from the filament IGM.
%Around the virial radii of galaxy clusters, the ICM transitions into the WHIM, and 
%The discovery of a radio bridge between merging clusters \textcolor{blue}{\citep{Govoni12019}} has confirmed that the outskirts of galaxy clusters are promising targets to look for diffuse synchrotron emission around the virial radii of galaxy clusters.
Therefore, we also show zoom-ins of three realisations from our MASSCW specific intensity function prior near three massive galaxy clusters: the Coma Cluster, the Hercules Cluster, and Abell 2199.
In \textbf{Figure}~\ref{fig:clusterFieldsSpecificIntensity150MHz}, we show $I_\nu$ over spherical domes with an angular radius of $12\degree$, whilst \textbf{Figure}~\ref{fig:clusterFieldsRedshiftWeighted150MHz} shows the specific-intensity-weighted mean redshift for the same realisations and sky regions.
For the Coma Cluster, two filaments are discernible: one in northeastern, and one running in western direction.
For the Hercules cluster, the realisations suggest one northern and two southbound filaments. %filamentary structures
Finally, for Abell 2199 a southbound filament is evident.\\
By inspecting the mean redshift functions, one can verify that the identified structures indeed lie at the cluster redshift, rather than being structures closer by or further away that appear connected to the clusters in chance alignments.

\subsection{MASSCW specific intensity function distribution: deep LOFAR HBA fields}
\begin{figure}
    \centering
    \begin{subfigure}{\columnwidth}
    \includegraphics[width=\textwidth]{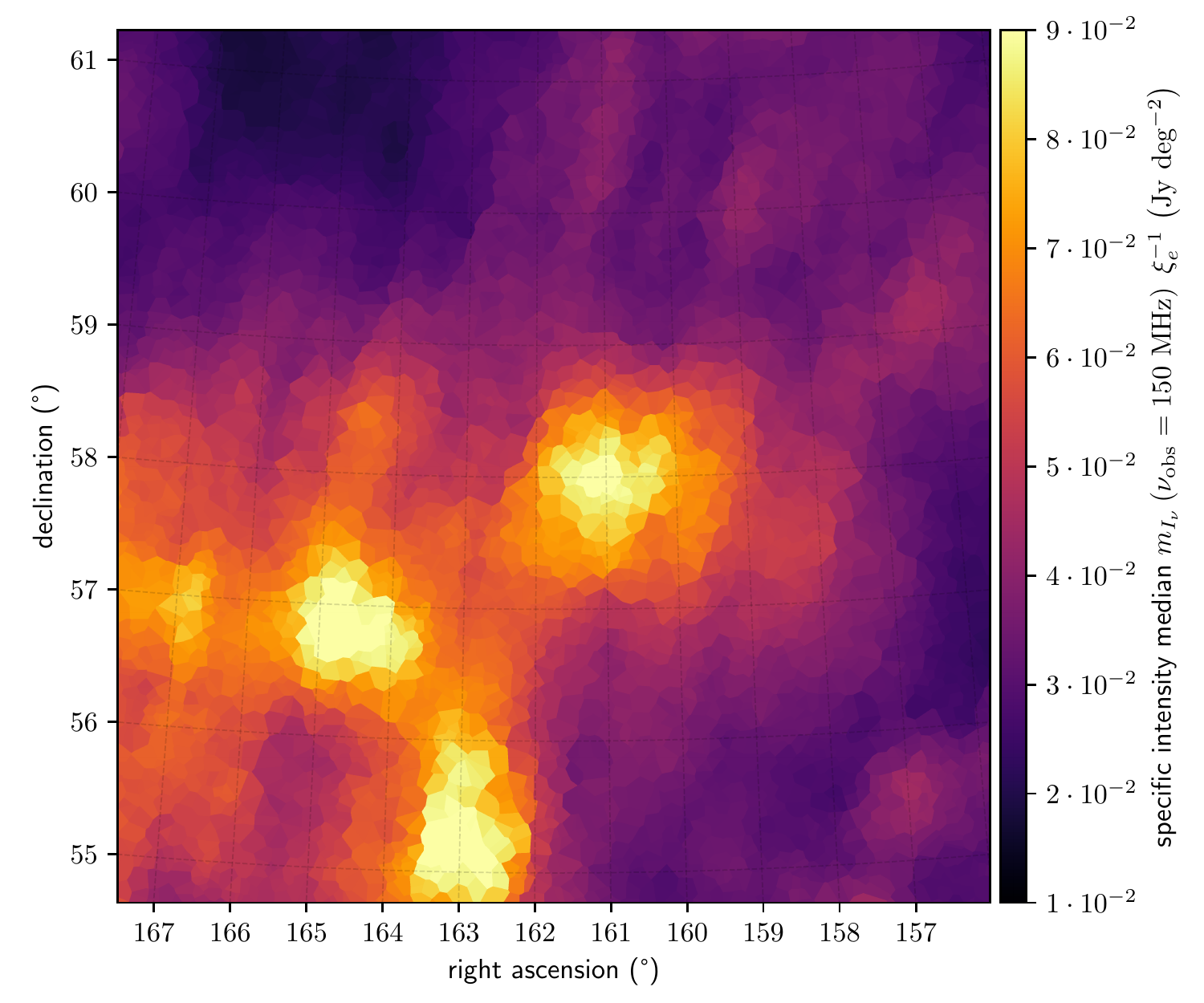}
    \end{subfigure}
    \begin{subfigure}{\columnwidth}
    \includegraphics[width=\textwidth]{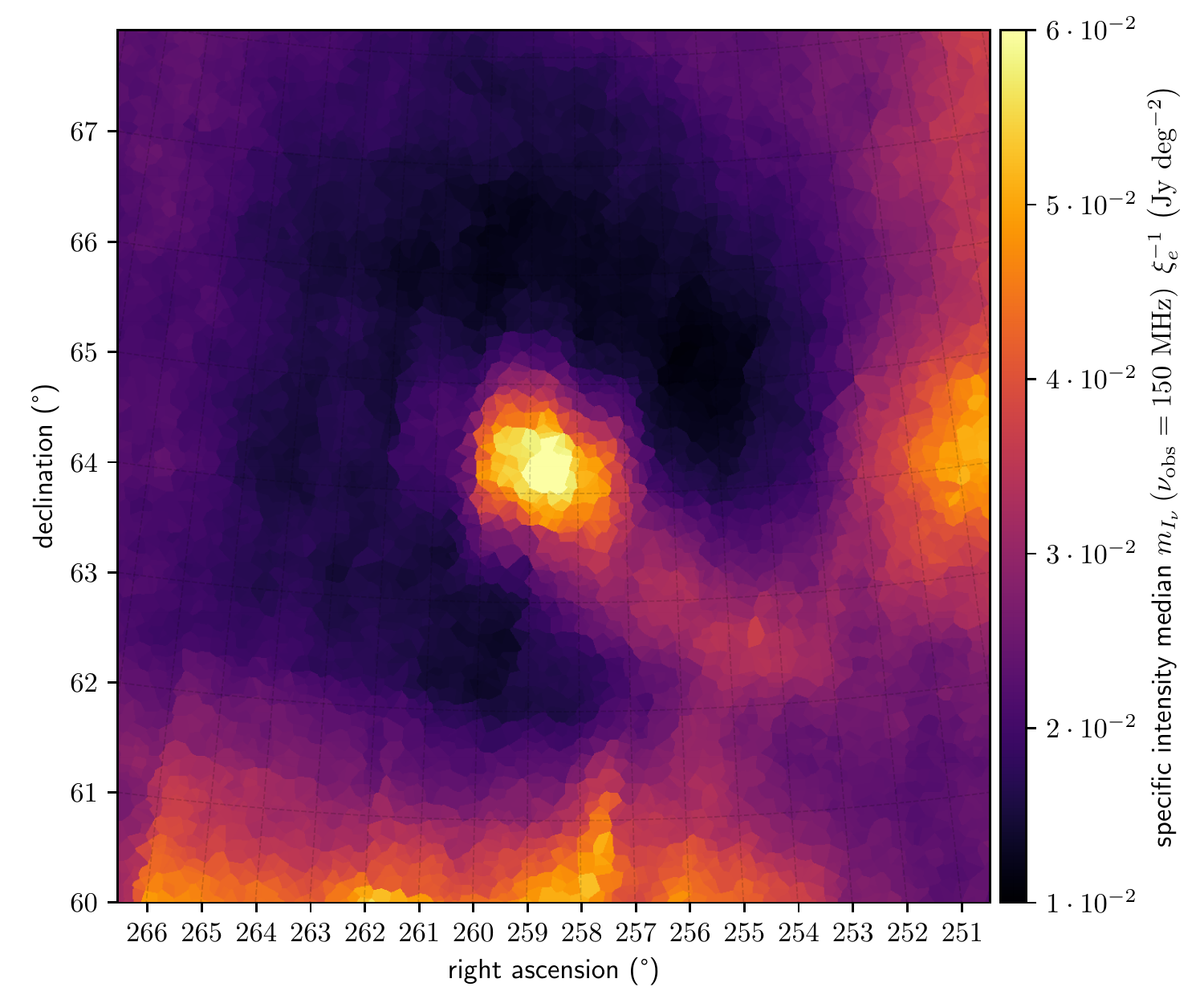}
    \end{subfigure}
    \begin{subfigure}{\columnwidth}
    \includegraphics[width=\textwidth]{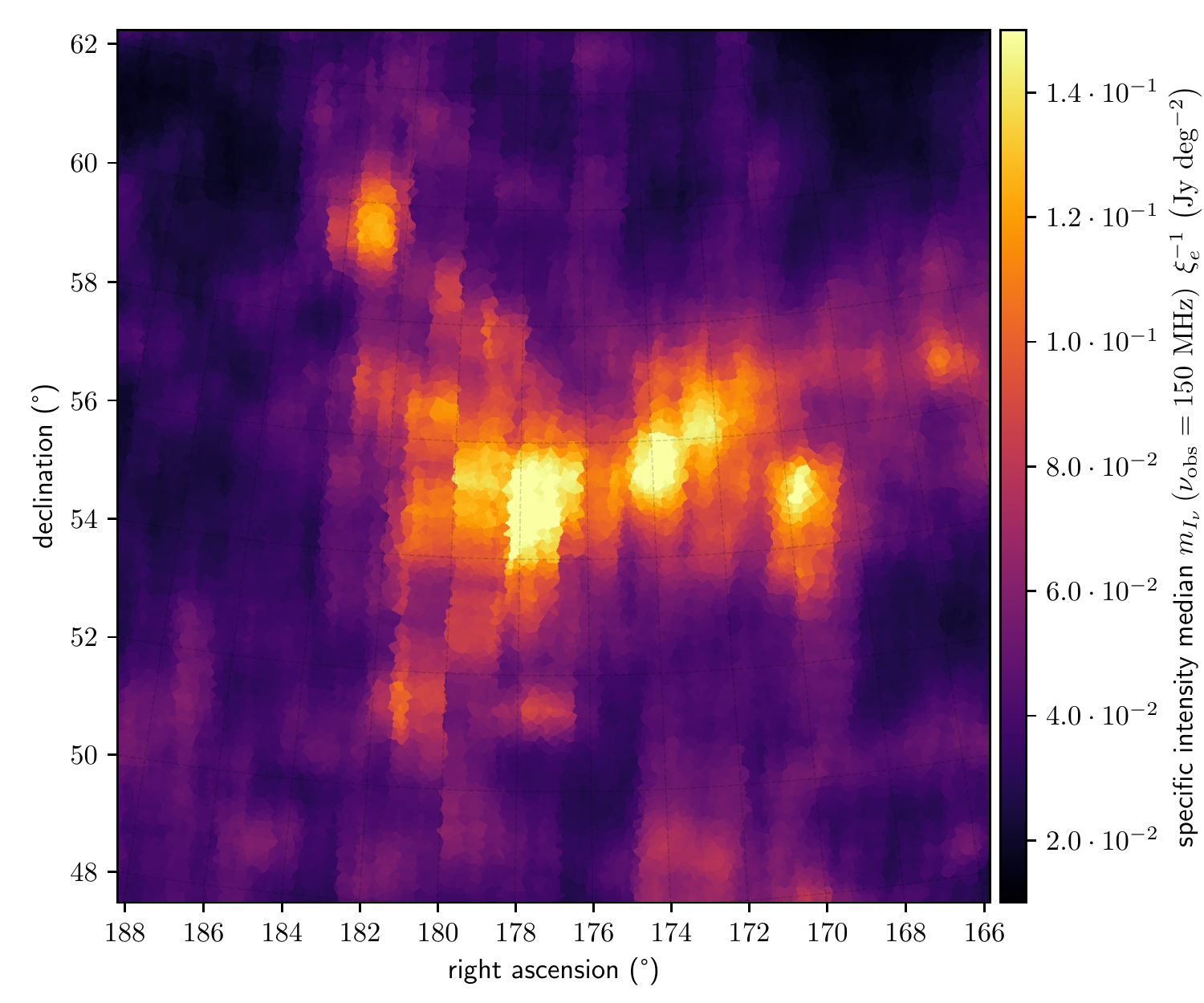}
    \end{subfigure}
    \caption{Merger- and accretion-shocked synchrotron Cosmic Web (MASSCW) specific intensity prior marginal medians at observing frequency $\nu_\mathrm{obs} = 150\ \mathrm{MHz}$ for three deep LOFAR HBA fields.
    Individual shocks should not be discernible in these statistical aggregates.
    The colour scales share the same lower bound.
    \textbf{Top:} Lockman Hole. \textbf{Middle:} Abell 2255. \textbf{Bottom:} Ursa Major Supercluster.}
    %\caption{Per-direction mean MASSCW specific intensity predictions at $\nu_\mathrm{obs} = 150\ \mathrm{MHz}$ for three deep LOFAR HBA fields. \textbf{Top:} Lockman Hole. \textbf{Middle:} Abell 2255. \textbf{Bottom:} Ursa Major Supercluster. The resolutions of the 3D and 2D MASSCW reconstructions are too low to recognise individual merger or accretion shocks.}
    \label{fig:LOFARDeepFields}
\end{figure}

With the LOFAR HBA, several deep ($\geq$ 50-hr) observations have been conducted that complement the wide-field, 8-hr approach of the LOFAR Two-metre Sky Survey (LoTSS) \textcolor{blue}{\citep{Shimwell12019}}.
Under thermal-noise-limited conditions, deep fields are of prime interest to search for a signature of the filament SCW.
For this reason, in \textbf{Figure}~\ref{fig:LOFARDeepFields}, we show the MASSCW specific intensity prior single-direction (i.e. marginal) medians for three such deep fields.\footnote{For these close-ups, we present the median instead of the mean to emphasise that our probabilistic approach enables the calculation of a variety of useful summary statistics.}
We generate imagery on arbitrary Flexible Image Transport System (FITS) grids by first calculating the marginal medians for each traced ray, and then applying a Voronoi tessellation to achieve a prediction for every pixel.\\
The Lockman Hole (for which a 100-hr dataset is available \textcolor{blue}{(Tasse et al., in prep.)}) is a field known for its relatively low Milky Way column densities of neutral hydrogen and dust \textcolor{blue}{\citep{Lockman11986}}, making it favourable for study in the extreme UV and soft X-ray bands, amongst others.
The potential for multi-wavelength synergy has made deep observations of the Lockman Hole a LOFAR HBA priority \textcolor{blue}{\citep{Mahony12016}}.
Abell 2255 (for which a 75-hr dataset is available \textcolor{blue}{(Botteon et al., in prep.)}) is a merging galaxy cluster \textcolor{blue}{\citep{Feretti11997, Botteon12020July}} at $z = 0.08$ that is part of the North Ecliptic Pole (NEP) Supercluster \textcolor{blue}{\citep{Mullis12001, Shim12011}}.
In this dynamic environment, merger and accretion shocks could light up the filaments connected to Abell 2255 \textcolor{blue}{\citep{Pizzo12008}}.
Lastly, the Ursa Major Supercluster (for which a 50-hr dataset is being assembled) at $z = 0.06$ stands out as the most prominent structure in our MASSCW specific intensity prior after the galaxy clusters shown in \textbf{Figure}~\ref{fig:clusterFieldsSpecificIntensity150MHz} and \ref{fig:clusterFieldsRedshiftWeighted150MHz}.
For this reason, it has been selected for deep LOFAR HBA observations.

\subsection{Specific-intensity-weighted mean redshift}
\begin{figure}
    \centering
    \begin{subfigure}{\columnwidth}
    \includegraphics[width=\textwidth]{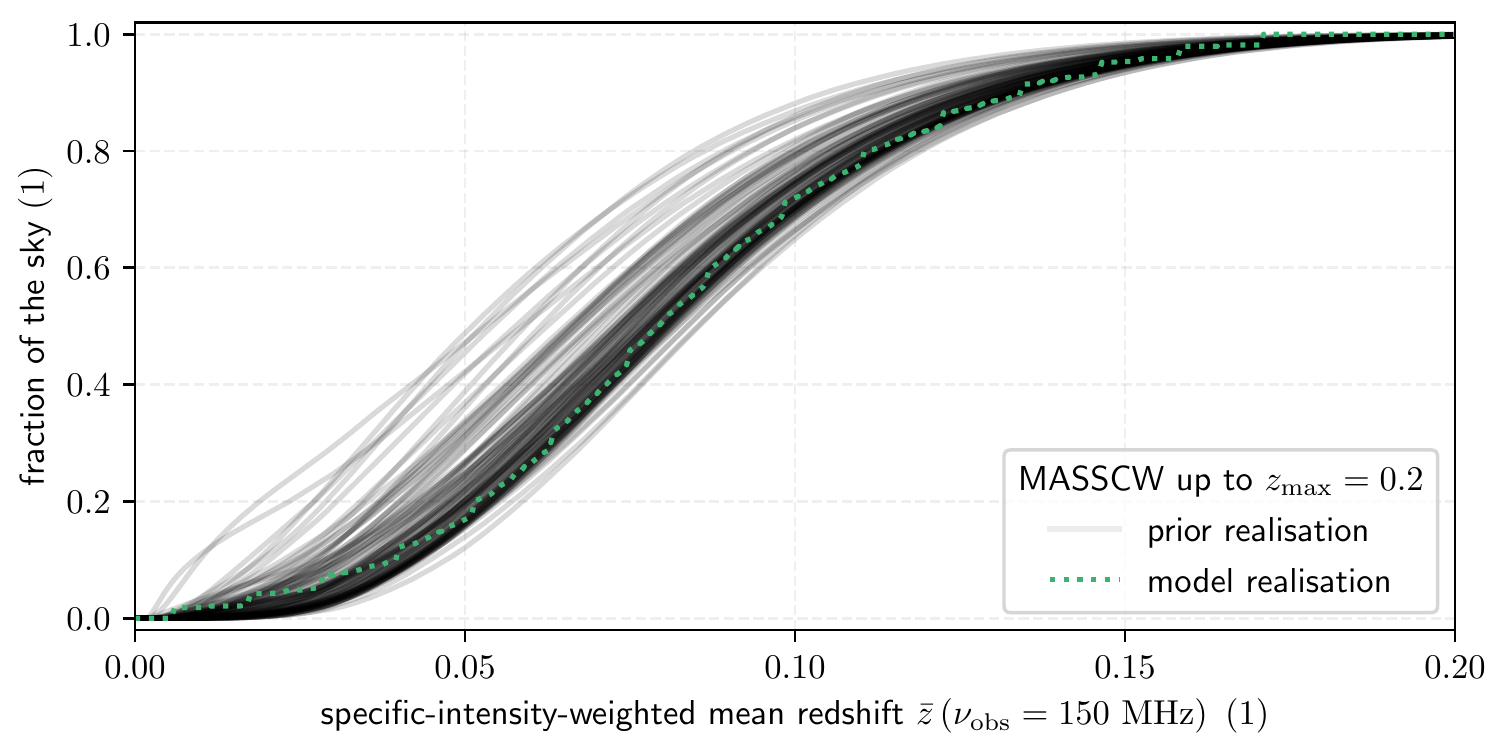}
    \end{subfigure}
    \begin{subfigure}{\columnwidth}
    \includegraphics[width=\textwidth]{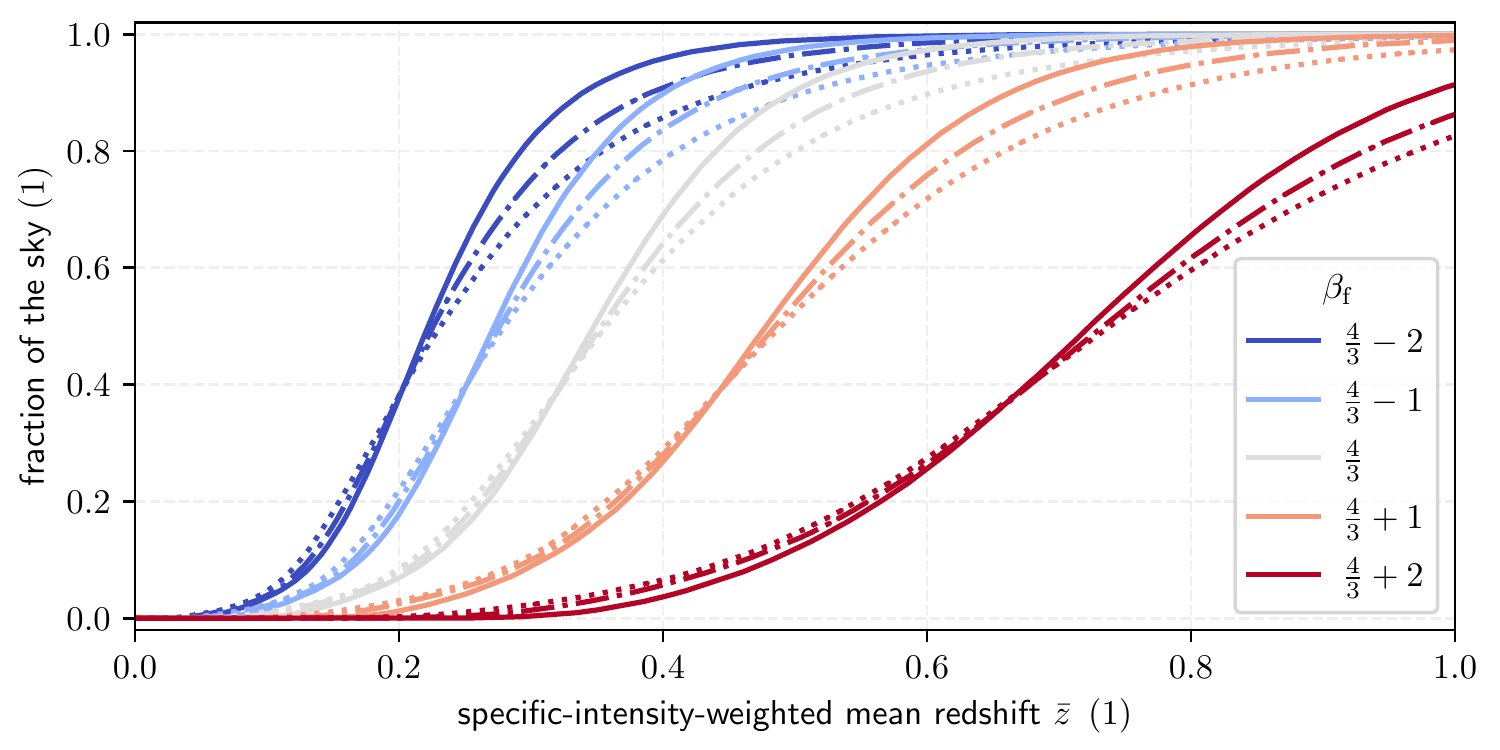}
    \end{subfigure}
    \caption{MASSCW specific-intensity-weighted mean redshift RV $\bar{Z}$ results.
    \textbf{Top:} distributions of $\bar{Z}$, showing 100 randomly selected ECDFs from our prior (grey curves), and a geometric model CDF (green curve).
    Both data and model consider LSS up to redshift $z_\mathrm{max} = 0.2$ only.
    %(dotted line), as well as one for when all LSS is included (dashed line).
    The other geometric model parameters are: $w_\mathrm{f} = 5\ \mathrm{Mpc}$, $l = 50\ \mathrm{Mpc}$, $d_1 = 25\ \mathrm{Mpc}$, $\mathcal{C} = 10^6$, $\alpha_\mathrm{f} = \alpha_\mathrm{c} = -\sfrac{3}{2}$, $\beta_\mathrm{f} = \sfrac{4}{3}$ and $\beta_\mathrm{c} = -\sfrac{4}{3}$.
    \textbf{Bottom:} geometric model CDFs of $\bar{Z}$ for filaments only ($\mathcal{C} = 0$), though including all of them ($z_\mathrm{max} = \infty$).
    By far the most influential parameter is also the most uncertain one; we therefore vary $\beta_\mathrm{f}$ over a plausible range.
    We also vary $l \in \{50\ \mathrm{Mpc}, 75\ \mathrm{Mpc}, 100\ \mathrm{Mpc}\}$, and by extension also $w_\mathrm{f}$ and $d_1$, by forcing $w_\mathrm{f} = \tfrac{1}{10}l$ and $d_1 = \tfrac{1}{2}l$.
    As $l$ increases, the unit cell number density decreases; we symbolise this with more sparsely drawn curves.
    We keep $\alpha_\mathrm{f} = -\sfrac{3}{2}$ constant.
    %The larger $l$, the lower the unit cell number density, and are represented by more sparsely drawn curves.
    %Apart from $z_\mathrm{max}$, the geometric model parameters are the same: $w_\mathrm{f} = 5\ \mathrm{Mpc}$, $l = 50\ \mathrm{Mpc}$, $d_1 = 25\ \mathrm{Mpc}$, $\mathcal{C} = 10^6$, $\alpha_\mathrm{f} = \alpha_\mathrm{c} = -\sfrac{3}{2}$, $\beta_\mathrm{f} = \sfrac{4}{3}$ and $\beta_\mathrm{c} = -\sfrac{4}{3}$.
    }
    \label{fig:BORGSkyAreaRedshift}
\end{figure}

In the top panel of \textbf{Figure}~\ref{fig:BORGSkyAreaRedshift}, we put our geometric model to the test by comparing a predicted specific-intensity-weighted mean redshift CDF $F_{\bar{Z}}\left(\bar{z}\right)$ (green dotted line) to $10^2$ randomly drawn ECDFs from our $10^3$ MASSCW prior realisations (translucent black lines), at $\nu_\mathrm{obs}= 150\ \mathrm{MHz}$.
Both prior realisations and the model reach a cumulative probability (`fraction of the sky') of $1$ at $\bar{z} = 0.2$: the BORG SDSS samples do not feature LSS beyond $z_\mathrm{max} = 0.2$ and the model is restricted accordingly.
The model CDF is constructed from $10^5$ draws from RV $\bar{Z}$ (see \textbf{Equation}~\ref{eq:redshiftWeightedSpecificIntensityRV}).\footnote{In a strict sense, given the finite number of realisations in our numerical approximation, this function is also an ECDF. However, for all practical purposes, it can be regarded as a CDF.}
We adopt parameters suggested by \textbf{Section}~\ref{sec:parameterEstimates}: $w_\mathrm{f} = 5\ \mathrm{Mpc}$, $l = 50\ \mathrm{Mpc}$, $d_1 = 25\ \mathrm{Mpc}$, $\mathcal{C} = 10^6$ at $\nu_\mathrm{ref} = 150\ \mathrm{MHz}$, $\alpha_\mathrm{f} = \alpha_\mathrm{c} = -\sfrac{3}{2}$, $\beta_\mathrm{f} = \sfrac{4}{3}$ and $\beta_\mathrm{c} = -\sfrac{4}{3}$.
Without further parameter tuning, the model CDF reproduces the trend revealed by the majority of prior ECDFs.
This correspondence is evidence that the geometric model \emph{in general} (and not just for some highly specific choice of parameters) captures the main features of the MASSCW, and motivates calculating the distribution of $\bar{Z}$ for the \emph{true} sky, which features LSS beyond $z_\mathrm{max} = 0.2$.
For the observational study of filaments, it is of greatest interest to calculate $F_{\bar{Z}}\left(\bar{z}\right)$ when the cluster contribution to the MASSCW is ignored, as it would otherwise dominate.
In the bottom panel of \textbf{Figure}~\ref{fig:BORGSkyAreaRedshift}, we therefore show $F_{\bar{Z}}\left(\bar{z}\right)$ according to the geometric model for $z_\mathrm{max} = \infty$ and $\mathcal{C} = 0$.
When $\mathcal{C} = 0$, $\alpha_\mathrm{c}$ and $\beta_\mathrm{c}$ become irrelevant.
Four parameters remain: $\alpha_\mathrm{f} + \beta_\mathrm{f}$, $l$, $w_\mathrm{f}$ and $d_1$.
The first of these has by far the most effect on $F_{\bar{Z}}\left(\bar{z}\right)$, as evidenced by the differently coloured curves (we keep $\alpha_\mathrm{f} = -\sfrac{3}{2}$ constant and vary $\beta_\mathrm{f}$, but varying $\alpha_\mathrm{f}$ and keeping $\beta_\mathrm{f}$ constant would lead to identical results).
We also vary the purely geometric parameters $l$, $w_\mathrm{f}$ and $d_1$, but in such a way that $w_\mathrm{f} = \tfrac{1}{10}l$ (as suggested by comparing the model's VFFs to those from cosmological simulations) and $d_1 = \tfrac{1}{2}l$ (i.e. the observer is always put at the centre of a cubic unit cell).
When $l \in \{50\ \mathrm{Mpc}, 75\ \mathrm{Mpc}, 100\ \mathrm{Mpc}\}$ is increased, and $w_\mathrm{f}$ and $d_1$ accordingly, modest changes in $F_{\bar{Z}}\left(\bar{z}\right)$ occur (dash-dotted and dotted lines): the distribution of $\bar{Z}$ attains larger spread.\\
$F_{\bar{Z}}\left(\bar{z}\right)$ provides detailed information.
For example, the model with $\beta_\mathrm{f} = \sfrac{4}{3}$ and $l = 50\ \mathrm{Mpc}$ predicts that for $\sim50\%$ of the sky (${\sim}21,000\ \mathrm{sq.\ deg.}$), the MASSCW signal has a mean redshift of $0.35$ or lower, whilst for $\sim80\%$ of the sky (${\sim}33,000\ \mathrm{sq.\ deg.}$), the mean redshift is $0.42$ or lower.
%First fixing $z_\mathrm{max} = 0.2$, we fit $w_\mathrm{f}$, $l$, $d_1$ and $\mathcal{C}$ to the ECDFs, yielding $w_\mathrm{f} = 5\ \mathrm{Mpc}$, $l = 50\ \mathrm{Mpc}$, $d_1 = 25\ \mathrm{Mpc}$ and $\mathcal{C} = 10^6$.
%The corresponding model is shown in \textbf{Figure}~\ref{fig:BORGSkyAreaRedshift} with a dotted green line.
%The remarkable correspondence between this low-parametric model and the prior realisations is evidence that the geometric model captures the main properties of the MASSCW, and motivates predicting the $\bar{z}$-distribution for the \emph{actual} sky, which features filaments beyond the BORG reconstruction limit of $z_\mathrm{max} = 0.2$.
%Keeping the best-fit parameters constant, we now set $z_\mathrm{max} = \infty$, and thus obtain a prediction for the actual MASSCW $\bar{z}$-distribution.
%This model is shown in \textbf{Figure}~\ref{fig:BORGSkyAreaRedshift} with a dashed green line. 
%This suggests that for $\sim50\%$ of the sky (${\sim}21,000\ \mathrm{sq.\ deg.}$), the MASSCW signal comes predominantly from a redshift of $0.15$ or lower, whilst for $\sim80\%$ of the sky (${\sim}33,000\ \mathrm{sq.\ deg.}$), the dominant redshift is $0.2$ or lower.

\subsection{Flux-density-weighted mean redshift}
\begin{figure}
    \centering
    \begin{subfigure}{\columnwidth}
    \includegraphics[width=\columnwidth]{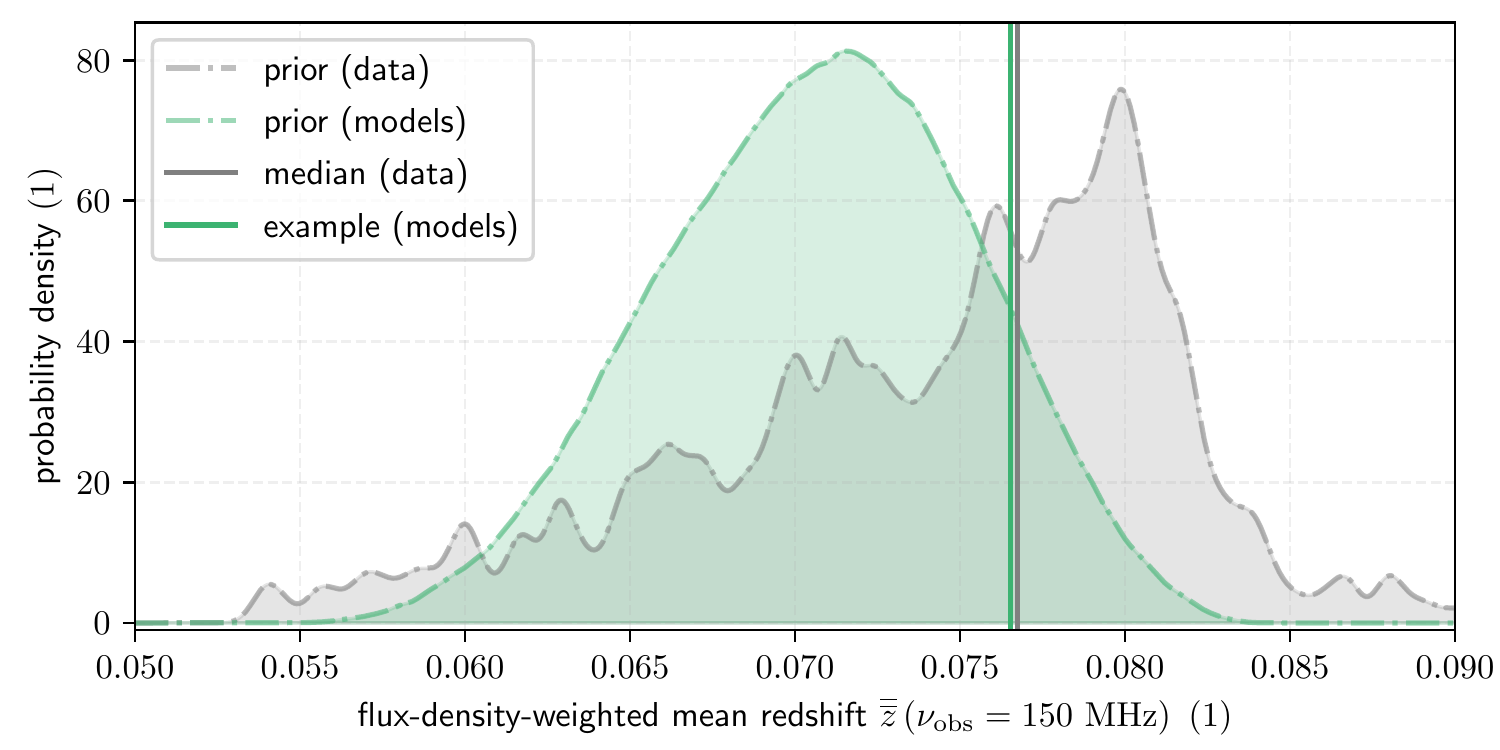}
    \end{subfigure}
    \begin{subfigure}{\columnwidth}
    \includegraphics[width=\columnwidth]{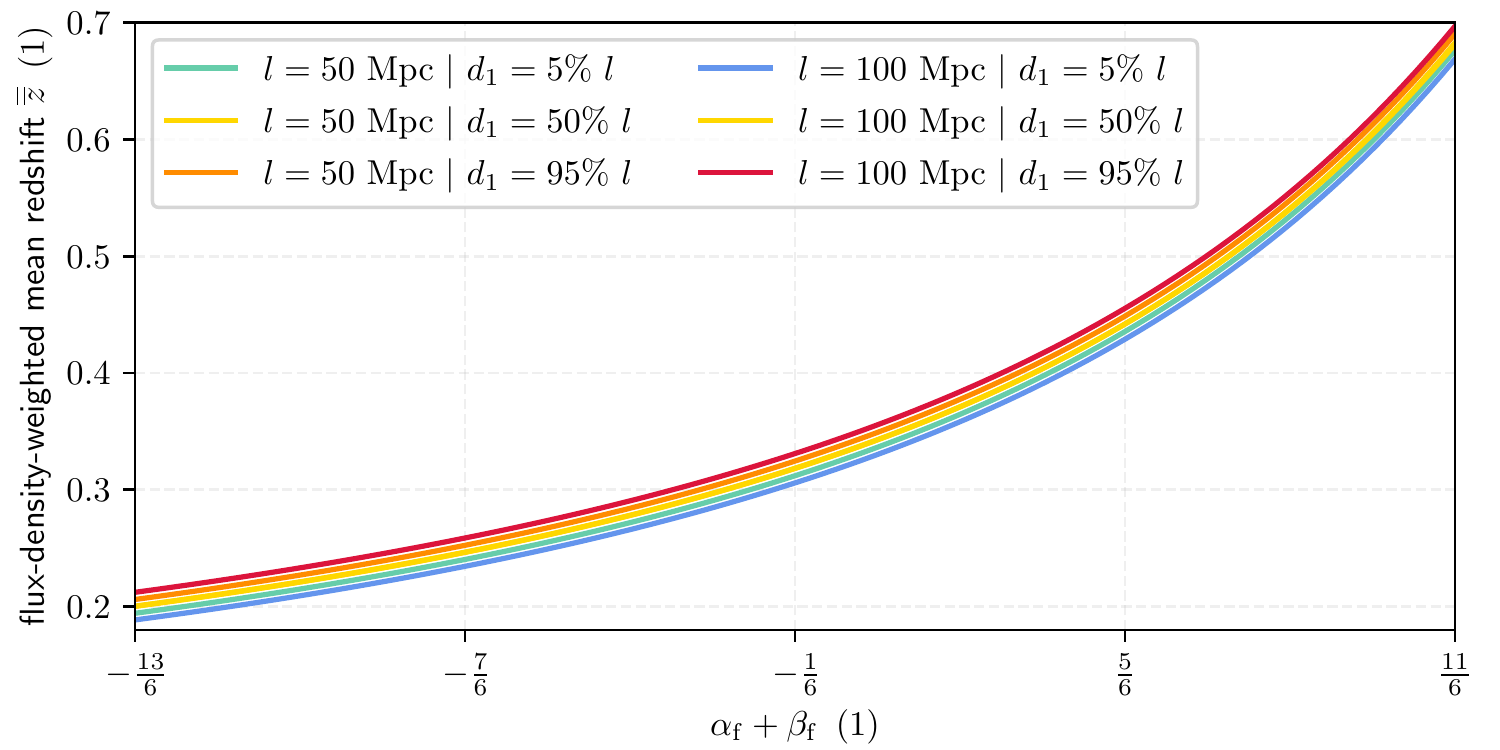}
    \end{subfigure}
    \caption{MASSCW flux-density-weighted mean redshift $\bar{\bar{z}}$ results.
    \textbf{Top:} distribution of $\bar{\bar{z}}$ calculated via KDE from our prior's $10^3$ realisations of $I_\nu$ and $\bar{z}$ (shaded grey), which are limited to $z_\mathrm{max} = 0.2$.
    The median is $\bar{\bar{z}} = 0.077$ (solid grey).
    Furthermore, we show the variety of geometric model predictions for $\bar{\bar{z}}$ when similarly limited to $z_\mathrm{max} = 0.2$, assuming flat priors on all model parameters (shaded green; see text for details).
    The median is $\bar{\bar{z}} = 0.072$.
    As a concrete example, we show the flux-density-weighted mean redshift of a single model that reproduces the data median (solid green).
    This model is given by $l = 75\ \mathrm{Mpc}$, $d_1 = 60\ \mathrm{Mpc}$ and $\beta_\mathrm{c} = -\sfrac{4}{3}$.
    %flux-density-weighted mean MASSCW redshifts $\bar{\bar{z}}$ from realisations of our detailed MASSCW prior, with the mean denoted in red.
    \textbf{Bottom:} predictions for $\bar{\bar{z}}$ considering filaments \emph{only}, varying all three relevant geometric model parameters: $\alpha_\mathrm{f} + \beta_\mathrm{f}$, $l$ and $d_1$.
    The two $d_1 = 50\%\ l$ curves are virtually indistinguishable.}
    %flux-density-weighted mean (i.e. sky-averaged) MASSCW redshift $\bar{\bar{z}}$ distribution under the geometric model, assuming flat priors $l \sim \mathrm{Uniform}\left(10\ \mathrm{Mpc}, 100\ \mathrm{Mpc}\right)$ and $d_1 \sim \mathrm{Uniform}\left(0, l\right)$, whilst $z_\mathrm{max}=0.2$.
    %Geometric model predictions for the flux-density-weighted mean (i.e. sky-averaged) MASSCW redshift, varying all three model parameters (excluding the parameters of the cosmological model $\mathfrak{M}$).
    %We show the predictions both for $z_\mathrm{max} = 0.2$ (to allow for comparison with our BORG-based MASSCW prior, which is limited to a redshift of $0.2$), and for $z_\mathrm{max} = \infty$, including LSS of all redshifts.}
    \label{fig:redshiftAllSky}
\end{figure}

The flux-density-weighted mean redshift $\bar{\bar{z}}$ constitutes the most concise measure of the typical MASSCW signal epoch.
In the top panel of \textbf{Figure}~\ref{fig:redshiftAllSky}, we show a distribution (shaded grey) over $\bar{\bar{z}}$ generated via KDE (Gaussian kernel, $\sigma_\mathrm{KDE} = 4 \cdot 10^{-4}$) from our $10^3$ MASSCW prior samples.
The median is $\bar{\bar{z}} = 0.077$ (solid grey); however, we stress that this is because only LSS up to $z_\mathrm{max} = 0.2$ is included.
We present an example model (solid green) that reproduces the median $\bar{\bar{z}}$, with main parameters $z_\mathrm{max} = 0.2$, $l = 75\ \mathrm{Mpc}$, $d_1 = 60\ \mathrm{Mpc}$, $\alpha_\mathrm{c} = -\sfrac{3}{2}$ and $\beta_\mathrm{c} = -\sfrac{4}{3}$.
(Adopting $\mathcal{C} = 10^6$ at $\nu_\mathrm{ref} = 150\ \mathrm{MHz}$ as suggested by \textbf{Section}~\ref{sec:parameterEstimates}, clusters dominate over filaments, and the other parameters play a very minor role.)
To explore the sensitivity of $\bar{\bar{z}}$ on the parameters, we generate $5\cdot10^5$ parameter sets by drawing from wide uniform distributions.
We draw $l \sim \mathrm{Uniform}\left(50\ \mathrm{Mpc}, 100\ \mathrm{Mpc}\right)$, $d_1 \sim \mathrm{Uniform}\left(0, l\right)$, $w_\mathrm{f} \sim \mathrm{Uniform}\left(5\ \mathrm{Mpc}, 10\ \mathrm{Mpc}\right)$, $\log_{10}\left(\mathcal{C}\right) \sim \mathrm{Uniform}\left(5, 7\right)$, $\alpha_\mathrm{f}, \alpha_\mathrm{c} \sim \mathrm{Uniform}\left(-\sfrac{3}{2} - \sfrac{1}{4}, -\sfrac{3}{2} + \sfrac{1}{4}\right)$, $\beta_\mathrm{f} \sim \mathrm{Uniform}\left(\sfrac{4}{3} - 2, \sfrac{4}{3} + 2\right)$ and $\beta_\mathrm{c} \sim \mathrm{Uniform}\left(-\sfrac{4}{3} - 2, -\sfrac{4}{3} + 2\right)$.
However, the resultant distribution for $\bar{\bar{z}}$ is mostly restricted to the relatively small range of $\left(0.06, 0.08\right)$.
We stress that the two distributions are conceptually distinct and are not meant to be compared directly.\\
In the bottom panel of \textbf{Figure}~\ref{fig:redshiftAllSky}, we show $\bar{\bar{z}}$ for filaments only ($\mathcal{C} = 0$) and $z_\mathrm{max} = \infty$.
The prediction then becomes a function of three parameters: $\alpha_\mathrm{f} + \beta_\mathrm{f}$, $l$ and $d_1$.
The dependencies on both $l$ and $d_1$ are weak, and $\bar{\bar{z}}$ is thus, to good approximation, determined by just a \emph{single} parameter.
Unfortunately, this parameter is currently highly uncertain, and should be constrained with upcoming cosmological simulations.
The various coloured curves suggest that the impact of $d_1$ on $\bar{\bar{z}}$ increases when $l$ is larger.
As the unit cells grow (so that their number density drops $\propto l^{-3}$), $\bar{\bar{z}}$ becomes more sensitive to contributions of individual boundary crossings (filaments), including the first.
Filament specific intensity contributions decrease quickly (especially for low values of $\alpha_\mathrm{f} + \beta_\mathrm{f}$) and monotonically with distance, so that the distance of the first filament to the observer (e.g. nearby, $d_1 = 10\%\ l$, or far away, $d_1 = 90\%\ l$) is of some importance.

%In \textbf{Figure}~\ref{fig:redshiftAllSky}, we show that $\bar{\bar{z}}$ varies weakly with $l$ over an order of magnitude from $10\ \mathrm{Mpc}$ to $100\ \mathrm{Mpc}$, whilst the comoving distance to the first boundary crossing $d_1$ also has minor impact.
%Understandably, as the unit cells grow larger (so that their number density drops $\propto l^{-3}$), $\bar{\bar{z}}$ becomes more sensitive to contributions of individual boundary crossings (filaments), including the first.
%Filament specific intensity contributions decrease quickly and monotonically with distance, so that the proximity of the first filament to the observer (e.g. nearby, $d_1 = 10\%\ l$, or far away, $d_1 = 90\%\ l$) is of importance.
%The saw-tooth feature seen for $z_\mathrm{max} = 0.2$ is due to the fact that $N$ decreases discontinuously (i.e. in integer steps) as $l$ increases.
%Notably, the model predictions are insensitive to variations in the weakly-constrained model parameters $d_1$ and $l$, and thus highly falsifiable.
%They dictate that our MASSCW prior realisations should yield $\bar{\bar{z}} \sim 0.07$ (see \textbf{Figure}~\ref{fig:redshiftAllSkyPriorWithDataZ02}).
%Meanwhile, the model predicts $\bar{\bar{z}} \sim 0.13$ for the actual Universe.
%\begin{figure}
%    \centering
%    \includegraphics[width=\columnwidth]{redshiftWeightedFluxDensityPriorWithData150MHz.pdf}
%    \caption{}
%    \label{fig:redshiftAllSkyPriorWithDataZ02}
%\end{figure}
%The correspondence with the mode of the geometric model distribution is striking; however, the result follows without parameter tuning.}
%between $10\ \mathrm{Mpc}$ and $100\ \mathrm{Mpc}$ for $l$
%l$, with $U \sim \mathrm{Uniform}\left(0, 1\right)$

\section{Discussion}
\label{sec:discussion}
\subsection{Inter-cluster radio bridges: evidence for turbulence rather than shocks}
\label{sec:bridges}
%low-frequency radio astronomers
In the last few years, LOFAR HBA observations around 140 MHz have revealed massive structures connecting clusters of galaxies at the onset of merging.
% their merging process.
These structures, or `bridges', likely are compressed filaments of the Cosmic Web.
%and checking their presence in our MASSCW prior constitutes a possible methodology validity test.\\
The first \emph{tentative} discovery (which has since been confirmed) of an inter-cluster bridge was in Abell 1758 (of length ${\sim}2\ \mathrm{Mpc}$) \textcolor{blue}{\citep{Botteon12018, Botteon12020December}}, whilst the first unambiguous detection was in Abell 399--401 (of length ${\sim}3\ \mathrm{Mpc}$) \textcolor{blue}{\citep{Govoni12019}}.
The latter ridge features higher densities (${\sim}3\cdot10^2\ \mathrm{m}^{-3}$), temperatures (${\sim}7\cdot10^7\ \mathrm{K}$), and magnetic field strengths (${\sim}1\ \mathrm{\mu G}$) than are expected to exist in typical filaments.\\
These two inter-cluster bridges cannot be, and are not expected to be, faithfully reproduced by our MASSCW prior.
Firstly, both bridges do not lie in the reconstructed volume: Abell 1758 lies at a redshift of $z = 0.279$, beyond the current redshift range ($z < 0.2$) of BORG SDSS reconstructions; Abell 399--401 has a favourable redshift of $z = 0.07$, but falls outside of the SDSS DR7 footprint.
Another complication is that bridges of $2 - 3\ \mathrm{Mpc}$ extent are smaller than the current BORG SDSS resolution of ${\sim}3\ \mathrm{Mpc}\ h^{-1}$, and would therefore only barely be identifiable if they fell within the reconstructed volume.\\
Notwithstanding practical difficulties, \textcolor{blue}{\citet{Govoni12019}} show that the \textcolor{blue}{\citet{Hoeft12007}} model alone cannot explain the LOFAR HBA data.
Significantly, this discovery has cast doubt on the widespread hypothesis --- also adopted in this article --- that accretion shocks generate the dominant contribution to the filaments' SCW signal.
Other mechanisms, like turbulence \textcolor{blue}{\citep{Brunetti12020}}, could play an important role.
%Detections of synchrotron radiation from the filament IGM are needed to establish the contribution of each radiation mechanism, as there is currently no evidence favouring one scenario over the other.\\
Cosmological simulations that model turbulence in the sense of \textcolor{blue}{\citep{Brunetti12020}} that would allow a recalculation of the SCW prior under this alternative emission mechanism, do not yet exist.
We stress that our methods, although presently employed to generate MASSCW predictions, are of general nature, and can be used to explore alternative (combinations of) SCW emission mechanisms in the future.

\subsection{Generalisation to other wavelength bands}
We have presented our methodology in the context of predicting the Cosmic Web's contribution to the \emph{radio} sky.
%low-frequency radio sky
However, the methodology does not rely on any special property of radio emission.
%, and can be generalised spectrally.
Apart from comparing different emission mechanisms for the \emph{same} wavelength band (as discussed in the previous subsection), the methodology could also be extended to predicting the Cosmic Web's contribution to the sky in \emph{other} wavelength bands.
%of interest.
With missions such as XRISM and Athena on the horizon, an extension into the X-ray window is of prime interest.
%to predict the best targets for direct direction of the WHIM seems 
%However, apart from exploring different emission mechanisms for a given wavelength band (as discussed in the previous subsection), the methodology can also be extended to predicting the Cosmic Web's contribution to the sky in other wavelength bands.
As a demonstration, we have tentatively calculated the thermal bremsstrahlung component of the Cosmic Web's contribution to the X-ray sky.
%, using Gaunt factor regimes from \textcolor{blue}{\citet{Novikov11973}}.
For plain simplicity, we do not invoke snapshots from cosmological simulations, but rather assume the proportionalities $\rho_\mathrm{BM}\left(\rho\right) = \frac{\Omega_\mathrm{BM,0}}{\Omega_\mathrm{M,0}}\rho$ and $T\left(\rho_\mathrm{BM}\right) = T_\mathrm{ref} \frac{\rho_\mathrm{BM}}{\rho_\mathrm{BM,ref}}$ (with $T_\mathrm{ref} = 10^6\ \mathrm{K}$ and $\rho_\mathrm{BM,ref} = m_p\ \mathrm{m^{-3}}$, where $m_p$ is the proton rest mass).
Furthermore, we assume a hydrogen-helium plasma of primordial chemical composition, and use the Gaunt factor regimes from \textcolor{blue}{\citet{Novikov11973}}.
%$T\left(\rho_\mathrm{BM}\right) = \max{\{ T_\mathrm{ref} \frac{\rho_\mathrm{BM}}{\rho_\mathrm{BM,ref}}, T_\mathrm{CMB}\}}$.
%As a first demonstration,
We show results in \textbf{Figure}~\ref{fig:LOFARDeepFieldsThermalBremsstrahlung}, for exactly the same sky regions as in \textbf{Figure}~\ref{fig:LOFARDeepFields}.
%In this case, the relation between total matter density $\rho$ and MEC $j_\nu$ 
Note that although the existence of a thermal bremsstrahlung component from the WHIM to the X-ray sky is uncontested and features relatively low uncertainties, it is likely that the oxygen line emission component dominates.
We have not considered this component due to its relative complexity and distance to our work's main focus, but propose its calculation as a promising direction for future research.
%We propose the calculation of this component as a direction for future research.

\subsection{Total matter density reconstructions: resolution, coverage and depth}
The voxelised nature of density reconstructions like the BORG SDSS causes large-scale (${\sim}10\degree$) blocky shapes in specific intensity and specific-intensity-weighted mean redshift function realisations that do not represent plausible real-life morphologies.
The angular scales at which these discontinuities occur, depend on the distances between the responsible (high-MEC) voxels and the observer.
Future density reconstructions that are run at higher resolution will contain fewer of these problematic patterns.
%alleviate this problem.
Furthermore, as hinted at in the previous paragraph, a modest resolution improvement (i.e. by a factor of order unity) would allow MASSCW predictions to contain inter-cluster bridges.\\
An improved resolution would also give relevance to the method of generating spatially correlated MEC fields as described in \textbf{Section}~\ref{sec:MEC}.
Currently, the BORG SDSS voxel length is larger than the shock correlation length $l_\mathrm{SE} = 2\ \mathrm{Mpc}$, so that the MEC fields obtained by our method are only marginally different from MEC fields where each voxel's MEC is drawn independently from the appropriate conditional probability distribution.\\
A promising idea that is not pursued in this work, is to interpolate the density fields (by e.g. doubling or tripling the number of voxels along each dimension), so that it becomes possible to generate merger and accretion shocks of the appropriate size.
Note that this approach would not add new small-scale structure to the density reconstructions, but merely ensures that the size of shocks generated stochastically are determined by the length scale $l_\mathrm{SE}$, instead of by the density reconstruction resolution.\\
A straightforward next step would be to use density reconstructions of the remaining $\sfrac{3}{4}$ of the sky to complete the MASSCW prior: generating predictions for the Southern Sky is relevant for SKA searches of the filament SCW.
This could be done with the already available BORG 2M++ \textcolor{blue}{\citep{Jasche12019}}, although these reconstructions remain shallower than those of the BORG SDSS, with $z_\mathrm{max} = 0.1$ instead of $z_\mathrm{max} = 0.2$.\\
Reconstructions that push beyond $z_\mathrm{max} = 0.2$ would also improve our predictions.
The yield of such an extension depends on the filament-only specific-intensity-weighted mean redshift CDF $F_{\bar{Z}}\left(\bar{z}\right)$ for $z_\mathrm{max}=\infty$.
However, the single most influential parameter governing $F_{\bar{Z}}\left(\bar{z}\right)$, $\alpha_\mathrm{f} + \beta_\mathrm{f}$, is currently ill-constrained.\footnote{Interestingly, its determination could be done with existing cosmological simulations.}
We therefore do not yet know to what extent deeper reconstructions would improve our MASSCW priors.
%but are expected to yield limited gain as the mean redshift of the full MASSCW signal is $\bar{\bar{z}} \sim 0.13$.
%Any BORG SDSS improvements necessary regarding 2LPT so that inter-cluster bridge reconstructions are reliable?
%The merits or uselessness of interpolation: just an aesthetic?

\subsection{Cosmological simulations}
The MASSCW predictions presented in this work are based on the statistical relationship between total matter density and the MASSCW MEC as inferred from Enzo cosmological simulations.
However, in absence of tight observational constraints, these cosmological simulations must assume one of many magnetogenesis scenarios.
The snapshots used in this work assume a primordial magnetogenesis scenario, starting from $z = 45$ with a seed magnetic field with a uniform comoving strength of $0.1\ \mathrm{nG}$.
More complex spectral energy distributions of primordial magnetic fields are however not excluded by present constraints from the CMB (e.g. \textcolor{blue}{\citet{Vazza12020}} and references therein).
Future work should explore the effect of different choices in the magnetogenesis scenario landscape on the MASSCW predictions, which could see a systematic change in specific intensity by an order of magnitude for filament-dominated directions.\\
A minor additional uncertainty comes from the fact that the $\left(100\ \mathrm{Mpc}\right)^3$ cube used in this work is not yet large enough to fully capture the density--MASSCW MEC relation in a statistically exhaustive manner.
This modest problem could be alleviated by appending the joint probability distribution from which the conditional probability distributions shown in \textbf{Figure}~\ref{fig:MHDPMFConditional} are derived with data from more simulation runs.\\
Finally, in this work, RAM limitations have necessitated discarding shocks with upstream Mach numbers below 2 in our shock identification procedure.
This Mach number cut means that the MECs assigned to voxels are lower bounds.
Future inclusion of the shocks now omitted could increase the MASSCW specific intensity functions by a (direction-dependent) factor of order unity.
%This might not be ideal, because the low-Mach shocks remain bright.
%We only test one magnetogenesis scenario... There are many!
%Improvements for ENZO or related simulations: Include more cubes to see if $2400^3$ is large enough to capture shocks statistically

%\subsection{Galactic foreground}
\subsection{Observational considerations}
Based on our predictions, what are the observational prospects of detecting the filament SCW --- and the filament \emph{MASSCW} in particular?
As evinced by radio bridge detections \textcolor{blue}{\citep{Botteon12018, Govoni12019, Botteon12020December}} and a statistical all-sky (or close to all-sky) detection \textcolor{blue}{\citep{Vernstrom12021}}, both special-geometry \emph{and} global observations of the filament SCW are already possible with modern low-frequency radio telescopes.
The more interesting question therefore is whether detections are possible on an \emph{intermediate} level --- that of individual regions of large-scale structure --- so that spatially resolved measurements of the intergalactic magnetic field strength $B_\mathrm{IGM}$ in filaments come within reach.
%in typical filaments
In this section, we therefore explore the observational prospects and challenges of detecting the filament SCW around individual massive galaxy clusters, such as the Hercules Cluster, the Coma Cluster, Abell 2199 and Abell 2255, and around larger LSS complexes, such as the cluster triple in the Lockman Hole and the Ursa Major Supercluster.
We assume that the MASSCW signal is the dominant contributor (in fact, the \emph{only} contributor) to the SCW signal of these LSS regions.\footnote{As discussed in \textbf{Section}~\ref{sec:bridges}, the MASSCW signal is unlikely to be the dominant contributor to the SCW signal \emph{in radio bridges}; in the case of the all-sky search of \textcolor{blue}{\citet{Vernstrom12021}}, the dominant emission mechanism remains unknown.}
%in an average, all-sky sense
For associated MASSCW predictions, see \textbf{Figures}~\ref{fig:clusterFieldsSpecificIntensity150MHz} and \ref{fig:LOFARDeepFields}.\\
Making firm observability forecasts is in the first place hampered by the fact that there are three major unknowns in MASSCW predictions: firstly, the strength of the seed magnetic field in the Early Universe combined with the dominant process by which this field has evolved into magnetic fields in filaments today (see \textbf{Section}~\ref{sec:magnetogenesis}); secondly, the filling factor of shocks of appropriate strength and obliquity to trigger DSA; and, thirdly, the magnitude of the electron acceleration efficiency $\xi_e$ in filaments (see \textbf{Section}~\ref{sec:HoeftBruggenModel}).
%the typical magnitude
%into those in filaments today
From \textbf{Figure}~\ref{fig:clusterFieldsSpecificIntensity150MHz}, we see that the specific intensity function in the direction of filaments around massive low-redshift galaxy clusters, at $\nu_\mathrm{obs} = 150\ \mathrm{MHz}$ and degree-scale resolution, reaches $I_\nu \sim 10^{-1}\ \mathrm{Jy\ deg^{-2}}$ for $\xi_e = 1$.
If $\xi_e \sim 10^{-2}$ in filaments, as in \textcolor{blue}{\citet{Keshet12004}}'s SNR shocks, the actual specific intensity is $I_\nu \sim 10^{-3}\ \mathrm{Jy\ deg^{-2}}$.
A conservative estimate of the uncertainty in $\xi_e$ is an order of magnitude, yielding a range of specific intensity estimates $I_\nu \sim 10^{-4}\text{--}10^{-2}\ \mathrm{Jy\ deg^{-2}}$.\\
%in each direction
%The more unresolved question therefore is whether detections are possible on an intermediate scale:
%And what are the prospects of detecting the MASSCW specifically?
In this work, we have considered the MASSCW signal without contaminants.
However, in most directions, the Milky Way is the dominant contributor to the specific intensity function at $\nu_\mathrm{obs}\ {\sim}10^2\ \mathrm{MHz}$, and actual observational attempts to detect the SCW should therefore avoid the Galactic Plane, the North Polar Spur \textcolor{blue}{\citep{Salter11983}}, and other bright synchrotron Milky Way features.
The extent to which the Milky Way hampers a SCW detection, depends on the typical angular scales of the synchrotron Milky Way, the SCW, and those measurable by the interferometer.
\textcolor{blue}{Oei et al., in prep.} have made compact-source--subtracted, low-resolution ($60''$ and $90''$) images with the LoTSS DR2 \textcolor{blue}{\citep{Shimwell12022}} that reveal the Milky Way's specific intensity function at $\nu_\mathrm{obs} = 144\ \mathrm{MHz}$ up to degree scales.
(By the lack of baselines shorter than 68 metres, \emph{larger} scales are resolved out.)
These images have most power on the degree scale and show specific intensity variations of $\sigma_{I_\nu} \sim 10^0\text{--}10^1\ \mathrm{Jy\ deg^{-2}}$ in the off-Galactic plane region.
%Compact-source-subtracted, low-resolution images \textcolor{blue}{(Oei et al., in prep.)} derived from the LoTSS DR2 \textcolor{blue}{(Shimwell et al., in prep.)} reveal that the Milky Way's specific intensity function at $\nu_\mathrm{obs} = 150\ \mathrm{MHz}$ has power on a degree scale, 
%${\sim}10^{-1} - 10^0\ \mathrm{deg}$
%By subtracting compact sources from the reimaging at low resolution
%Compact-source-subtracted, low-resolution LoTSS DR2 \textcolor{blue}{(Shimwell et al., in prep.)} images reveal that the Milky Way's specific intensity function contribution has power on scales ${\sim}10^{-1} - 10^0\ \mathrm{deg}$ \textcolor{blue}{(Oei et al., in prep.)}, with larger scales resolved out.
Meanwhile, the scales at which a sky region's SCW specific intensity function has most power depend on the distances to the region's most massive large-scale structures and vary per dominant emission mechanism assumed, with power on larger scales for turbulence compared to merger and accretion shocks.\\
A discrepancy in dominant scales between the specific intensity functions of the synchrotron Milky Way and the SCW can be leveraged to bolster the prospects of a SCW detection.
Doing so appears important, because already at the 8-hr depth of the LoTSS DR2, the Milky Way's specific intensity dominates over the thermal noise.
%the Milky Way--induced SCW detection noise
By using an inner $(u,v)$-cut that \emph{removes} most Milky Way emission (but that \emph{retains} most SCW emission), Milky Way--induced specific intensity variations can be reduced by one to two orders of magnitude,\footnote{This estimate follows from considering the spherical harmonics angular power spectrum $C_\ell$ of the interferometrically observed (and thus large-angular-scale--deprived) synchrotron Milky Way.
We model $C_\ell$ as a power law \textcolor{blue}{\citep{LaPorta12008}} from degree $\ell_0$ onwards:
\begin{align}
    C_\ell\left(\ell\right) = \begin{cases}
    0 & \ell < \ell_0;\\
    C_\ell\left(\ell_0\right)\left(\frac{\ell}{\ell_0}\right)^\beta & \ell \geq \ell_0,
    \end{cases}
\end{align}
where $\beta < -1$.
Generally, given an angular power spectrum, the total power $P$ is
\begin{align}
    P = \sum_{\ell=0}^\infty C_\ell\left(\ell\right).
\end{align}
We model the imposition of an additional inner $(u,v)$-cut as the removal of power on all angular scales up to (but excluding) $\ell_0'$, with $\ell_0' \geq \ell_0$.
%represented by $\ell_0 \leq \ell < \ell_0'$, with $\ell_0' > \ell_0$.
The (negative) relative change in total power caused by the $(u,v)$-cut is
\begin{align}
    \frac{P'-P}{P} =\frac{\sum_{\ell_0'}^\infty \ell^\beta}{\sum_{\ell_0}^\infty \ell^\beta}-1=\frac{\zeta\left(-\beta, \ell_0'\right)}{\zeta\left(-\beta,\ell_0\right)}-1,
    %= \frac{E'}{E}-1
\end{align}
where $\zeta$ is the Hurwitz zeta function.
For $\beta \in [-3, -2]$ \textcolor{blue}{\citep[e.g.][]{LaPorta12008, Ghosh12012, Sims12016}}, $\ell_0 \sim 10^2$ and $\frac{\ell_0'}{\ell_0} \sim 10^0$, we find relative total power changes of $-90\%$ to $-99\%$.
} leaving a Milky Way contamination of $\sigma_{I_\nu}\sim 10^{-2}\text{--}10^0\ \mathrm{Jy\ deg^{-2}}$.
(The optimal $(u,v)$-cut choice for a particular sky region can be derived by comparing the angular power spectrum of the synchrotron Milky Way with that of the region's SCW predictions.)
For depths such that the Milky Way remains the dominant noise source, the signal-to-noise ratio for a solid angle of a square degree centred around a massive filament thus is $10^{-4}$ in a very pessimistic case, $10^{-1}$ in a fairly optimistic case, and $10^0$ in a very optimistic one.
The signal-to-noise ratio grows with the square root of the number of such solid angles considered, and thus linearly with the angular diameter of the observed region.
Notably, in the fairly optimistic case, we need to observe a region of $10^1$ degree diameter to achieve a signal-to-noise ratio of order unity (e.g. three).
For example, the Ursa Major Supercluster (see the bottom panel of \textbf{Figure}~\ref{fig:LOFARDeepFields}) is a region of roughly the required extent.
In case the MASSCW specific intensity is an order of magnitude weaker, i.e. of order $I_\nu \sim 10^{-3}\ \mathrm{Jy\ deg^{-2}}$, the sky region required to detect the MASSCW is roughly the entire sky.\\
%Thus, the fact that \textcolor{blue}{\citet{Vernstrom12021}} claim an all-sky SCW detection implies that $I_\nu \gtrsim 10^{-3}\ \mathrm{Jy\ deg^{-2}}$.\\
%The claimed detection by \textcolor{blue}{\citet{Vernstrom12021}} thus implies that $I_\nu \gtrsim 10^{-3}\ \mathrm{Jy\ deg^{-2}}$.\\
Galaxies populate filaments and generate synchrotron radiation.
If they are not masked or removed from the observed imagery, their presence could mimic a SCW detection signal in cross-correlation experiments with low-spatial-resolution MASSCW predictions.
(However, at high spatial resolution, \textcolor{blue}{\citet{Hodgson12021}} show that the SCW and synchrotron emission from galaxies can be separated, as these signals trace the LSS matter distribution in different ways.)
%the SCW traces the LSS matter distribution in a different way than synchrotron emission from galaxies can be separated.
%This issue underlines the importance of deep observations that feature low thermal noise levels, although not for the usual reason of minimising random errors.
This issue underlines the importance of deep observations that feature low thermal noise levels, although --- in contrast to the usual situation --- in order to minimise \emph{systematic} rather than random errors.
%not in order to minimise random errors, but systematice rrors.
As noted before, for LoTSS DR2 observations, Milky Way contamination dominates over thermal noise, rendering thermal noise largely irrelevant --- at least prior to the suppression of degree scales.
%However, the lower the thermal noise level, the larger the fraction of the galaxy population in filaments that is detected and that can be subtracted to control; the thermal noise level
However, low thermal noise levels do allow for a more thorough subtraction of the galaxy population in filaments and thus help control an important systematic effect.
%Deeper observations lower the thermal noise, and alhtough
%through supernova remnants and AGN ejecta.
%Compact-source
%A further complication is formed the specific intensity contributions f

\subsection{Independence of random fields}
In this work, we have outlined how a specific intensity random field $I_\nu\left(\hat{r},\nu_\mathrm{obs}\right)$ can be generated from a percentile random field $\mathcal{P}\left(\mathbf{r}\right)$ (or, equivalently, a Gaussian random field $\mathcal{Z}\left(\mathbf{r}\right)$) and a total matter density random field $\rho\left(\mathbf{r}\right)$; $\hat{r} \in \mathbb{S}^2$, $\nu_\mathrm{obs} \in \mathbb{R}_{> 0}$, and $\mathbf{r} \in \mathcal{R}$.
The function $f$ that maps the two input random fields to the output random field is deterministic, informed by conditional probability distributions extracted from cosmological simulations, and non-linear in both arguments.
Symbolically, $I_\nu = f\left(\mathcal{P}, \rho\right)$.
Implicitly, our approach has been to sample $\rho$ from the BORG SDSS posterior, and $\mathcal{P}$ from another, \emph{independent} distribution (which is fixed by the distribution of $\mathcal{Z}$, and thus specified by the covariance function $K_\mathrm{SE}$).
Under what physical scenario is this justified?
Consider the joint distribution for the input random fields $P\left(\mathcal{P}, \rho\right)$, which can be written as the product of a conditional and marginal: $P\left(\mathcal{P}, \rho\right) = P\left(\mathcal{P}\ \vert\ \rho\right) P\left(\rho\right)$.
Thus, the central assumption that underlies our sampling approach is that $P\left(\mathcal{P}\ \vert\ \rho\right) \approx P\left(\mathcal{P}\right)$, so that $P\left(\mathcal{P},\rho\right) \approx P\left(\mathcal{P}\right) P\left(\rho\right)$.
Our approximation is thus that the density field does not inform where, given a set of locations \emph{with the same density}, high (or, equivalently, low) shock emission is more likely to occur.\\
%Our approximation is thus that the density field is not informative as to where above-average (as well as below-average) shock emission occurs at a set of locations \emph{with the same density}.\\
The real world will violate this assumption to some extent.
For example, a point in the outskirts of a galaxy cluster could be as dense as a point along the central axis of a prominent filament; however, the cluster point still likely has a different MASSCW MEC probability distribution than the filament point.
One reason could be the presence of passing merger shocks in cluster outskirts; another could be the higher typical speed by which accretion shocks crash onto clusters, compared to the typical speed by which they hit filaments.

\subsection{Spectral indices in 3D and 2D}
In \textbf{Section}~\ref{sec:observersSCWSpecificIntensity}, we calculate $I_\nu$ by generating $j_\nu$ at different emission frequencies with the same percentile random field.
This approach implicitly assumes that if shocks were to be ordered by their MEC, the ordering remains invariant over emission frequency range $[\nu_\mathrm{obs}, \nu_\mathrm{obs}\left(1+z_\mathrm{max}\right)]$.
Does this assumption correspond to a plausible physical scenario?
%In what physical scenario would this be true?
Let the MECs of two shocks at some reference emission frequency $\nu = \nu_\mathrm{ref}$ be $j_1$ and $j_2$, and let $j_1 > j_2$ without loss of generality.
Assume the existence of some function $\alpha = \alpha\left(j_\nu\right)$, that assigns integrated spectral indices to shocks based on their MEC at $\nu = \nu_\mathrm{ref}$.
At emission frequency $\nu$, the MEC ranks of the shocks are the same as at $\nu_\mathrm{ref}$ if and only if
\begin{align}
    j_1\left(\frac{\nu}{\nu_\mathrm{ref}}\right)^{\alpha\left(j_1\right)} > j_2\left(\frac{\nu}{\nu_\mathrm{ref}}\right)^{\alpha\left(j_2\right)}\ \mathrm{for\ all}\ j_1, j_2 \in \mathbb{R}_{\geq 0},\ j_1 > j_2.
\end{align}
The rewritten inequality
\begin{align}
    \frac{j_1}{j_2} > \left(\frac{\nu}{\nu_\mathrm{ref}}\right)^{\alpha\left(j_2\right) - \alpha\left(j_1\right)}\ \mathrm{for\ all}\ j_1, j_2 \in \mathbb{R}_{\geq 0},\ j_1 > j_2
\end{align}
suggests $\alpha\left(j_2\right) - \alpha\left(j_1\right) \leq 0$, or $\alpha\left(j_1\right) \geq \alpha\left(j_2\right)$ for $\nu > \nu_\mathrm{ref}$: $\alpha\left(j_\nu\right)$ must be a monotonically \emph{increasing} function.
Analogously, for $\nu < \nu_\mathrm{ref}$, we find that $\alpha\left(j_\nu\right)$ must be a monotonically \emph{decreasing} function.
These scenarios are visualised in \textbf{Figure}~\ref{fig:assumptionSpectralIndices}.
Using the Enzo simulation data used in this work, we explore the spectral index - MEC relation at $\nu_\mathrm{ref} = 180\ \mathrm{MHz}$, and find a general downward trend for spectral index as a function of MEC at this frequency.
However, the relation is scattery and therefore not fully described by a monotonically decreasing function $\alpha\left(j_\nu\right)$; this implies that shock MEC ranks \emph{do} change when varying the emission frequency.
Although our approach to determining spectral indices remains approximate, compared to other uncertainties in our methodology, the error thus introduced is likely of minor importance.\footnote{A conceptually correct way to address this problem would be to realise both a spectral index and a MEC at some fixed emission frequency ($\nu = \nu_\mathrm{obs}$, say), using $P\left(\alpha, j_\nu\ \vert\ \rho\right) = P\left(\alpha\ \vert\ j_\nu, \rho\right) P\left(j_\nu\ \vert\ \rho\right)$. The conditional probability distribution $P\left(\alpha\ \vert\ j_\nu, \rho\right)$ could be learnt from the Enzo simulation data used in this work, too.}\\
Our formalism allows for the generation of the function $I_\nu\left(\hat{r},\nu_\mathrm{obs}\right)$ at two (or more) different observing frequencies.
In turn, this enables the calculation of spectral indices for specific intensity rather than MEC (i.e. `in 2D' instead of `in 3D'), emulating the type of spectral analysis routinely performed by observational astronomers.
We caution that the procedure for MEC spectral index assignment used in this work does not respect the full diversity of spectral behaviour present in the Enzo simulation, and instead forces MEC spectral indices to approach the MEC-weighted mean.
In turn, this also causes specific intensity spectral index variations to be biased low.
Future work should adapt this procedure so that a plausible specific intensity spectral index prior can be added to the potent suite of predictions that follow simultaneously from our methodology.
%For this reason, specific intensity spectral indices are currently not reliable either.
%In this work, we have not considered the 
%Would the Milky Way change our decision where to look?
%If the emission is small scale, would we resolve out the Milky Way? Does this change for turbulence, which maybe has a larger large-scale component, mixing with the Milky Way contribution?

%\subsection{Analytical model extensions}
%An infinite number of independent cells is not realistic

%\subsection{Uncertainty of predictions and practical usefulness}
%The utility of a forecast with extreme uncertainties
%What is a good spot: high mean brightness, or a lower brightness with less variability?
%And the role of uncertainty of the emission mechanism (for example: perhaps less variation expected under the turbulence Ansatz)

\section{Conclusions}
\label{sec:conclusions}
   \begin{enumerate}
   % What is the methodology?
   \item In this work, we describe and implement the first methodology to produce a (prior) probability distribution over specific intensity functions representing the synchrotron cosmic web (SCW) of the Local Universe.
   We assume merger and accretion shocks to be the main generators of the SCW, and assume a primordial magnetogenesis scenario for the evolution of magnetic fields in the IGM.
   However, the methodology is general enough to explore alternative physical hypotheses in the future.
   Our prior can be used to guide and verify observational attempts to detect the SCW with low-frequency radio telescopes such as the LOFAR and the SKA.
   % What are the computational results?
   \item Using BORG SDSS total matter density reconstructions and Enzo cosmological simulations, we build a prior distribution that is informative over half of the Northern Sky, and that has a ${\sim} 0.6\degree$ resolution for LSS at $z = 0.1$.
   Although not a fundamental limitation of the methodology, the current resolution is not high enough to resolve individual merger and accretion shocks.
   Typically, filaments near massive structures give $I_\nu\ \xi_e^{-1} \sim 10^{-1}\ \mathrm{Jy\ deg^{-2}}$ at $\nu_\mathrm{obs} = 150\ \mathrm{MHz}$;  $\xi_e$ is the highly uncertain electron acceleration efficiency.
   Even at the ${\sim}3\ \mathrm{Mpc}\ h^{-1}$ reconstruction resolution, our merger and accretion shock SCW (MASSCW) prior indicates that the specific intensity for a given direction is highly uncertain (with a typical standard deviation being ${\sim}100\%$ of the mean) due to uncertainty regarding the presence and highly variable nature of shock emission along the line-of-sight.
   We present (marginal median) MASSCW specific intensity predictions for three deep LOFAR HBA fields: the Lockman Hole, Abell 2255, and the Ursa Major Supercluster.
   % What are the theoretical results?
   \item With a simple geometric model of cubic unit cells, we calculate both the distribution of the specific-intensity-weighted mean redshift RV $\bar{Z}$, as well as the flux-density-weighted mean redshift $\bar{\bar{z}}$ of the MASSCW signal for large-scale structure (LSS) reconstructions up to $z_\mathrm{max} = 0.2$.
   We obtain results that closely resemble those found numerically from our data-driven MASSCW prior, whose construction is much more involved.
   %\item With a simple geometric model of cubic unit cells, we calculate the flux-density-weighted mean redshift $\bar{\bar{z}}$ of the MASSCW signal for large-scale structure (LSS) reconstructions up to $z_\mathrm{max} = 0.2$.
   %We find $\bar{\bar{z}} = 0.07$, and obtain the same result numerically from our MASSCW prior, whose construction is much more involved.
   Encouraged, we present filament-only geometric model predictions for $\bar{\bar{z}}$ that include \emph{all} LSS (i.e. $z_\mathrm{max} = \infty$).
   %The geometric model predicts that, upon inclusion of all LSS (i.e. $z_\mathrm{max} = \infty$), $\bar{\bar{z}} = 0.13$.
   These predictions are highly insensitive to plausible variations in model parameters $l$ and $d_1$, demonstrating that $\bar{\bar{z}}$ is effectively determined by a single parameter: the sum of the typical MEC-weighted filament spectral index $\alpha_\mathrm{f}$ and the MEC - cosmological redshift power law exponent $\beta_\mathrm{f}$.
   Its future determination will characterise the completeness of the MASSCW predictions put forth in this work.
   In an optimistic case, our prior already reveals a great deal about filamentary baryons, and where to find them.
   %The radio quest for filamentary baryons
   %However, at this point, 
   %$\left(1+z\right)$
   %This prediction is highly insensitive to plausible variations in the two remaining model parameters $l$ and $d_1$.
   %We conclude that the MASSCW is a phenomenon to be searched for at low redshifts.
      %\item The anatomy of the Universe at the largest scales changes with cosmic time, with four structural elements canonically identified: voids, walls, filaments and clusters. Together, these structures give rise to an intricate web-like morphology. The past- and present-day filaments of the cosmic web contain a large fraction of the Universe's baryons; ... and ... respectively. Magnetic fields in filaments, at least in regions far away from galaxies, remain undisturbed for long periods of time, retaining the memory of their formation. Filaments therefore constitute an ideal target to investigate the origins of large-scale magnetism in the Universe.
      %\item As part of the LOFAR Deep Fields campaign, we analyse 100 hours of observations from the Lockman Hole, a famous window into the extragalactic Universe.
      %\item Although not obvious \textit{a prima facie}, we emphasise that the correlation result depends strongly on the choice of SCW predictive model. However, it is unclear what predictive model yields the most accurate results, as the physics of the SCW are not yet well-understood. Thus, all SCW searches are model-dependent; ours assumes that DSA is the dominant synchrotron emission mechanism.
      %\item We constrain the filamentary magnetic field strength to ... .
   \end{enumerate}

\begin{acknowledgements}
% The following acknowledgement is according to Reinout's December 11, 12019 e-mail:
M.S.S.L. Oei and R.J. van Weeren acknowledge support from the VIDI research programme with project number 639.042.729, which is financed by the Netherlands Organisation for Scientific Research (NWO).
F. Vazza acknowledges support from the ERC STG MAGCOW (714196) from the H2020.
F. Leclercq acknowledges funding from the Imperial College London Research Fellowship Scheme.\\
The cosmological simulations used in this work were produced with the Enzo code (\textcolor{blue}{enzo-project.org}) and run on the Piz-Daint supercluster at LSCS (Lugano) under project `s701' with F. Vazza as P.I..\\\\
M.S.S.L. Oei warmly thanks Andrea Botteon, Matthias Hoeft, Vincent Icke, Josh Albert, Lara Anisman, Jacob Bakermans and Jesse van Oostrum for helpful discussions.\\
\end{acknowledgements}

\bibliographystyle{aa}
\bibliography{cite}

\appendix

\section{Additional figures}
\label{ap:additionalFigures}

\begin{figure*}
    \centering
    \begin{subfigure}{\textwidth}
    \includegraphics[width=\textwidth]{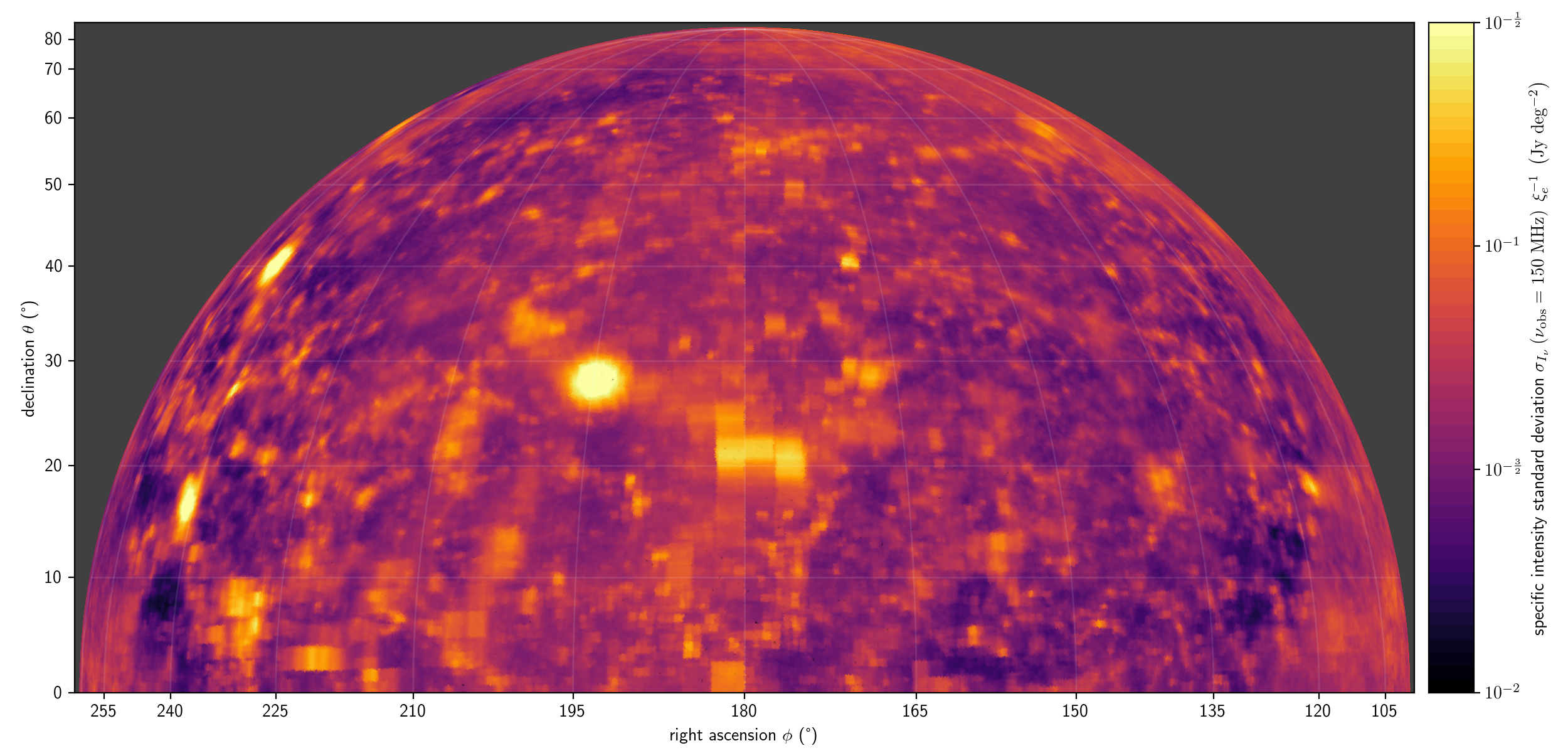}
    \caption{}
    \label{fig:specificIntensitySDSphere150MHz}
\end{subfigure}
\begin{subfigure}{\textwidth}
    \includegraphics[width=\textwidth]{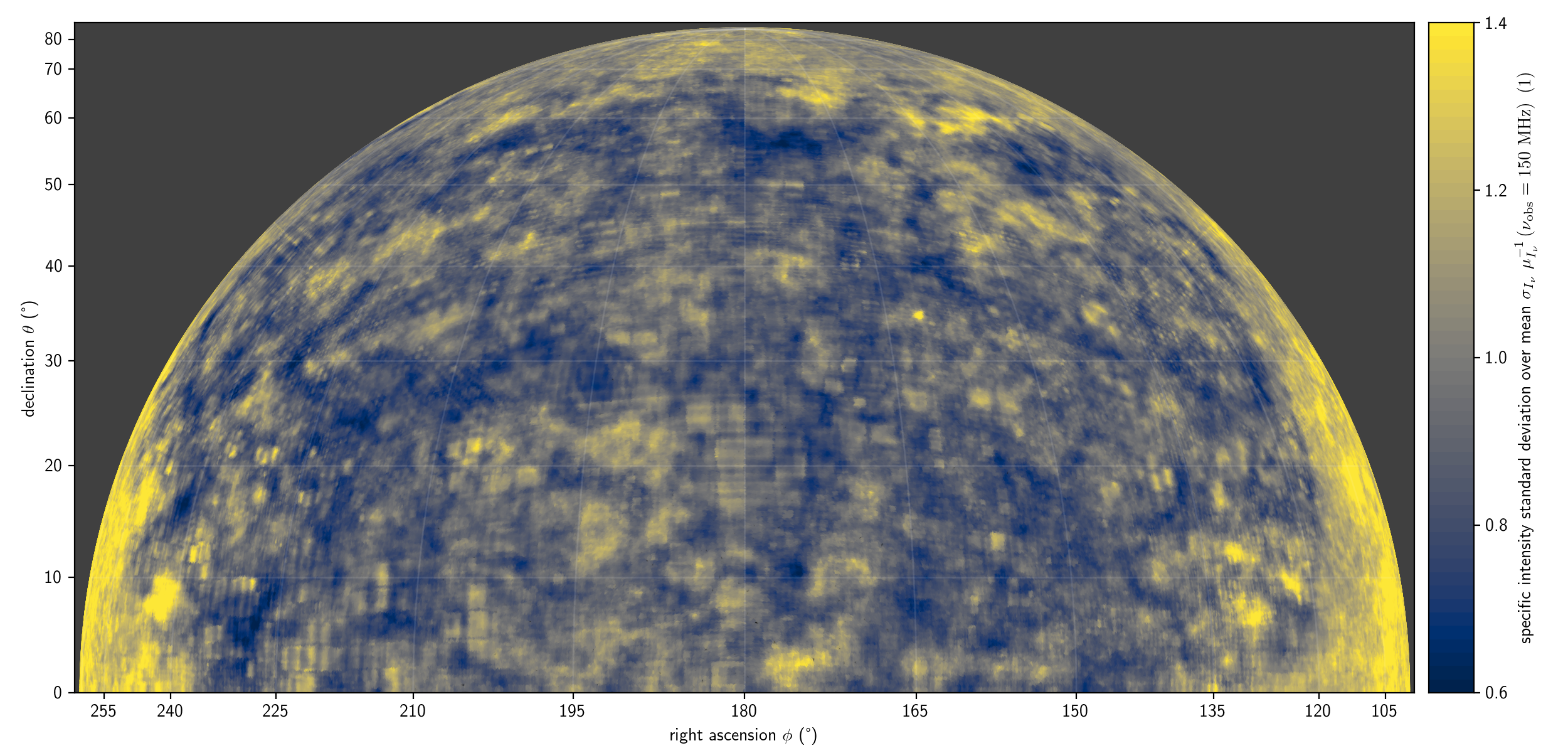}
    \caption{}
    \label{fig:specificIntensitySDOverMeanSphere150MHz}
\end{subfigure}
\caption{The MASSCW priors allow for a quantification of prediction uncertainty.
Here we show both an absolute and a relative measure of spread for the single-direction specific intensity distributions (i.e. marginals) at $\nu_\mathrm{obs} = 150\ \mathrm{MHz}$. For each direction, we discard data outside the $1 - 99\%$ percentile range. \textbf{Top:} marginal standard deviation (absolute uncertainty), which closely resembles the marginal mean of \textbf{Figure}~\ref{fig:specificIntensityMeanSphere150MHz}. \textbf{Bottom:} marginal standard deviation over mean (relative uncertainty), which reveals an inverted trend.}
\end{figure*}

\begin{figure*}
    \centering
    \includegraphics[width=.91\textwidth]{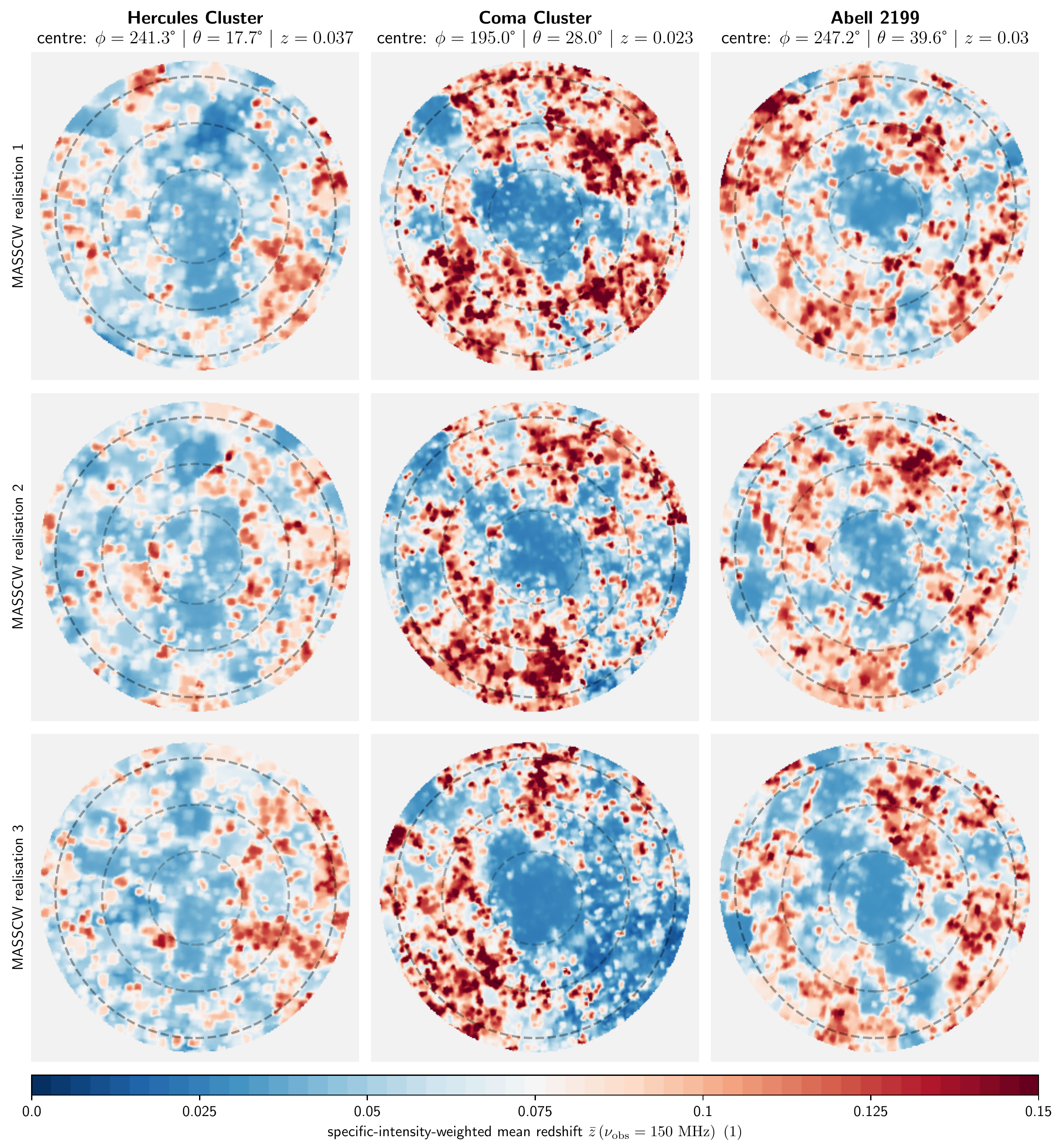}
    \caption{As \textbf{Figure}~\ref{fig:clusterFieldsSpecificIntensity150MHz}, but now for specific-intensity-weighted mean redshift instead of specific intensity.}
    \label{fig:clusterFieldsRedshiftWeighted150MHz}
\end{figure*}

\begin{figure}
    \centering
    \begin{subfigure}{\columnwidth}
    \includegraphics[width=\textwidth]{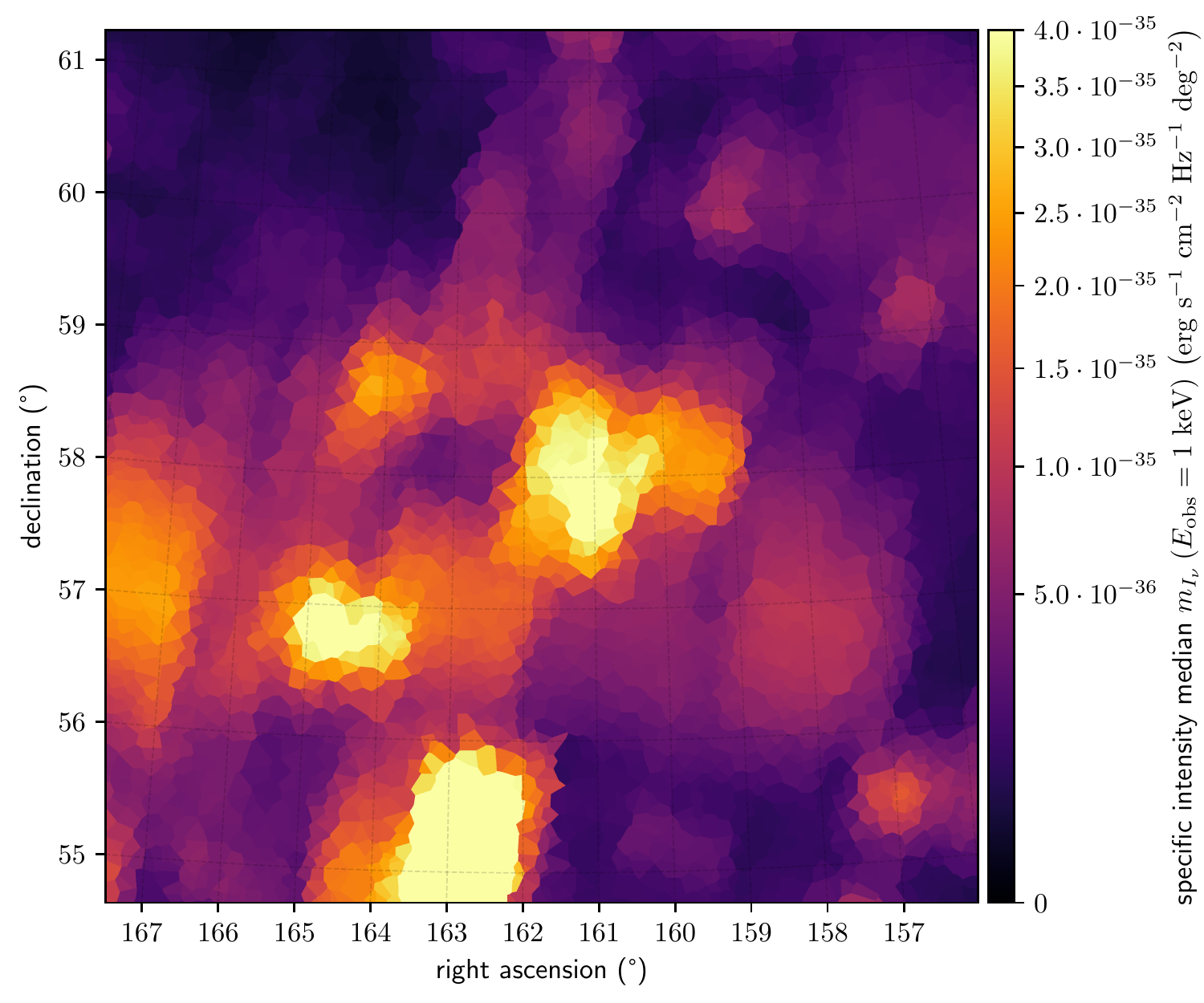}
    \end{subfigure}
    \begin{subfigure}{\columnwidth}
    \includegraphics[width=\textwidth]{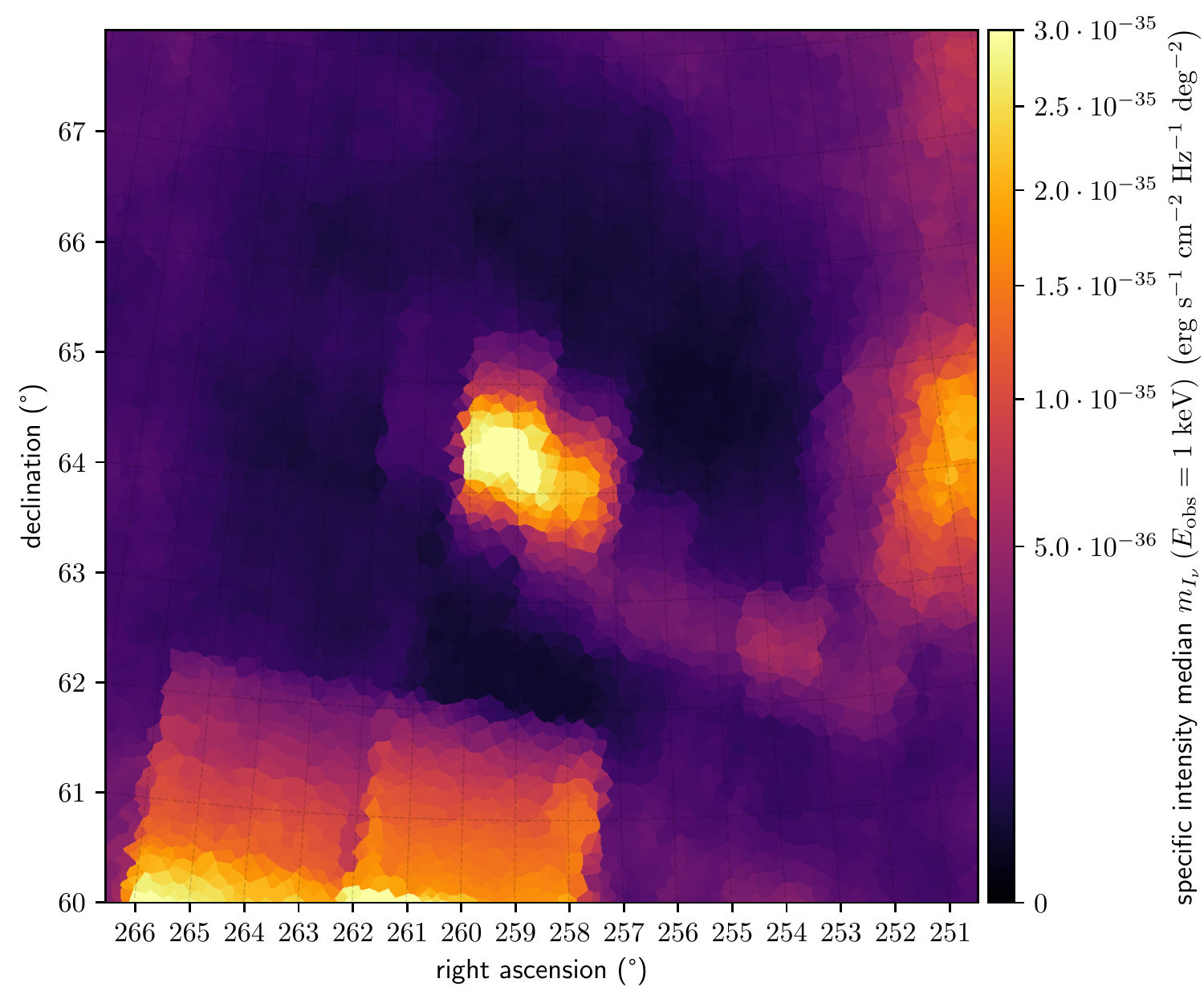}
    \end{subfigure}
    \begin{subfigure}{\columnwidth}
    \includegraphics[width=\textwidth]{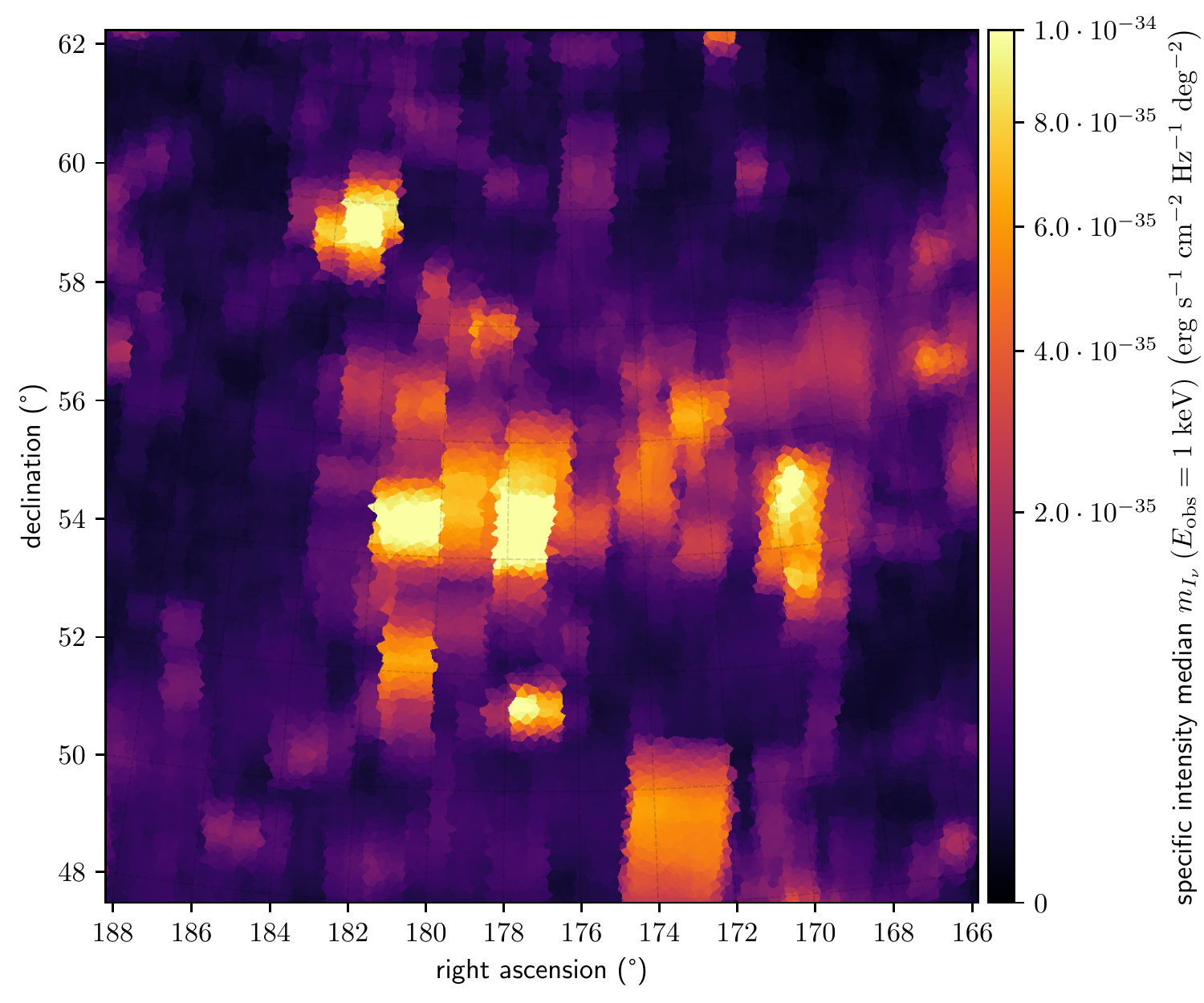}
    \end{subfigure}
    \caption{Thermal bremsstrahlung specific intensity prior marginal medians at $E_\mathrm{obs} = 1\ \mathrm{keV}$ for three deep LOFAR HBA fields. For merger and accretion shock synchrotron predictions of the same regions, see \textbf{Figure}~\ref{fig:LOFARDeepFields}.
    \textbf{Top:} Lockman Hole. \textbf{Middle:} Abell 2255. \textbf{Bottom:} Ursa Major Supercluster.}
    \label{fig:LOFARDeepFieldsThermalBremsstrahlung}
\end{figure}

\begin{figure*}
    \centering
    \includegraphics[width=\textwidth]{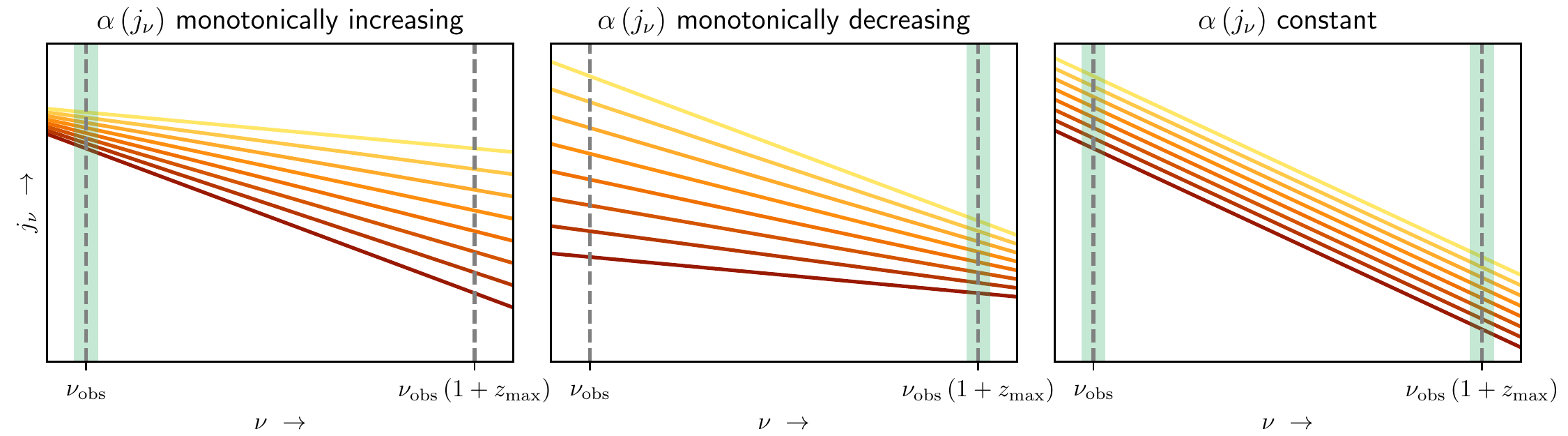}
    \caption{Due to cosmological redshifting, the calculation of the specific intensity $I_\nu$ at observing frequency $\nu_\mathrm{obs}$ necessitates knowing the monochromatic emission coefficient (MEC) $j_\nu$ at a range of emission frequencies $\nu$ (see \textbf{Equation}~\ref{eq:specificIntensity}). In this work, we generate $j_\nu$ at emission frequencies $\nu = \nu_\mathrm{obs}$ and $\nu = \nu_\mathrm{obs}\left(1 + z_\mathrm{max}\right)$, where $z_\mathrm{max}$ is the maximum cosmological redshift of the large-scale structure reconstructions. We do so using the same Gaussian random field realisation $\mathcal{Z}$ (and thus percentile random field realisation $\mathcal{P}$), implicitly assuming that shocks retain their MEC percentile rank over the frequency range $[\nu_\mathrm{obs}, \nu_\mathrm{obs}\left(1+z_\mathrm{max}\right)]$. This assumption holds in three scenarios assuming simple power-law synchrotron spectra for shocks (sketched above). Each solid line represents a shock with constant percentile rank (over the range shown, at least); the graphs are drawn with logarithmic scaling. Regarding the sign of the spectral index, we use the convention $j_\nu \propto \nu^\alpha$. \textbf{Left:} the integrated spectral index $\alpha$ is a monotonically \emph{increasing} function of $j_\nu$ at $\nu = \nu_\mathrm{obs}$ (`brighter shocks have flatter spectra'). \textbf{Middle:} the integrated spectral index $\alpha$ is a monotonically \emph{decreasing} function of $j_\nu$ at $\nu = \nu_\mathrm{obs}\left(1 + z_\mathrm{max}\right)$ (`brighter shocks have steeper spectra'). \textbf{Right:} the integrated spectral index $\alpha$ is a $j_\nu$-independent constant (at all frequencies); this is a limiting case of the previous two.}
    \label{fig:assumptionSpectralIndices}
%Due to cosmological redshifting, the calculation of the specific intensity $I_\nu$ at observing frequency $\nu_\mathrm{obs}$ necessitates the calculation of the monochromatic emission coefficient (MEC) $j_\nu$ at two emission frequencies $\nu$: $\nu = \nu_\mathrm{obs}$ and $\nu = \nu_\mathrm{obs}\left(1 + z_\mathrm{max}\right)$, where $z_\mathrm{max}$ is the maximum cosmological redshift of the large-scale structure reconstructions used. In this work, we generate $j_\nu\left(\nu = \nu_\mathrm{obs}\right)$ and $j_\nu\left(\nu = \nu_\mathrm{obs}\left(1 + z_\mathrm{max}\right)\right)$ using the same Gaussian random field realisation (and thus percentile random field realisation), implicitly assuming that shocks retain their MEC percentile rank over the frequency range $[\nu_\mathrm{obs}, \nu_\mathrm{obs}\left(1+z_\mathrm{max}\right)]$. 
\end{figure*}

In this appendix, we list four figures referenced in the main text, but relegated for structural clarity.

\section{Single-shock synchrotron MEC - total matter density scaling relation}
\label{ap:scalingRelation}
In this appendix, we derive a synchrotron power density - total matter density scaling relation for individual shocks with high upstream Mach numbers in cluster outskirts and filaments, assuming the \textcolor{blue}{\citet{Hoeft12007}} model and $\gamma = \sfrac{5}{3}$.
Because the power density and MEC of a single shock are proportional, this immediately also yields the desired single-shock synchrotron MEC - total matter density scaling relation.
% Note: the adiabatic index assumption sets the integrated spectral index, which sets the MEC - magnetic field strength scaling.
\subsection{Temperature and the speed of sound}
Note that $\mathcal{M}_\mathrm{u} \coloneqq \frac{v_\mathrm{u}}{c_\mathrm{s,u}}$, where $v_\mathrm{u}$ is the shock velocity relative to the upstream plasma, and $c_\mathrm{s,u}$ is the speed of sound in the upstream plasma.
The Newton--Laplace equation for an ideal gas predicts $c_\mathrm{s,u} \propto \sqrt{T_\mathrm{u}}$, where $T_\mathrm{u}$ is the upstream plasma temperature.\footnote{For example, the speed of sound is $10$ times higher in the $10^8\ \mathrm{K}$ ICM than in the $10^6\ \mathrm{K}$ WHIM, and $100$ times higher in the $10^9\ \mathrm{K}$ ICM than in the $10^5\ \mathrm{K}$ WHIM.}
The upstream Mach number of a shock incident on the WHIM would therefore be higher than that of a shock incident on the ICM if the shocks arrive at the same velocity $v_\mathrm{u}$ relative to these media.
%For an ensemble of cosmological structure formation shocks, some incident on the WHIM, some on the ICM, that collide with these media at some fixed relative velocity $v_\mathrm{u}$, we thus find the scaling relation $\mathcal{M}_\mathrm{u} \propto T_\mathrm{u}^{\sfrac{-1}{2}}$.
%; it is even
%\textcolor{red}{For an ensemble of cosmological structure formation shocks, some incident on the WHIM, some on the ICM, that collide with these media at some fixed relative velocity $v_\mathrm{u}$, we thus find the scaling relation $\mathcal{M}_\mathrm{u} \propto T_\mathrm{u}^{\sfrac{-1}{2}}$.
%Numerical simulations validate these general considerations: \textcolor{blue}{\citet{Ryu12003}} calculate that $\mathcal{M}_\mathrm{u} \sim 10^0 - 10^1$ for the ICM, whilst finding $\mathcal{M}_\mathrm{u} \sim 10^1 - 10^2$ for the WHIM.}
\subsection{The filament regime: low magnetic field strengths}
One of the prime reasons for pursuing SCW detections is to gauge the unknown strength of the Universe's largest magnetic fields.
Numerical simulations by \textcolor{blue}{\citet{Vazza12015a, Vazza12017}} that reproduce the observed magnetic field strengths in galaxy clusters, predict magnetic field strengths in filaments that depend strongly on the magnetogenesis scenario considered, ranging between $10^{-1} - 10^2\ \mathrm{nG}$.
%Nevertheless, the results indicate a strength of $10^{-1} - 10^2\ \mathrm{nG}$.
For the purposes of finding a power density - matter density scaling relation, the relevant quantity to compare the filament IGM magnetic field strength $B_\mathrm{IGM}$ with at cosmological redshift $z$, is the CMB magnetic field strength $B_\mathrm{CMB}\left(z\right)$.\\\\
As the CMB is well modelled by a blackbody\footnote{In fact, it is the most accurate blackbody ever observed!}, the CMB magnetic field strength $B_\mathrm{CMB}\left(z\right)$ is derived by equating the electromagnetic energy density $u_\mathrm{EM}$ of a blackbody of temperature $T$ to the electromagnetic energy density of a magnetic field of magnitude $B$:
\begin{align}
    \frac{4\sigma}{c} T^4 = u_\mathrm{EM} = \frac{1}{2\mu_0} B^2,
\end{align}
where $\sigma$ is the Stefan--Boltzmann constant and $\mu_0$ is the vacuum permeability.
Upon rearranging, and for $T = T_\mathrm{CMB}$ and $B = B_\mathrm{CMB}$, we find
\begin{align}
    B_\mathrm{CMB}\left(z\right) = \sqrt{\frac{8\mu_0\sigma}{c}}T_\mathrm{CMB}^2\left(z\right).
\end{align}
Let $a$ be the scale factor and let $a_0$ be its present-day value.
As $T_\mathrm{CMB}^4 \propto u_\mathrm{EM} \propto a^{-4} = a_0^{-4}\left(1 + z\right)^4$ due to the expansion of the Universe, it follows that $T_\mathrm{CMB} \propto 1 + z$, and thus
\begin{align}
    B_\mathrm{CMB}\left(z\right) &= \sqrt{\frac{8\mu_0\sigma}{c}}T_\mathrm{CMB}^2\left(0\right)\left(1+z\right)^2 \nonumber\\
    &= B_\mathrm{CMB}\left(0\right)\left(1+z\right)^2.
\end{align}
Using $T_\mathrm{CMB}\left(0\right) = 2.725\ \mathrm{K}$ yields $B_\mathrm{CMB}\left(0\right) = 3.238\ \mathrm{\mu G}$.\\\\
Thus, under all plausible scenarios of magnetogenesis, $B_\mathrm{IGM}\left(z\right) \ll B_\mathrm{CMB}\left(0\right) \leq B_\mathrm{CMB}\left(z\right)$.

\subsection{Magnetic field strength and baryon density}
A scaling relation between $B$ and $\rho_\mathrm{BM}$ follows from considering the conservation of magnetic flux as the Universe expands.
The magnetic flux through a surface is the product of the surface area and the magnetic field strength (and the cosine of the angle between the surface normal and the magnetic field).
Over time, the surface area increases $\propto a^2$, so that the magnetic field strength must follow $\propto a^{-2}$ if conservation of magnetic flux is to hold.
Finally, as $\rho_\mathrm{BM} \propto a^{-3}$, one obtains $B \propto \rho_\mathrm{BM}^{\sfrac{2}{3}}$.
\textbf{Figure}~4 of \textcolor{blue}{\citet{Vazza12017}} compares this scaling relation with simulated magnetic field strengths and baryon densities under various scenarios of magnetogenesis.

% Use microGauss (deprecated, though common) or Tesla (SI, though unconventional)?
%\cdot10^{-10}\ \mathrm{T}

\subsection{The power density expression simplifies}
The \textcolor{blue}{\citet{Hoeft12007}} power density folded into \textbf{Equation}~\ref{eq:properMonochromaticEmissionCoefficient} simplifies appreciably if high Mach numbers and low magnetic field strengths are assumed.
In such a regime, $\alpha \approx -1$ and $\Psi \approx 1$, while $B_\mathrm{d}^2 \ll B_\mathrm{CMB}^2$.
Thus,
\begin{align}
    P_\nu &\propto \rho_{\mathrm{BM,d}} \cdot T_\mathrm{d}^\frac{3}{2} \cdot B_\mathrm{d}^2 \nonumber\\
    &= \frac{\rho_{\mathrm{BM,d}}}{\rho_{\mathrm{BM,u}}} \rho_{\mathrm{BM,u}} \cdot \left(\frac{T_\mathrm{d}}{T_\mathrm{u}} T_\mathrm{u}\right)^\frac{3}{2} \cdot \left(\frac{B_\mathrm{d}}{B_\mathrm{u}} B_\mathrm{u}\right)^2.
\end{align}
From the Rankine--Hugoniot jump conditions, one can derive that the compression factor (\textbf{Figure}~\ref{fig:spectralIndex}, central panel), as a function of $\mathcal{M}_\mathrm{u}$ and $\gamma$, is
\begin{align}
    \frac{\rho_\mathrm{BM,d}}{\rho_\mathrm{BM,u}}\left(\mathcal{M}_\mathrm{u},\gamma\right) = \frac{\left(\gamma + 1\right) \mathcal{M}_\mathrm{u}^2}{\left(\gamma-1\right)\mathcal{M}_\mathrm{u}^2+2} = \left(\frac{B_\mathrm{d}}{B_\mathrm{u}}\left(\mathcal{M}_\mathrm{u},\gamma\right)\right)^\frac{3}{2}.
\end{align}
The same equations dictate that the temperature increase (\textbf{Figure}~\ref{fig:spectralIndex}, bottom panel), as a function of $\mathcal{M}_\mathrm{u}$ and $\gamma$, is
\begin{align}
    %\frac{T_\mathrm{d}}{T_\mathrm{u}} = \frac{1}{\left(\gamma+1\right)^2}\left(2\gamma\left(\gamma-1\right)\mathcal{M}^2 + 4\gamma-\left(\gamma-1\right)^2-\frac{2\left(\gamma-1\right)}{\mathcal{M}^2}\right).
    \frac{T_\mathrm{d}}{T_\mathrm{u}}\left(\mathcal{M}_\mathrm{u},\gamma\right) = \frac{2\gamma\left(\gamma-1\right)\mathcal{M}_\mathrm{u}^2 + 4\gamma-\left(\gamma-1\right)^2-2\left(\gamma-1\right)\mathcal{M}_\mathrm{u}^{-2}}{\left(\gamma+1\right)^2}.
\end{align}
Thus, for $\mathcal{M}_\mathrm{u} \gg 1$, $\tfrac{\rho_\mathrm{BM,d}}{\rho_\mathrm{BM,u}} \propto 1$ and $\tfrac{T_\mathrm{d}}{T_\mathrm{u}} \propto \mathcal{M}_\mathrm{u}^2$.
Returning to the power density scaling relation, we find
\begin{align}
    P_\nu &\propto \rho_{\mathrm{BM,u}} \cdot \left(\mathcal{M}_\mathrm{u}^2\  T_\mathrm{u}\right)^\frac{3}{2} \cdot B_\mathrm{u}^2 \nonumber \\
    &= \rho_{\mathrm{BM,u}} \cdot v_\mathrm{u}^3\cdot B_\mathrm{u}^2.
\end{align}
To arrive at the second line, we use the definition of the upstream Mach number and the upstream sound speed - upstream plasma temperature scaling relation.
%that holds for shocks of some fixed velocity.

\subsection{Upstream velocity and total density}
We investigate the upstream velocity - total (i.e. dark and baryonic matter) density relation for three simple geometries.
We invoke Gauss' law for gravity to find expressions for the gravitational field, derive the gravitational potential using the fact that the gravitational force is conservative, and equate, for a test particle, the loss in gravitational potential energy to the gain in kinetic energy.
We assume that the structures considered have hard edges and are equidense (with total matter density $\rho \coloneqq \rho_\mathrm{BM,u} + \rho_\mathrm{DM}$) within.\\
Upon impact, the velocity of a test particle starting from rest at a distance $d$, and falling onto an isolated spherical galaxy cluster with radius $R$, is
\begin{align}
    v_\mathrm{u} = \sqrt{\frac{8\pi G}{3} \rho \left(1 - \frac{R}{d}\right)}\ R.
\end{align}
Upon impact, the velocity of a test particle starting from rest at a distance $d$, and falling onto an isolated cylindrical filament with radius $R$, is
\begin{align}
    v_\mathrm{u} = \sqrt{4\pi G \rho \ln{\frac{d}{R}}}\ R.
\end{align}
Upon impact, the velocity of a test particle starting from rest at a distance $d$, and falling onto an isolated, thick planar sheet of half-thickness $R$, is
\begin{align}
    v_\mathrm{u} = \sqrt{8 \pi G \rho R \left(d - R\right)}.
\end{align}
In all three cases, $R$ determines the size of the structure types.
Assuming no relation between total density $\rho$ and $R$, we find $v_\mathrm{u} \propto \sqrt{\rho}$, irrespective of the geometry.

%The monochromatic emission coefficient scaling relation becomes
%\begin{align}
%    j_\nu \propto \rho_\mathrm{BM,u}^\frac{7}{3}.
%\end{align}

\subsection{Baryon density and dark matter density}
Structure formation theory predicts that after decoupling, and if gas pressure is ignored, $\rho_\mathrm{BM,u} \propto \rho_\mathrm{DM}$.
\emph{Including} gas pressure, this proportionality is expected to remain valid on large scales only.
%; filaments are structures with volumes of many $\mathrm{Mpc}^3$.
For filaments in particular, \textbf{Figure}~6 of \textcolor{blue}{\citet{Gheller12016}} shows that the baryon fraction in filaments remains close to $f_\mathrm{cosmic} \coloneqq \Omega_{\mathrm{BM},0} \left(\Omega_{\mathrm{BM},0} + \Omega_{\mathrm{DM},0}\right)^{-1} \left(= 0.167\right)$ over four orders of magnitude of total baryonic mass, and for redshifts from $1$ to $0$; this is consistent with $\rho_\mathrm{BM,u} \propto \rho_\mathrm{DM}$.\\\\
Using $B_\mathrm{u} \propto \rho_\mathrm{BM,u}^{\sfrac{2}{3}}$, $v_\mathrm{u} \propto \sqrt{\rho}$ and $\rho_\mathrm{BM,u} \propto \rho_\mathrm{DM}$, the final scaling relation becomes
\begin{align}
    j_\nu \propto P_\nu \propto \rho^\frac{23}{6}.
    %j_\nu \propto \rho_\mathrm{DM}^\frac{7}{3}.
\end{align}
Note that, to obtain a scaling relation for the \emph{total} (rather than single-shock) MEC, one should also consider how the shock number density relates to the total matter density.
The total MEC - matter density and the single-shock MEC - matter density scaling relation exponents are only the same when no such relationship exists.

\section{Ray tracing in the cosmological setting}
\label{ap:rayTracing}
Projecting a ray's 4D null geodesic in a pure Friedmann--Lema\^itre--Robertson--Walker (FLRW) metric onto 3D comoving space results in a straight line.
This follows readily from the FLRW metric in hyperspherical coordinates
\begin{align}
    \mathrm{d}s^2 = -c^2 \mathrm{d}t^2 + \left(\frac{a(t)}{a_0}\right)^2\left(\mathrm{d}r^2 + S_k^2\left(r\right)\mathrm{d}\Omega^2\right),
\end{align}
where $\mathrm{d}s^2$ is the spacetime line element, $c$ is the speed of light \textit{in vacuo}, $t$ is physical time since the Big Bang, and $a\left(t\right)$ is the scale factor (with $a_0 \coloneqq a\left(t_0\right)$ being its present-day ($t = t_0$) value).
%$S_k\left(r\right) \coloneqq r\ \mathrm{sinc}\left(r\sqrt{k}\right)$
Also, $r$ is the radial comoving distance, $k$ is the Universe's Gaussian curvature (with SI units $\mathrm{m}^{-2}$), $S_k\left(r\right) \coloneqq r\ \mathrm{sinc}\left(r\sqrt{k}\right)$ is the transverse comoving distance and $\mathrm{d}\Omega^2 \coloneqq \mathrm{d}\theta^2 + \cos^2\theta\mathrm{d}\phi^2$.
(The sinc function follows the mathematical (unnormalised) convention.)
Finally, let the location of present-day Earth be the spatial origin.
% It seems that the assumption of a flat universe is not even needed here, as $\mathrm{d}\Omega^2 = 0$ anyways...
An initially radial ($\mathrm{d}\Omega^2 = 0$) null ($\mathrm{d}s^2 = 0$) geodesic thus satisfies% in a flat ($k = 0$, so $S_k^2\left(r\right) = r^2$) universe satisfies
\begin{align}
    %c \mathrm{d}t =\ \pm \frac{a_0}{1 + z} \mathrm{d}r,
    c \mathrm{d}t =\ \pm \frac{\mathrm{d}r}{1 + z},
\label{eq:timeAndComovingRadialDistance}
\end{align}
regardless of $k$.
%Thus, as time progresses, such rays maintain their direction and only change in $r$; this justifies considering the path $\mathcal{L}$ of a light ray with direction $\hat{r}$ in comoving space as the set of points $\mathcal{L}\left(\hat{r}\right) \coloneqq \{\mu\ \hat{r} \in \mathbb{R}^3\ |\ \mu \in \mathbb{R}_{\geq 0}\}$.\\
As time progresses, such rays maintain their direction and only change in $r$; this justifies considering the path $\mathcal{L}$ of a light ray with direction $\hat{r}$ in comoving space as the set of points $\mathcal{L}\left(\hat{r}\right) \coloneqq \{r\ \hat{r} \in \mathbb{R}^3\ |\ r \in \mathbb{R}_{\geq 0}\}$.

\section{Observer's specific intensity}
\label{ap:SIFormula}
Our aim is to derive an expression for the specific intensity in direction $\hat{r}$ at observing frequency $\nu_\mathrm{obs}$.
We follow \textbf{Chapter}~12 of \textcolor{blue}{\citet{Peacock11999}}, but generalise to arbitrary $\Lambda$CDM models (by allowing $\Lambda \neq 0$), and recast the results in terms of the MEC instead of the emissivity.\\
As in \textbf{Appendix}~\ref{ap:rayTracing}, consider a FLRW metric with arbitrary Gaussian curvature $k$.
A comoving volume element $\mathrm{d}V_\mathrm{c}$ and the corresponding proper volume element $\mathrm{d}V_\mathrm{p}$ at comoving radial distance $r$ and cosmological redshift $z = z\left(r\right)$ that cover a solid angle $\mathrm{d}\Omega$ on the sky are given by
\begin{align}
    \mathrm{d}V_\mathrm{c} &= S_k\left(r\right)^2\ \mathrm{d}\Omega \mathrm{d}r,\\
    \mathrm{d}V_\mathrm{p} &= S_k\left(r\right)^2\ \mathrm{d}\Omega \mathrm{d}r \left(1+z\right)^{-3}.
\end{align}
Recall that we have defined $j_\nu$ as the \emph{proper} MEC, and assume that the filament IGM radiates isotropically.
The luminosity density $\mathrm{d}L_\nu$ of the volume, seen in direction $\hat{r}$ and at \emph{emission} frequency $\nu = \left(1+z\right)\nu_\mathrm{obs}$, then equals
\begin{align}
    \mathrm{d}L_\nu\left(\hat{r},\nu\right) = 4\pi j_\nu\left(r\ \hat{r}, z, \nu\right) \mathrm{d}V_\mathrm{p}.
\end{align}
The corresponding observer's flux density $\mathrm{d}F_\nu$ of the volume in direction $\hat{r}$ at \emph{observing} frequency $\nu_\mathrm{obs}$ is
\begin{align}
    \mathrm{d}F_\nu\left(\hat{r},\nu_\mathrm{obs}\right) = \frac{\mathrm{d}L_\nu\left(\hat{r},\nu\right)}{4\pi S_k\left(r\right)^2\left(1+z\right)},
\end{align}
and so the observer's specific intensity $\mathrm{d}I_\nu$ in direction $\hat{r}$ at observing frequency $\nu_\mathrm{obs}$ is
\begin{align}
    &\mathrm{d}I_\nu\left(\hat{r},\nu_\mathrm{obs}\right) \coloneqq \frac{\mathrm{d}F_\nu\left(\hat{r},\nu_\mathrm{obs}\right)}{\mathrm{d}\Omega}\\
    &= \frac{4\pi j_\nu\left(r\ \hat{r}, z, \nu\right) S_k\left(r\right)^2 \mathrm{d}\Omega \mathrm{d}r}{4\pi S_k\left(r\right)^2 \left(1+z\right) \mathrm{d}\Omega \left(1+z\right)^3} = \frac{j_\nu\left(r\ \hat{r},z,\nu\right)\ \mathrm{d}r}{\left(1+z\right)^4}.
\end{align}
We neglect absorption so that the specific intensity of the ray only accumulates as the ray travels through LSS to the observer: the Universe is mostly optically thin for $\nu < 1\ \mathrm{GHz}$; we assume this holds perfectly.
So, by collecting all contributions along the ray's path, we obtain the equivalent of \textbf{Equation}~12.12 of \textcolor{blue}{\citet{Peacock11999}}:
\begin{align}
    I_\nu\left(\hat{r}, \nu_\mathrm{obs}\right) = \int_0^\infty \frac{j_\nu\left(r\ \hat{r},z\left(r\right),\nu_\mathrm{obs}\left(1+z\left(r\right)\right)\right)}{\left(1+z\left(r\right)\right)^4} \mathrm{d}r.
\label{eq:specificIntensityComovingRadialDistance}
\end{align}
%We compute \emph{today's} quantity $I_\nu$ by summing up contributions for all times, yielding
%\begin{align}
    %I_{\nu, \mathrm{obs}}\left(\hat{r}, \nu\right) &= \int j_{\nu, \mathrm{obs}}\left(\mu\left(l\right)\hat{r}, z\left(l\right), \nu\right) \ \mathrm{d}l\\
    %&= \int_0^{t_0} \frac{j_\nu\left(\mu\left(t\right)\hat{r}, z\left(t\right), \nu\left(1+z\left(t\right)\right)\right)}{\left(1 + z\left(t\right)\right)^3}\ c\ \mathrm{d}t.
%    I_\nu\left(\hat{r}, \nu_\mathrm{obs}\right) = \int_0^{t_0} \frac{j_\nu \left(r\left(t\right)\hat{r}, z\left(t\right), \nu_\mathrm{obs}\left(1+z\left(t\right)\right)\right)}{\left(1 + z\left(t\right)\right)^3}\ c\ \mathrm{d}t.
%\label{eq:specificIntensityTime}
%\end{align}
Alternatively, one can express $I_\nu\left(\hat{r},\nu_\mathrm{obs}\right)$ as an integral over $z$.
Because
\begin{align}
    r\left(z\right) = \frac{c}{H_0} \int_0^z \frac{\mathrm{d}z'}{E\left(z'\right)},
    \label{eq:comovingRadialDistanceRedshift}
\end{align}
with the dimensionless Hubble parameter $E\left(z\right) \coloneqq \frac{H\left(z\right)}{H_0}$ being $E\left(z\right) = \sqrt{\Omega_{\mathrm{R},0}\left(1+z\right)^4 + \Omega_{\mathrm{M},0}\left(1+z\right)^3 + \Omega_{\mathrm{K},0}\left(1+z\right)^2 + \Omega_{\Lambda,0}}$ and today's curvature density parameter being $\Omega_{\mathrm{K},0} = 1 - \Omega_{\mathrm{R},0} - \Omega_{\mathrm{M},0} - \Omega_{\Lambda,0}$, we have
\begin{align}
%&E\left(z\right) = \sqrt{\Omega_{\mathrm{R},0}\left(1+z\right)^4 + \Omega_{\mathrm{M},0}\left(1+z\right)^3 + \Omega_{\mathrm{K},0}\left(1+z\right)^2 + \Omega_{\Lambda,0}}\\
%    &\textrm{and}\ \Omega_{\mathrm{K},0} = 1 - \Omega_{\mathrm{R},0} - \Omega_{\mathrm{M},0} - \Omega_{\Lambda,0},\\
%    &\textrm{we find}\ \mathrm{d}r = \frac{c}{H_0}\frac{\mathrm{d}z}{E\left(z\right)}.
\mathrm{d}r = \frac{c}{H_0}\frac{\mathrm{d}z}{E\left(z\right)}.
\label{eq:comovingRadialDistanceRedshiftDifferential}
\end{align}
Combining \textbf{Equations}~\ref{eq:specificIntensityComovingRadialDistance}, \ref{eq:comovingRadialDistanceRedshift} and \ref{eq:comovingRadialDistanceRedshiftDifferential},
\begin{align}
    I_\nu\left(\hat{r},\nu_\mathrm{obs}\right) = \frac{c}{H_0} \int_0^\infty \frac{j_\nu\left(r\left(z\right) \hat{r},z,\nu_\mathrm{obs}\left(1+z\right)\right)}{\left(1+z\right)^4 E\left(z\right)} \mathrm{d}z.
\label{eq:specificIntensityRedshift}
\end{align}
(Barring notational differences, the $\Omega_{\Lambda,0} = 0$ (and $\Omega_{\mathrm{R},0} = 0$) limit of this formula is \textbf{Equation}~12.10 of \textcolor{blue}{\citet{Peacock11999}}.)
Finally, we can read off that
\begin{align}
\frac{\mathrm{d}I_\nu}{\mathrm{d}z}\left(\hat{r},\nu_\mathrm{obs},z\right) = \frac{c}{H_0}\frac{j_\nu\left(r\left(z\right)\hat{r},z,\nu_\mathrm{obs}\left(1+z\right)\right)}{\left(1+z\right)^4 E\left(z\right)}.
\end{align}
%We prefer to work with an integral over $z$, and in a FLRW universe with only matter and a cosmological constant, the appropriate coordinate transformation (\textbf{Equation}~7.64 of \textcolor{blue}{\citet{Longair12008}}) is
%We prefer to work with an integral over $z$, and hence perform a coordinate transformation.
%In a FLRW universe with matter and a cosmological constant, we have (\textbf{Equation}~7.64 of \textcolor{blue}{\citet{Longair12008}}):
%\begin{align}
%    \mathrm{d}t = - \frac{\mathrm{d}z}{\left(1+z\right)H_0\sqrt{\left(1+z\right)^2\left(1+\Omega_{\mathrm{M},0}z\right) - \Omega_{\Lambda,0} z \left(2 + z\right)}}.
%\end{align}
%Upon changing integration variables from $t$ to $z$, and swapping the limits, we obtain\footnote{Barring notational differences, the $\Omega_{\Lambda,0} = 0$ limit of this formula is \textbf{Equation}~12.10 of \textcolor{blue}{\citet{Peacock11999}}.}
%The Peacock formula assumes a universe with only matter, but which is not necessarily flat. The Omega refers to (today's?) matter density parameter.
%\begin{align}
%\boxed{
%&I_\nu\left(\hat{r}, \nu_\mathrm{obs}\right) = \int_0^\infty \frac{\mathrm{d}I_\nu}{\mathrm{d}z}\left(\hat{r},\nu_\mathrm{obs},z\right)\ \mathrm{d}z \nonumber \\
%&= \int_0^\infty \frac{c\ H_0^{-1}\ j_\nu \left(r\left(z\right)\hat{r}, z, \nu_\mathrm{obs}\left(1 + z\right)\right)}{(1 + z)^4 \sqrt{\left(1+z\right)^2\left(1 + \Omega_{\mathrm{M},0} z\right) - \Omega_{\Lambda,0} z \left(2 + z\right)}}\ \mathrm{d}z.
%\end{align}}

\section{Volume-filling fractions}
\label{ap:VFFs}
\begin{table*}[h]%bp
\caption[]{Comparison between cosmic web structure type volume-filling fractions (VFFs) predicted by the cubic unit cell model and those obtained by \textcolor{blue}{\citet{Forero-Romero12009}} from cosmological simulations for eigenvalue threshold $\lambda_\mathrm{th} = 1$ and two effective smoothing scales $R_\mathrm{eff}$.
The VFF ratio columns give the simulation VFFs of the preceding column divided by the cubic unit cell VFFs of the first column.}
\label{tab:VFFs}
$$ 
         \begin{array}{l | l | l | l | l | l }%p{\columnwidth}l}
            \hline
            \noalign{\smallskip}
            \textrm{cosmic web} & \textrm{cubic unit cells} & \textrm{\textcolor{blue}{\citet{Forero-Romero12009}}} & \mathrm{VFF} & \textrm{\textcolor{blue}{\citet{Forero-Romero12009}}} & \mathrm{VFF}\\
             \textrm{structure type}& \frac{w_\mathrm{f}}{l} = 10^{-1} & \lambda_\mathrm{th} = 1, R_\mathrm{eff} = 0.88\ h^{-1}\ \mathrm{Mpc} & \textrm{ratio} & \lambda_\mathrm{th} = 1, R_\mathrm{eff} = 2.05\ h^{-1}\ \mathrm{Mpc} & \textrm{ratio}\\
            \noalign{\smallskip}
            \hline
            \noalign{\smallskip}
            \textbf{\textrm{voids}} & 72.9\% & 76\% & 1.04 & 82\% & 1.12\\
            \textbf{\textrm{sheets}} & 24.3\% & 18\% & 0.74 & 14\% & 0.58\\
            \textbf{\textrm{filaments}} & 2.7\% & 5\% & 1.85 & 4\% & 1.48\\
            \textbf{\textrm{clusters}} & 0.1\% & 0.5\% & 5.00 & 0.28\% & 2.80\\
            \noalign{\smallskip}
            \hline
         \end{array}
$$ 
\end{table*}

\begin{figure}
    \centering
    \includegraphics[width=\columnwidth]{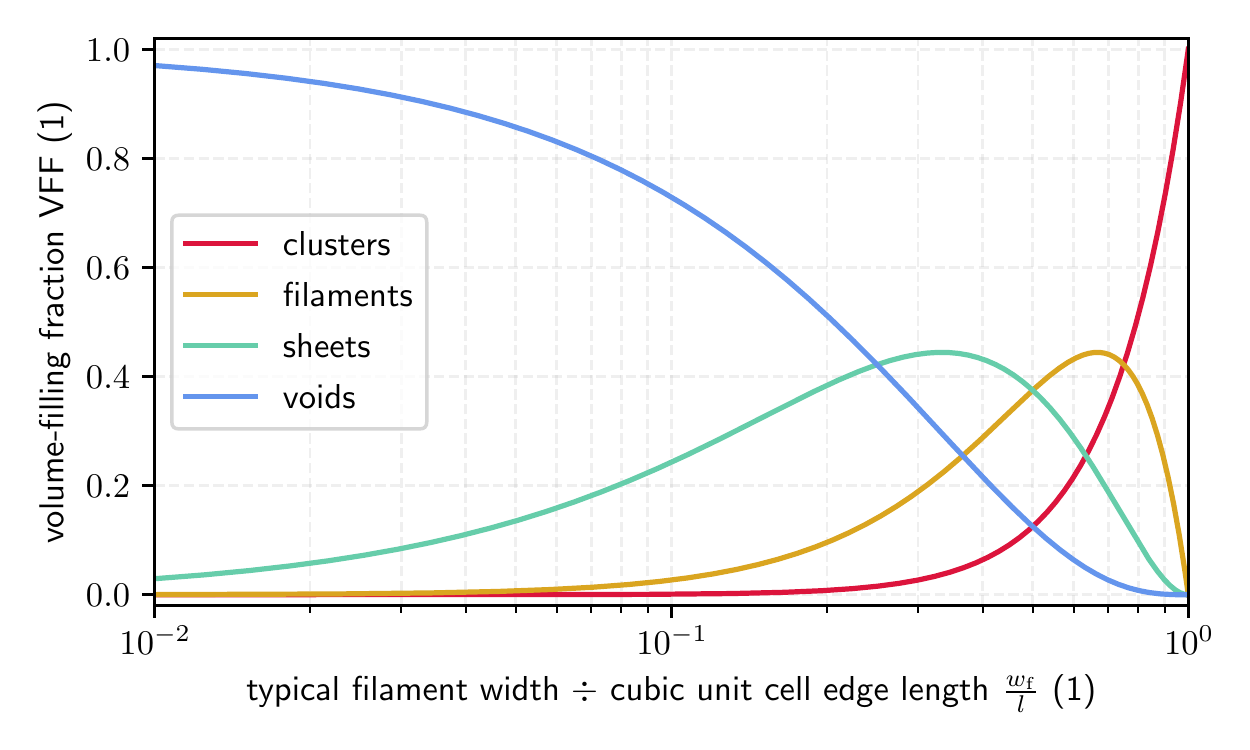}
    \caption{The cubic unit cell model of \textbf{Section}~\ref{sec:model} predicts cosmic web structure type volume-filling fractions (VFFs) that depend on just one parameter: the ratio between the (comoving) filament width and the (comoving) cubic unit cell edge length $\frac{w_\mathrm{f}}{l}$.}
%Volume-filling factors of Cosmic Web structure types according to the cubic unit cell model of \textbf{Section}~\ref{sec:model}.}
    \label{fig:VFFs}
\end{figure}
In this appendix, we compute the volume-filling fractions (VFFs) of the four canonical structure types (clusters, filaments, sheets and voids) as predicted by the cubic unit cell geometric model developed in \textbf{Section}~\ref{sec:model}.\\
A single cubic unit cell features two typical lengthscales: a large scale ($l_\mathrm{f}$), and a small scale ($w_\mathrm{f}$), which can be interpreted as the typical filament length and width, respectively.
The cube obtained by raising the large scale to the third power, represents a void.
Similarly, the three rectangular cuboids obtained by taking the product of the square of the large scale, and the small scale, resemble three sheets.
The three rectangular cuboids obtained by taking the product of the large scale, and the square of the small scale, resemble three filaments.
Finally, the cube obtained by raising the small scale to the third power, resembles a cluster.
The natural volume-filling fractions (VFFs) suggested by this geometry are thus
\begin{align}
    \mathrm{VFF}_\mathrm{c} \left(\frac{w_\mathrm{f}}{l}\right) &= \left(\frac{w_\mathrm{f}}{l}\right)^3 \\
    \mathrm{VFF}_\mathrm{f} \left(\frac{w_\mathrm{f}}{l}\right) &= 3 \left(\frac{w_\mathrm{f}}{l}\right)^2 \left(1 - \frac{w_\mathrm{f}}{l}\right) \\
    \mathrm{VFF}_\mathrm{s} \left(\frac{w_\mathrm{f}}{l}\right) &= 3 \frac{w_\mathrm{f}}{l} \left(1 - \frac{w_\mathrm{f}}{l}\right)^2 \\
    \mathrm{VFF}_\mathrm{v} \left(\frac{w_\mathrm{f}}{l}\right) &= \left(1 - \frac{w_\mathrm{f}}{l}\right)^3
\end{align}
See \textbf{Figure}~\ref{fig:VFFs}.
To see that the VFFs sum to 1, we rewrite 1 using the binomial theorem:
\begin{align}
    1 = \left(\frac{w_\mathrm{f}}{l} + 1 - \frac{w_\mathrm{f}}{l}\right)^3 = \sum_{n=0}^3 \binom{3}{n} \left(\frac{w_\mathrm{f}}{l}\right)^n \left(1 - \frac{w_\mathrm{f}}{l}\right)^{3-n},
\end{align}
and recognise the VFFs as the four terms in this expansion.\\
The VFFs obtained from this \emph{one}-parameter model for $\frac{w_\mathrm{f}}{l} = 10^{-1}$, are similar to those retrieved by \textcolor{blue}{\citet{Forero-Romero12009}} from cosmological simulations for eigenvalue threshold $\lambda_\mathrm{th} = 1$ and effective smoothing scales $R_\mathrm{eff} \sim 10^0\ \mathrm{Mpc}$ (see \textbf{Table}~\ref{tab:VFFs}).

\section{Notation}
This paper adopts the SI system of units (for formulae concerning electromagnetism), and the following symbols.
We list dimensionalities in SI base units with the radian `rad' appended.
Current-day quantities are subscripted with a zero: e.g. $a_0$ is today's scale factor, while $a$ is the scale factor for arbitrary times.
Upstream and downstream quantities are subscripted with a `u' or `d', respectively: e.g. $T_\mathrm{u}$ is the upstream plasma temperature, whilst $T$ is the general plasma temperature.

\[
\begin{tabularx}{\columnwidth}{llX}
$\alpha$ & $\mathrm{1}$ & integrated synchrotron spectral index\\
$\alpha_\mathrm{c}$ & $\mathrm{1}$ & typical cluster $\alpha$\\
$\alpha_\mathrm{f}$ & $\mathrm{1}$ & typical filament $\alpha$\\
$\bar{\alpha}$ & $\mathrm{1}$ & MEC-weighted mean $\alpha$\\
%\end{tabularx}
%\]
%\[
%\begin{tabularx}{\columnwidth}{llX}
$\beta_\mathrm{c}$ & $\mathrm{1}$ & typical cluster MEC $\left(1+z\right)$ power law exponent\\
$\beta_\mathrm{f}$ & $\mathrm{1}$ & typical filament MEC $\left(1+z\right)$ power law exponent\\
$\gamma$ & $\mathrm{1}$ & adiabatic index\\
$\theta$ & $\mathrm{rad}$ & declination (J2000)\\
$\mu_0$ & $\mathrm{kg\ m\ s^{-2}\ A^{-2}}$ & vacuum permeability\\
$\nu$ & $\mathrm{s^{-1}}$ & emission frequency\\
$\nu_\mathrm{obs}$ & $\mathrm{s^{-1}}$ & observing frequency\\
$\nu_\mathrm{ref}$ & $\mathrm{s^{-1}}$ & reference frequency (see $\mathcal{C}$)\\
$\xi_e$ & 1 & electron acceleration efficiency\\
$\Xi_\rho$ & 1 & CDF of RV $j_\nu\ \vert\ \rho$\\
$\rho$ & $\mathrm{kg\ m^{-3}}$ & total matter density\\
$\rho_\mathrm{BM}$ & $\mathrm{kg\ m^{-3}}$ & baryonic matter density\\
$\rho_\mathrm{c}$ & $\mathrm{kg\ m^{-3}}$ & critical density\\
$\rho_\mathrm{DM}$ & $\mathrm{kg\ m^{-3}}$ & dark matter density\\
$\sigma$ & $\mathrm{kg\ s^{-3}\ K^{-4}}$ & Stefan--Boltzmann constant\\
$\phi$ & $\mathrm{rad}$ & right ascension (J2000)\\
$\Phi$ & 1 & CDF of standard normal RV\\
$\Psi$ & 1 & as in \textcolor{blue}{\citet{Hoeft12007}}\\
$\Omega_{\Lambda}$ & $\mathrm{1}$ & dark energy density parameter\\
$\Omega_{\mathrm{BM}}$ & $\mathrm{1}$ & baryonic matter density parameter\\
$\Omega_{\mathrm{DM}}$ & $\mathrm{1}$ & dark matter density parameter\\
$\Omega_{\mathrm{K}}$ & $\mathrm{1}$ & curvature density parameter\\
$\Omega_{\mathrm{M}}$ & $\mathrm{1}$ & matter density parameter\\
$\Omega_{\mathrm{R}}$ & $\mathrm{1}$ & relativistic particle (i.e. photon and neutrino) density parameter\\
\end{tabularx}
\]
\[
\begin{tabularx}{\columnwidth}{llX}
$a$ & $\mathrm{1}$ & scale factor\\
$A$ & $\mathrm{m}^2$ & shock surface area\\
$B$ & $\mathrm{kg\ s^{-2}\ A^{-1}}$ & proper magnetic field strength\\
$B_\mathrm{CMB}$ & $\mathrm{kg\ s^{-2}\ A^{-1}}$ & CMB magnetic field strength\\
$c$ & $\mathrm{m\ s^{-1}}$ & speed of light \textit{in vacuo}\\
$c_\mathrm{s}$ & $\mathrm{m\ s^{-1}}$ & speed of sound
\end{tabularx}
\]
\[
\begin{tabularx}{\columnwidth}{llX}
$\mathcal{C}$ & 1 & typical cluster-to-filament synchrotron MEC ratio at $\nu=\nu_\mathrm{ref}$ and $z=0$\\
$d$ & $\mathrm{m}$ & initial distance between test particle and equidense cluster centre, filament axis or sheet midplane\\
$d_n$ & $\mathrm{m}$ & comoving distance to the $n$-th unit cell boundary crossing\\
$\mathrm{d}\Omega^2$ & $\mathrm{rad^2}$ & solid angle element\\
$\mathrm{d}s^2$ & $\mathrm{m^2}$ & spacetime line element\\
$E$ & $1$ & dimensionless Hubble parameter\\
$F_{\bar{Z}}$ & 1 & CDF of $\bar{Z}$\\
$G$ & $\mathrm{kg^{-1}\ m^3\ s^{-2}}$ & Newton's gravitational constant\\
$h$ & 1 & Hubble constant divided by $100\ \mathrm{km\ s^{-1}\ Mpc^{-1}}$\\
$H$ & $\mathrm{s^{-1}}$ & Hubble parameter\\
$I_{\nu}$ & $\mathrm{kg\ s^{-2}\ rad^{-2}}$ & specific intensity (observed)\\%(now, Earth-centred)
$j_\nu$ & $\mathrm{kg\ m^{-1}\ s^{-2}\ rad^{-2}}$ & proper (not comoving) MEC\\
$\mathbf{k}$ & $\mathrm{m}^{-1}$ & Fourier dual of $\mathbf{r}$\\
$k$ & $\mathrm{m^{-2}}$ & FLRW (Gaussian) curvature of the Universe\\
$l$ & $\mathrm{m}$ & comoving cubic unit cell edge length\\
$l_\mathrm{f}$ & $\mathrm{m}$ & comoving filament length\\
$l_\mathrm{SE}$ & $\mathrm{m}$ & SE kernel lengthscale\\
$\mathcal{L}$ & - & light ray path\\
$m_r$ & $\mathrm{1}$ & Petrosian r-band apparent magnitude\\
$M$ & 1 & number of sightlines\\
$\mathfrak{M}$ & - & cosmological model parameter tuple\\
$\mathcal{M}_\mathrm{u}$ & $\mathrm{1}$ & upstream shock Mach number\\
$n_e$ & $\mathrm{m^{-3}}$ & electron number density\\
$N$ & 1 & number of unit cell boundary crossings considered\\
$p_\mathrm{c-f}$ & 1 & probability to pierce through a cluster, then a filament\\
$p_\mathrm{f}$ & 1 & probability to pierce through a filament only\\
$p_\mathrm{s-v}$ & 1 & probability to pierce through a sheet, then a void\\
$P_\nu$ & $\mathrm{kg\ m^{2}\ s^{-2}}$ & proper power density\\
$\mathcal{P}$ & 1 & percentile random field\\
$\mathbf{r}$ & $\mathrm{m}$ & comoving position vector\\
$r$ & $\mathrm{m}$ & radial comoving distance\\
$\hat{r}$ & $\mathrm{1}$ & sky direction unit vector\\
$\hat{r}_i$ & $\mathrm{1}$ & sky direction unit vector of ray $i$\\
$R$ & $\mathrm{m}$ & equidense cluster radius, filament radius or sheet half-width\\
$\mathcal{R}$ & - & comoving reconstruction region; subset of $\mathbb{R}^3$\\
$S_k$ & $\mathrm{m}$ & transverse comoving distance
\end{tabularx}
\]
\[
\begin{tabularx}{\columnwidth}{llX}
$t$ & $\mathrm{s}$ & physical time since the Big Bang\\
$T$ & $\mathrm{K}$ & proper plasma temperature\\
$T_\mathrm{CMB}$ & $\mathrm{K}$ & CMB temperature\\
$u_\mathrm{EM}$ & $\mathrm{kg\ m^{-1}\ s^{-2}}$ & electromagnetic energy density\\
$v$ & $\mathrm{m\ s^{-1}}$ & shock or test particle velocity\\
$V$ & $\mathrm{m}^3$ & shock effective volume\\
$w_\mathrm{f}$ & $\mathrm{m}$ & typical comoving filament width\\
$X_n$ & 1 & relative specific intensity contribution of the $n$-th newly-entered unit cell\\
$X_{nm}$ & 1 & relative specific intensity contribution of the $n$-th newly-entered unit cell for the $m$-th ray\\
$\langle y \rangle$ & $\mathrm{m}$ & shock effective width\\
$z$, $z_\mathfrak{M}$ & $\mathrm{1}$ & cosmological redshift (under cosmological model $\mathfrak{M}$)\\
$z_\mathrm{max}$ & 1 & cosmological redshift up to which LSS is considered\\
$z_n$ & $\mathrm{1}$ & cosmological redshift of $n$-th unit cell boundary crossing\\
$\bar{z}$ & 1 & specific-intensity-weighted mean redshift\\
$\bar{\bar{z}}$ & 1 & flux-density-weighted mean redshift\\
$Z$ & 1 & cosmological redshift RV\\
$\bar{Z}$ & 1 & specific-intensity-weighted mean redshift RV\\
$\mathcal{Z}$ & 1 & standard normal GRF
\end{tabularx}
\]

\end{document}